\documentclass[aps,showpacs,rmp,twocolumn,superscriptaddress, floatfix]{revtex4-2}

\usepackage[colorlinks,urlcolor=blue,linkcolor=blue,anchorcolor=blue,citecolor=blue]{hyperref}

\usepackage{amsmath}
\usepackage{graphicx}
\usepackage{amsthm}

\usepackage{color}
\usepackage{amsfonts}
\usepackage{tikz}
\usepackage{esint}
\usepackage{amssymb}
\usepackage{mathrsfs}
\usepackage{braket}
\usepackage{lettrine}
\usepackage[shortlabels]{enumitem}
\usepackage{lipsum}
\usepackage{array}
\usepackage{dsfont}
\usepackage[ruled,vlined]{algorithm2e}

\usepackage{caption}
\usepackage{subcaption}
\DeclareMathOperator{\Tr}{Tr}
\begin{document}
\title{Quantum Monte Carlo in the Age of Many-Body Quantum Information}

\author{Yi-Ming Ding}
\email{dingyiming@westlake.edu.cn}
\affiliation{Department of Physics, School of Science and Research Center for Industries of the Future, Westlake University, Hangzhou 310030,  China}
\affiliation{Institute of Natural Sciences, Westlake Institute for Advanced Study, Hangzhou 310024, China}
\affiliation{State Key Laboratory of Surface Physics and Department of Physics, Fudan University, Shanghai 200438, China}

\author{Bin-Bin Mao}
\email{maobinbin@uhrs.edu.cn}
\affiliation{Center for Quantum Sciences and School of Physics, Northeast Normal University, Changchun, Jilin, 130024, China}

\author{Zheng Yan}
\email{zhengyan@westlake.edu.cn}
\affiliation{Department of Physics, School of Science and Research Center for Industries of the Future, Westlake University, Hangzhou 310030,  China}
\affiliation{Institute of Natural Sciences, Westlake Institute for Advanced Study, Hangzhou 310024, China}

\begin{abstract}
Quantum Monte Carlo (QMC) methods are among the central numerical tools for studying strongly correlated quantum many-body systems, particularly in higher dimensions. As quantum information has introduced new information-theoretic perspectives and diagnostics into many-body physics, QMC methods have accordingly been extended beyond the measurement of conventional linear observables. This review summarizes recent progress in adapting QMC to many-body quantum-information, focusing on qubit or spin-$1/2$ systems as a concrete setting while keeping the discussion broadly applicable to qudit and bosonic systems.
We present a unified perspective on the extraction of nonlinear diagnostics, including entanglement entropies and entanglement spectra, Rényi negativities for mixed-state entanglement, stabilizer entropies for quantum magic, and decoherence-driven phenomena such as the interplay between imaginary-time evolution and decoherence and strong-to-weak spontaneous symmetry breaking.
\end{abstract}

\maketitle
\tableofcontents

\section{Introduction}
The past two decades have witnessed a deep and fruitful interplay between quantum information science and quantum many-body physics. Concepts originally developed in the context of quantum computation and information, such as entanglement, have become indispensable tools, reshaping our understanding of collective phenomena and establishing a new paradigm for investigating many-body systems~\cite{Amico2008ent-review,Eisert2010ent-review,zengb2019quantummatter,Laflorencie2016review}. 
This conceptual unification has been accompanied and reinforced by rapid advances in experimental platforms, which enable the simulation and discovery of novel quantum phases of matter~\cite{Bloch2012quantum-sim,Georgescu2014quantum-sim,Kaufman2016thermalization,Brydges2019renyi,Shaw2024rydberg-entanglement}.

In this context, numerical simulation serves as a crucial bridge between theoretical concepts and experimental observations. While analytical approaches provide essential insights, the complexity of interacting quantum many-body systems, particularly in strongly correlated regimes, often necessitates large-scale numerical simulations.
Among these, tensor network approaches, notably the density matrix renormalization group (DMRG) formulated in terms of matrix product states (MPS), have achieved remarkable success~\cite{White1992dmrg,White1993dmrg,Ostlund1995mps,Schollwock2011dmrg,Cirac2021mps-peps,Vidal2003slightly,Vidal2004tebd,Verstraete2004dmrg-mps}. By exploiting the area-law entanglement structure of ground states of local Hamiltonians, they provide efficient representations of many-body states and are particularly powerful for studying ground-state and low-energy properties in $(1+1)$D systems. 
Despite these successes, the applicability of tensor network methods can become more challenging in higher dimensions or in situations involving substantial entanglement growth. In such regimes, complementary numerical approaches are essential.

Quantum Monte Carlo (QMC) methods extend the statistical-sampling philosophy of
classical Monte Carlo (MC) to quantum many-body systems, providing a powerful and broadly applicable framework for numerically exact simulations of interacting quantum many-body systems, up to controllable systematic and statistical errors. A key advantage of QMC is that, its efficiency is not intrinsically limited by spatial dimensionality or by the growth of entanglement, enabling controllable and scalable numerical studies in two and higher dimensions. The primary limitation is instead the sign problem~\cite{Troyer2005sign,Henelius2000sign,Li2019signfree,Pan2024sign}, which restricts the class of models that can be simulated efficiently. 
Nevertheless, for qubit, qudit, and bosonic systems, there exists a broad and physically important class of stoquastic Hamiltonians that admit sign-problem-free QMC representations~\cite{Bravyi2008stoquastic,Bravyi2006stoqma}. These models already realize a wide range of quantum phases and phase transitions, including conventional symmetry-breaking orders, topological and symmetry-protected phases, as well as quantum critical points described by conformal field theories (CFTs). 
% Over the past decades, alongside advances in computational capabilities, QMC has become an indispensable tool for the quantitative study of quantum many-body phenomena. In particular, it enables unbiased and high-precision determination of critical properties, including critical exponents and conformal data such as central charges.

In the age of many-body quantum information, the scope of physical observables of interest has significantly expanded beyond conventional quantities. In particular, diagnosing quantum phases and dynamics increasingly relies on nonlinear probes of the density matrix, such as entanglement entropies, which provide refined characterizations of quantum states beyond traditional order parameters~\cite{Amico2008ent-review,Eisert2010ent-review,Laflorencie2016review,
Calabrese2009ent-cft,Horodecki2009ent-review,Latorre2009ent-spin-review,DeChiara2018correlations-review,Frerot2023quantum-correlations,winter2022mbm}. 
Meeting this demand has driven significant developments in QMC methods over the past decades.

New formulations based on replica constructions, partition-function-ratio estimators, and quantum-inspired techniques have substantially broadened the range of accessible observables. These developments will be reviewed in detail in the following sections. At the same time, these developments are not limited to new observables, but also extend to new classes of quantum states, including replicated states and decohered states. These developments signal a transition toward a modern formulation of QMC, in which traditional statistical sampling is seamlessly combined with quantum information concepts to access increasingly sophisticated probes of many-body systems.

A central question is how far these extensions can be developed while
preserving the main strengths of QMC. A complete answer is not yet available,
partly because different QMC formulations can differ substantially.
We therefore clarify the scope of this review. We focus on QMC formulations for
qubit, or equivalently spin-$1/2$, systems, which provide a natural setting for
quantum-information diagnostics. This focus is mainly a matter of presentation:
most of the ideas discussed below extend, with suitable modifications, to more
general qudit and bosonic systems. Fermionic QMC formulations are beyond the scope of this review. 
In this setting, the target quantum state or density matrix is
represented through statistical weights generated by Hamiltonians, projectors,
replicas, operator insertions, quantum channels, or related generalized
constructions, as discussed in Sec.~\ref{sec:qmc_form}. Variational MC
approaches are not covered here.

% We hope the review can help understand which nonlinear observables admit
% efficient and stable QMC estimators, and how their statistical properties scale
% with system size, inverse temperature, and the structure of the underlying
% quantum state. When these observables are applied beyond equilibrium Gibbs
% states or ground states, for example to decohered mixed states, the efficient
% preparation and representation of the corresponding state within QMC become
% additional nontrivial issues, as discussed in Sec.~\ref{sec:state_constrct}.

% A key requirement is that these extensions should not introduce new
% prohibitive costs. In particular, they should avoid effective sign problems,
% severe variance growth, or exponential degradation of signal-to-noise ratios in
% otherwise sign-problem-free settings. They should also admit efficient and
% ergodic Monte Carlo update schemes, so that the gain from new estimators is not
% lost to long autocorrelation times or poorly scaling sampling algorithms. 

In line with the above background and motivation, we now outline the
organization of this review. In Sec.~\ref{sec:qmc_form}, we introduce QMC
representations of several classes of quantum states and density matrices,
including conventional Gibbs states, ground-state projector constructions,
reduced density matrices, replicated states, operator-inserted states, and
states transformed by quantum channels. 
The general open-boundary framework introduced in
Sec.~\ref{sec:open-general} provides a unified viewpoint for these constructions,
organizing them in terms of configuration spaces, boundary conditions, and
generalized partition functions. This viewpoint will be used throughout the
review as the basis for formulating general nonlinear information-theoretic diagnostics.

Since the choice of basis strongly affects how easily different observables can be evaluated within QMC, selecting an appropriate basis can make certain observables computationally tractable. In Sec.~\ref{sec:special_basis}, we review two representative basis choices beyond the standard local-site basis. One is based on the valence-bond representation within projector QMC (see Sec.~\ref{sec:vb_basis}), while the other is a quantum-inspired approach based on Bell sampling from quantum computation (see Sec.~\ref{sec:bell_basis}). 
Both methods enable direct estimators for SWAP operators, whose expectation values yield the Rényi-2 entropy (see Sec.~\ref{sec:swap}). The latter further allows for sampling of the Pauli spectrum, which is essential for the characterization of quantum magic.

After that, we turn to QMC techniques that are less tied to a particular basis
choice. In Secs.~\ref{sec:pfr}, we introduce one of the
most general frameworks currently available for evaluating nonlinear
observables: the formulation in terms of generalized partition-function ratios.
Within this framework, a broad class of linear and nonlinear observables can be
recast as ratios, or differences of logarithms of ratios, between suitably
defined QMC partition functions.
For concreteness, we use the Rényi-2 entropy as a guiding example. We review
several techniques developed for such ratio estimations, including direct
estimators, reweighting--annealing methods,
thermodynamic integration, and nonequilibrium extensions based on the
Jarzynski equality. We also discuss extended-ensemble methods and the
incremental (or ratio-decomposition) trick, which are designed to mitigate the
overlap problem and improve sampling efficiency.

Building on the techniques developed in the preceding sections, Sec.~\ref{sec:nonlinear} provides an application map of QMC methods for many-body quantum information. We review how the constructions and estimators introduced earlier can be used to access a broad range of information-theoretic diagnostics and discuss the classes of many-body problems to which these methods can be applied.

Finally, in Sec.~\ref{sec:summary}, we summarize the main developments and discuss future directions. Our aim is to provide a coherent overview of this rapidly evolving field while highlighting its central ideas, methodological advances, and outstanding challenges. Given the breadth of the subject, the present review is necessarily not exhaustive. We hope that the perspective developed here will serve as a useful guide and entry point for readers and help stimulate further progress at the interface of QMC and many-body quantum information.

\section{QMC Representations of Quantum States}\label{sec:qmc_form}

\subsection{Conventions and notation}
\label{sec:conventions}

Before proceeding, we summarize the main conventions and notation used
throughout this review. Readers may return to this section whenever necessary.

\begin{itemize}
     \item $H$ denotes a local many-body Hamiltonian defined on a finite system of qubits, and $N$ denotes the total number of qubits. For a finite lattice model in spatial dimension $d$, we write
    $N \propto L^d$,
    where $L$ is the linear system size. When one qubit is associated with each lattice site, $N$ also coincides with the total number of sites.

    \item We denote the Hilbert space of the full system by
    \begin{equation}
        \mathcal H
        =
        \bigotimes_{i=1}^N \mathcal H_i
        =
        \mathcal H_1\otimes\cdots\otimes\mathcal H_N ,
    \end{equation}
    where $\mathcal H_i\simeq\mathbb C^2$ is the local Hilbert space of qubit
    $i$. Pure states are denoted by kets $\ket{\Psi}\in\mathcal H$ satisfying
    $\langle\Psi|\Psi\rangle=1$. The space of bounded linear operators acting on
    $\mathcal H$ is denoted by $\mathcal B(\mathcal H)$. Thus, a density matrix
    of the full system is an operator $\rho\in\mathcal B(\mathcal H)$ satisfying
    $\rho\ge 0$ and $\Tr(\rho)=1$.

    \item Unless otherwise specified, we use the local computational basis, namely
    the eigenbasis of $\sigma^z_i$ on each qubit,
    \begin{equation}
        \ket{s}
        =
        \bigotimes_{i=1}^N \ket{s_i}
        =
        \ket{s_1}\otimes\cdots\otimes\ket{s_N}.
    \end{equation}
    Here $s_i\in\{0,1\}$ labels the local state of qubit $i$, with the convention
    \begin{equation}
        \ket{0}\equiv\ket{\uparrow},
        \qquad
        \ket{1}\equiv\ket{\downarrow},
    \end{equation}
    such that
    \begin{equation}
        \sigma_i^z\ket{0_i}=\ket{0_i},
        \qquad
        \sigma_i^z\ket{1_i}=-\ket{1_i}.
    \end{equation}
    Other basis choices will be introduced explicitly when needed, such as the
    valence-bond basis in Sec.~\ref{sec:vb_basis} and the Bell basis in
    Sec.~\ref{sec:bell_basis}.

    \item  Since QMC samples configurations with weights proportional to a
    normalization factor, the density operators or states appearing in the
    simulation are often unnormalized. We denote such an unnormalized density
    operator by $\tilde{\rho}$, with
    \begin{equation}
        \tilde{\rho}\geq 0,
        \qquad
        \Tr(\tilde{\rho})>0 .
    \end{equation}
    The corresponding normalized density matrix is
    \begin{equation}
        \rho
        =
        \frac{\tilde{\rho}}{\Tr(\tilde{\rho})} .
    \end{equation}
   Similarly, for an unnormalized state vector $\ket{\tilde{\Psi}}$, we denote the
corresponding normalized state by
\begin{equation}
    \ket{\Psi}
    =
    \frac{
        \ket{\tilde{\Psi}}
    }{
        \sqrt{\braket{\tilde{\Psi}|\tilde{\Psi}}}
    } .
\end{equation}

    \item When discussing a bipartite system, we partition the full system into
two subsystems $A$ and $B$, denoted by $A\cup B$. The Hilbert space then
factorizes as $\mathcal H
    =
    \mathcal H_A\otimes\mathcal H_B$,
where
\begin{equation}
    \mathcal H_A=\bigotimes_{i\in A}\mathcal H_i,
    \qquad
    \mathcal H_B=\bigotimes_{i\in B}\mathcal H_i .
\end{equation}

For a normalized density matrix $\rho\equiv\rho_{AB}$, the corresponding
reduced density matrices (RDMs) are
\begin{equation}
    \rho_A=\Tr_B(\rho),
    \qquad
    \rho_B=\Tr_A(\rho),
\end{equation}
which are automatically normalized,

Similarly, for an unnormalized density matrix
$\tilde\rho\equiv\tilde\rho_{AB}$, we define the corresponding unnormalized
RDMs as
\begin{equation}
    \tilde\rho_A=\Tr_B(\tilde\rho),
    \qquad
    \tilde\rho_B=\Tr_A(\tilde\rho).
\end{equation}
Their traces satisfy
\begin{equation}
    \Tr_A(\tilde\rho_A)
    =
    \Tr_B(\tilde\rho_B)
    =
    \Tr(\tilde\rho).
\end{equation}
Consequently, the RDMs can be normalized using the same factor as the full density matrix:
\begin{equation}
    \rho_A
    =
    \frac{\tilde\rho_A}{\Tr(\tilde\rho)},
    \qquad
    \rho_B
    =
    \frac{\tilde\rho_B}{\Tr(\tilde\rho)}.
\end{equation}

    \item For a bipartite system $A\cup B$, we denote the partial transpose with respect to subsystem $B$ by
$\rho^{T_B}$ or $\rho_{AB}^{T_B}$.
In a product basis, the density matrix can be written as
\begin{equation}
    \rho_{AB}
    =
    \sum_{\substack{s_A,s_B\\ s_A',s_B'}}
    \rho_{s_A s_B,\,s_A' s_B'}
    \ket{s_A,s_B}\bra{s_A',s_B'}.
\end{equation}
The partial transpose with respect to subsystem $B$ is defined by keeping the matrix elements unchanged while exchanging the subsystem-$B$ basis states between the ket and bra:
\begin{equation}
    \rho_{AB}^{T_B}
    =
    \sum_{\substack{s_A,s_B\\ s_A',s_B'}}
    \rho_{s_A s_B,\,s_A' s_B'}
    \ket{s_A,s_B'}\bra{s_A',s_B}.
\end{equation}
Thus, the partial transpose leaves the subsystem-$A$ indices unchanged and exchanges the subsystem-$B$ indices between the ket and bra. One can similarly discuss an unnormalized density matrix $\tilde{\rho}$.

\item For a subsystem $A$, we denote by $|A|$ the number of qubits, or lattice
sites, contained in $A$. We denote its boundary by $\partial A$, and
$|\partial A|$ denotes the number of boundary sites.

    \item Each QMC formulation samples an effective ensemble specified by a
normalization factor, or \emph{generalized partition function}. For a simulated
density matrix $\tilde\rho$ or wave function $\ket{\tilde\Psi}$, we denote this
normalization by $Z(\tilde\rho)$ or $Z(\ket{\tilde\Psi})$, respectively. For replicated
density matrices, we use the convention
\begin{equation}
    Z^{(\alpha)}(\tilde\rho)
    \equiv
    Z(\tilde\rho^\alpha),
    \qquad
    \alpha\in\mathbb N^+ .
\end{equation}
For example, when the normalization is the trace of a replicated state, one has
$Z^{(\alpha)}(\tilde\rho)=\Tr(\tilde\rho^\alpha)$.

    \item In a given QMC representation, a generalized partition function takes
    the form
    \begin{equation}
        Z
        =
        \sum_{\mathcal C\in\Omega}
        W(\mathcal C),
    \end{equation}
    where $\mathcal C$ denotes a configuration, $\Omega$ is the configuration
    space, and $W(\mathcal C)$ is the configuration weight. A sign-problem-free
    probabilistic interpretation requires
    \begin{equation}
        W(\mathcal C)\in\mathbb{R}^+,
        \qquad
        \forall\,\mathcal C\in\Omega .
    \end{equation}
    If no such representation is available, the formulation is said to 
    suffer from a sign problem.

   \item For a physical observable $O$, its expectation value is denoted by
$\langle O\rangle$. In QMC, it is estimated using a configuration-dependent
estimator $\hat O(\mathcal C)$, which is generally not unique. With the
normalized sampling probability
\begin{equation}
    P(\mathcal C)
    =
    \frac{W(\mathcal C)}{Z},
\end{equation}
the expectation value is written as
\begin{equation}
    \langle O\rangle
    =
    \sum_{\mathcal C\in\Omega}
    P(\mathcal C)\hat O(\mathcal C)
    =
    \langle \hat O\rangle_{\mathrm{MC}} .
\end{equation}
When no confusion can arise, we also write
$\langle \hat O\rangle\equiv\langle \hat O\rangle_{\mathrm{MC}}$. The variance
of the estimator is denoted by
\begin{equation}
    \mathrm{Var}(\hat O)
    =
    \left\langle
        \hat O^2
    \right\rangle
    -
    \left\langle
        \hat O
    \right\rangle^2 .
\end{equation}
\end{itemize}

% :::::::::::::::::::::::::::::::::::::::::::::::::
%   NEW SUBSECTION
% :::::::::::::::::::::::::::::::::::::::::::::::::
\subsection{General configuration-space representations}\label{sec:general-rho}

\subsubsection{Open-time-boundary density matrices}\label{sec:open-general}
\begin{figure}[ht]
    \centering
    \begin{subfigure}[b]{0.45\linewidth}
        \centering
        \includegraphics[width=\linewidth]{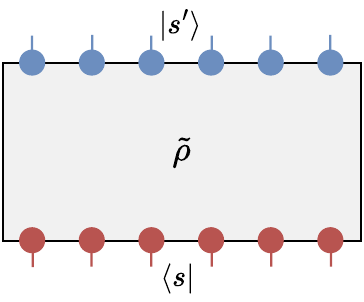}
        \caption{}
    \end{subfigure}
    \hspace{0.01\textwidth}
    \begin{subfigure}[b]{0.45\linewidth}
        \centering
        \includegraphics[width=\linewidth]{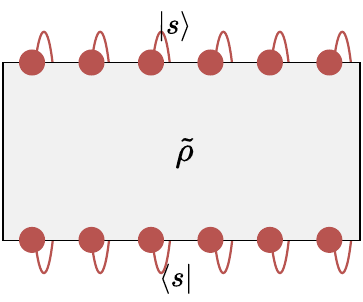}
        \caption{}
    \end{subfigure}
    \caption{
Diagrammatic illustration of a density matrix and its trace.
(a) An open-time-boundary matrix element $\langle s|\tilde\rho|s'\rangle$, where $\tilde\rho$
connects the ket boundary state $\ket{s'}$ to the bra boundary state
$\bra{s}$ along the transfer direction.
(b) The trace $\Tr(\tilde\rho)$, obtained by identifying $s=s'$, closing the boundary
legs, and summing over the common boundary state.
Only the boundary states are shown explicitly; representation-dependent auxiliary variables, such as SSE operator strings or path-integral imaginary-time histories, are omitted.
}
    \label{fig:rho}
\end{figure}

Consider a general density matrix such as the Gibbs operator $\tilde\rho=e^{-\beta H}$, which will be discussed in Sec.~\ref{sec:Gibbs-state}. In an orthonormal
computational basis $\{\ket{s}\}$, it can be expanded as
\begin{equation}\label{eq:rho-expd}
    \tilde\rho
    =
    \sum_{s,s'}
    \tilde\rho_{s,s'}\ket{s}\bra{s'},
    \qquad
    \tilde\rho_{s,s'}
    =
    \langle s|\tilde\rho|s'\rangle .
\end{equation}
The matrix element $\langle s|\tilde\rho|s'\rangle$ may be viewed
diagrammatically as a transfer process: $\tilde\rho$ connects the ket boundary state
$\ket{s'}$ to the bra boundary state $\bra{s}$, as illustrated in
Fig.~\ref{fig:rho}(a). 
Here the word ``boundary'' refers to the two ends of this time-like transfer direction, not to a physical spatial boundary of the lattice. This transfer direction need not be identified with imaginary time, as $\tilde\rho$ can be a general operator rather than a Gibbs operator.
Thus a general matrix element $\tilde\rho_{s,s'}$ has two open boundary indices, corresponding to the ket and bra states.

The trace is obtained by identifying these boundary indices and summing over the common boundary state,
\begin{equation}\label{eq:zrho-diag}
    Z(\tilde\rho)
    :=
    \Tr(\tilde\rho)
    =
    \sum_s
    \langle s|\tilde\rho|s\rangle ,
\end{equation}
as shown in Fig.~\ref{fig:rho}(b). For $\tilde\rho=e^{-\beta H}$, this reduces to
the usual canonical partition function. 

The partition function $Z(\tilde\rho)=\Tr(\tilde\rho)$ is the standard starting point for
conventional QMC simulations. Sampling $Z(\tilde\rho)$ gives a stochastic
representation of the diagonal matrix elements $\langle s|\tilde\rho|s\rangle$ in a
chosen basis, and the efficiency of this representation determines which
observables can be measured with controlled statistical errors.

This perspective, however, naturally extends beyond diagonal matrix elements.
By sampling off-diagonal matrix elements, one can construct a stochastic
representation of the density matrix itself, rather than only of its partition
function. 
Although this possibility has long been implicit in path-integral formulations~\cite{Ceperley1995path-int,militzer2003off-diagonal-path}, it has only recently been exploited systematically as a practical route to many-body studies. In particular, recent works have shown that this density-matrix-level viewpoint enables access to information-theoretic quantities beyond the scope of simulations based solely on $Z(\tilde\rho)$ \cite{mao2025samplingrdm,WangTT2025rdm}. 
It is therefore useful to organize the discussion according to the different
classes of quantum states and generalized density matrices that admit QMC
representations in the following sections.

Specifically, QMC avoids the explicit construction of $\tilde\rho$, whose storage
cost grows exponentially with system size, by introducing an enlarged
configuration space $\Omega$ in which the relevant matrix elements are
decomposed into products of local weights. For example, an open-time-boundary matrix
element may admit a representation (Fig.\ref{fig:rho} (a))
\begin{equation}
    \langle s|\tilde\rho|s'\rangle
    =
    \sum_{\mathcal C\in\Omega(s,s')}
    W(\mathcal C),
\end{equation}
where $\Omega(s,s')$ is the set of configurations compatible with the fixed time-boundary
states $(s,s')$. Closing the boundary indices (Fig.\ref{fig:rho} (b)) gives
\begin{equation}
    Z(\tilde\rho)\equiv \Tr(\tilde\rho)
    =
    \sum_s
    \sum_{\mathcal C\in\Omega(s,s)}
    W(\mathcal C)
    \equiv
    \sum_{\mathcal C\in\Omega}
    W(\mathcal C).
\end{equation}
Here $\mathcal C$ contains not only the boundary basis states, but also the auxiliary degrees of freedom introduced by the chosen representation. 
We define $\Omega \equiv \bigcup_s \Omega(s,s)$ as the full configuration space obtained after closing the boundary indices, namely, the set of all configurations that contribute to $\Tr(\tilde\rho)$.
For Gibbs states, for example, these auxiliary variables will appear as
imaginary-time histories or operator strings (see Sec.~\ref{sec:Gibbs-state}).

\subsubsection{Sign problems and probabilistic interpretations}

The key requirement for QMC sampling is that the weights admit a probabilistic
interpretation, namely $W(\mathcal C)\in\mathbb{R}^+$, $\forall \mathcal{C}\in\Omega$. If no such representation is
available, we say that the construction suffers from a \emph{sign problem} in the chosen
basis and representation. One may formally sample $|W(\mathcal C)|$ and include the sign or phase by reweighting, but the average sign typically decays exponentially with system size or inverse temperature, leading to an exponentially large statistical cost~\cite{Troyer2005sign}.

Unless otherwise specified, we assume throughout this review that the
Hamiltonians and generalized state constructions considered below admit
sign-problem-free QMC representations. Since the sign problem itself is not the
main focus here, we refer the reader to
Refs.~\cite{Troyer2005sign,Henelius2000sign,wu2005sufficient,wei2016majorana,Pan2024sign,Li2019signfree}
for comprehensive reviews.

\subsubsection{Estimators for linear observables}
Once a positive-real-weight configuration-space representation has been constructed,
linear observables can be evaluated as statistical averages over
configurations. For an observable $O$,
\begin{equation}
    \langle O\rangle
    =
    \frac{1}{Z}
    \sum_{\mathcal C\in\Omega}
    W(\mathcal C)\,
    \hat O(\mathcal C),
\end{equation}
where $\hat O(\mathcal C)$ is an estimator associated with the
configuration $\mathcal C$. The estimator is not unique; different choices may
lead to different variances and computational efficiencies.

In practice, $\langle O \rangle$ is estimated by sampling configurations $\{\mathcal{C}_i\}\subseteq \Omega$ drawn from the distribution $P(\mathcal{C}) \equiv W(\mathcal{C})/Z(\tilde\rho)$, 
\begin{equation}
    \langle O \rangle
    \approx
    \frac{1}{N_{\mathrm{MC}}}
    \sum_{i=1}^{N_{\mathrm{MC}}}
    \hat{O}(\mathcal{C}_i).
\end{equation}
where $N_{\mathrm{MC}}$ is the number of MC samples. In the limit
$N_{\mathrm{MC}}\to\infty$, this sample average converges to the ensemble
average $\langle \hat O\rangle_{\mathrm{MC}}=\langle O\rangle$.

The statistical error of this estimator is governed by its variance and the number of effectively independent samples. For uncorrelated samples, one has
\begin{equation}
    \Delta O
    \sim
    \sqrt{\frac{\mathrm{Var}(\hat{O})}{N_{\mathrm{MC}}}},
\end{equation}
implying algebraic convergence with the number of samples. 
In practice, configurations are sampled by constructing a Markov chain, and correlations between successive configurations give rise to an autocorrelation time $\tau_{\mathrm{auto}}$. Consequently, the effective number of statistically independent samples is reduced to$N_{\mathrm{eff}} \sim {N_{\mathrm{MC}}}/{\tau_{\mathrm{auto}}}$.
Efficient QMC simulations therefore require update schemes that reduce autocorrelation times while preserving ergodicity.

As a simple example, consider the two-point correlation function
\begin{equation}
    O = \sigma^z_i \sigma^z_j
\end{equation}
in a qubit system in the computational basis ($\sigma^z$-basis).
Since this operator is diagonal in the chosen basis, for a configuration $\mathcal{C}$ containing a basis state $s$, with $s_k = \pm 1$ the eigenvalue of $\sigma^z_k$, one has
\begin{equation}\label{eq:diag-O-example}
    \hat{O}(\mathcal{C}) = s_i s_j .
\end{equation}
The variance of this estimator is
\begin{equation}
    \mathrm{Var}(\hat{O})
    =
     \langle (\sigma^z_i \sigma^z_j)^2 \rangle - \langle \sigma^z_i \sigma^z_j \rangle^2
     =
     1 - \langle \sigma^z_i \sigma^z_j \rangle^2
     ,
\end{equation}
which is bounded by $0 \le \mathrm{Var}(\hat{O}) \le 1$. Consequently, the statistical error scales as $\Delta O \sim {1}/{\sqrt{N_{\mathrm{eff}}}},$
with a prefactor that remains finite and controlled.

\subsubsection{Direct sampling of density-matrix data}\label{sec:sampling-rdm-data}
As mentioned above, another important aspect of the open-time-boundary representation
in Eq.~\eqref{eq:rho-expd} is that one is not restricted to the diagonal
ensemble in Eq.~\eqref{eq:zrho-diag}. Instead, one can define an open-boundary
ensemble,
\begin{equation}\label{eq:open-ensemble}
    Z_{\mathrm{open}}(\tilde\rho)
    := 
    \sum_{s,s'}
    \langle s|\tilde\rho|s'\rangle
    =
    \sum_{s,s'}
    \sum_{\mathcal C\in\Omega(s,s')}
    W(\mathcal C)
     .
\end{equation}
Here the boundary states $(s,s')$ themselves become part of the sampled
configuration.
Again, this open-boundary ensemble is useful only when the summed open-boundary
weights satisfy $W(\mathcal{C})\in\mathbb{R}^+$ in the chosen representation.

In particular, the marginal probability distribution of the boundary
states satisfies
\begin{equation}
    P(s,s')
    =
    \frac{
        \langle s|\tilde\rho|s'\rangle
    }{
        Z_{\mathrm{open}}(\tilde\rho)
    },
\end{equation}
where
\begin{equation}
    Z_{\mathrm{open}}(\tilde\rho)
    =
    \sum_{s,s'}
    \langle s|\tilde\rho|s'\rangle .
\end{equation}
Thus, by simulating the open-boundary ensemble and recording the frequency with
which each pair of boundary states $(s,s')$ appears, one can stochastically
reconstruct the matrix elements of $\tilde\rho$ in the chosen basis, denoted by $\tilde\rho_{\mathrm{MC}}$. The overall
normalization can then be fixed by imposing $\Tr(\rho_{\mathrm{MC}})=1$ for the
normalized density matrix $\rho_{\mathrm{MC}}=\tilde\rho_{\mathrm{MC}}/\Tr(\tilde\rho_{\mathrm{MC}})$.

Both the storage and the sampling complexity of reconstructing
$\tilde\rho_{\mathrm{MC}}$ grow exponentially with system size, because the full density matrix contains exponentially many matrix elements.
Thus reconstructing the full density matrix is
only feasible for small systems.
This perspective is nevertheless useful when one is interested in RDMs of small subsystems. 
The direct QMC sampling of RDMs was first demonstrated in
Ref.~\cite{mao2025samplingrdm}, and has subsequently been developed and applied
in Refs.~\cite{WangTT2025rdm,Hari2025rdm-rom,Aditya2026rdm,Lyu2025rdm,
Mao2026rdm-tos}. More recently, the same framework has been extended to
generalized operator-inserted density matrices (see
Sec.~\ref{sec:rho-insert}) in Ref.~\cite{wangzy2026rdm}.
 We will return to
these applications in Sec.~\ref{sec:rdm-data-app}.

%=======================================================
%   NEW SUBSECTION
% ======================================================
\subsection{Thermal Gibbs states}\label{sec:Gibbs-state}
The general configuration-space viewpoint introduced above becomes concrete
once a specific representation of the density matrix is chosen. For thermal
equilibrium problems, the central object is the Gibbs operator $\tilde\rho = e^{-\beta H}$,
where $\beta$ is the inverse temperature and $H$ is the many-body Hamiltonian.
Its trace gives the canonical partition function
\begin{equation}
    Z
    =
    \Tr(e^{-\beta H})
    =
    \sum_s
    \langle s|e^{-\beta H}|s\rangle .
\end{equation}

Two widely used QMC representations realize this Gibbs partition function in different
ways: the imaginary-time path-integral
formulation~\cite{Suzuki1976-quantum-to-classical,Prokofev1998wordline,Kawashima2004worldline,Beard1996continuous-qmc,Alet2005continous,Evertz01012003loop}
and the stochastic series expansion (SSE)~\cite{Handscomb1962TheMC,Sandvik1992sse,Sandvik1999sse,Sandvik2003sse,Syljuasen2002sse,Melko2013sse,Melko2005sse,Yan2019sweeping,Yan2022sweeping,Sandvik2019sse}. In both cases, the Hamiltonian can be 
first decomposed into local terms,
\begin{equation}\label{eq:local-decomp}
    H = -\sum_b H_b ,
\end{equation}
where the local operators $H_b$ generally do not commute. The resulting QMC
configurations then encode different representations of the transfer process
generated by the Gibbs operator $e^{-\beta H}$.

\subsubsection{Imaginary-time path-integral representation}
In the path-integral approach, the Gibbs operator $e^{-\beta H}$ is viewed as
an imaginary-time evolution operator acting on basis states. One first
partitions the imaginary-time interval into $M$ steps,
\begin{equation}
    e^{-\beta H}
    =
    \left(e^{-\Delta\tau H}\right)^M,
    \qquad
    \beta = M\Delta\tau .
\end{equation}
Using the local decomposition~\eqref{eq:local-decomp}, each short-time
propagator is approximated by a product of local propagators through a
Trotter--Suzuki decomposition~\cite{Trotter1959product,Suzuki1976}. For example, the first-order decomposition gives
\begin{equation}
    e^{-\Delta\tau H}
    =
    e^{\Delta\tau \sum_b H_b}
    =
    \prod_b e^{\Delta\tau H_b}
    +
    \mathcal{O}(\Delta\tau^2),
\end{equation}
where the error arises from the noncommutativity of different local terms
$H_b$. Higher-order Trotter--Suzuki decompositions replace this simple ordered
product by more symmetric or nested products of local propagators and can
reduce the discretization error to higher powers of $\Delta\tau$~\cite{Childs2021trotter-error}. In what
follows, we use the first-order form for notational simplicity.

With the first-order decomposition, the partition function becomes
\begin{equation}
    Z
    =
    \Tr\left[
    \left(
    \prod_b e^{\Delta\tau H_b}
    \right)^M
    \right]
    +
    \mathcal{O}(\Delta\tau).
\end{equation}
Here the total error in $Z$ is of order $\mathcal{O}(\Delta\tau)$ after
accumulating the local Trotter error over $M=\beta/\Delta\tau$ time slices. In
practice, one chooses $M$ sufficiently large so that the discretization error is
negligible, or extrapolates to the continuous imaginary-time limit.

One then inserts complete sets of basis states between all elementary local
propagators. This gives a path-integral representation of the form
\begin{equation}\label{eq:path-integral-formulation}
    Z
    \approx
    \sum_{\{s_{\tau,\ell}\}}
    \prod_{\tau=0}^{M-1}
    \prod_{\ell}
    \bra{s_{\tau,\ell+1}}
    e^{\Delta\tau H_{b_\ell}}
    \ket{s_{\tau,\ell}},
\end{equation}
where $\tau$ labels the imaginary-time slice, $\ell$ labels the position inside
the local Trotter sequence, and $b_\ell$ denotes the local operator appearing at
position $\ell$. The trace imposes periodic boundary conditions along the
imaginary-time direction: the
last time slice is identified with the first one.

Thus a path-integral configuration can be viewed as an imaginary-time history,
or \emph{worldline} configuration,
\begin{equation}
    \mathcal C=\{s_{\tau,\ell}\},
\end{equation}
consisting of the intermediate basis states generated along the transfer
direction.
For a local Hamiltonian, each matrix element
\begin{equation}
    \bra{s_{\tau,\ell+1}}
    e^{\Delta\tau H_{b_\ell}}
    \ket{s_{\tau,\ell}}
\end{equation}
depends only on the degrees of freedom acted on by the local operator
$H_{b_\ell}$.

\subsubsection{Stochastic-series-expansion representation}\label{sec:sse}

Alternatively, SSE avoids an explicit
imaginary-time discretization by expanding the Gibbs operator directly in a
Taylor series,
\begin{equation}
    e^{-\beta H}
    =
    \sum_{n=0}^{\infty}
    \frac{(-\beta H)^n}{n!}.
\end{equation}
The partition function becomes
\begin{equation}
    Z
    =
    \sum_{n=0}^{\infty}
    \frac{\beta^n}{n!}
    \sum_s
    \bra{s}(-H)^n\ket{s}.
\end{equation}

Using the local decomposition~\eqref{eq:local-decomp}, each power of $-H$ can
be written as a sum over ordered sequences of local operators, usually called
\emph{operator strings},
\begin{equation}
    (-H)^n
    =
    \sum_{b_0,\dots,b_{n-1}}
    H_{b_{n-1}}\cdots H_{b_1}H_{b_0}.
\end{equation}
Thus, in contrast to the path-integral representation, SSE samples the operator products directly rather than discretizing imaginary time into Trotter slices.

Inserting complete sets of basis states between successive operators gives
\begin{equation}\label{eq:sse-formulation}
    Z
    =
    \sum_{n=0}^{\infty}
    \sum_{\{b_p\}}
    \sum_{\{s_p\}}
    \frac{\beta^n}{n!}
    \prod_{p=0}^{n-1}
    \bra{s_{p+1}}H_{b_p}\ket{s_p},
\end{equation}
with the periodic boundary condition $s_n=s_0$ imposed by the trace. 

In practical simulations, one introduces a sufficiently large cutoff $\Lambda$
for the expansion order, chosen so that contributions with $n>\Lambda$ are
negligible. To work with operator strings of fixed length $\Lambda$, a string
with expansion order $n<\Lambda$ is padded by inserting $(\Lambda-n)$ identity
operators, often called null operators and denoted by
$H_{-1}$.

It is therefore useful to introduce a fixed-length operator string
\begin{equation}
    S_\Lambda
    =
    (b_0,b_1,\ldots,b_{\Lambda-1}),
    \qquad
    b_p\in\{-1\}\cup\{b\}.
\end{equation}
Here $p=0,\ldots,\Lambda-1$ labels the position in the operator string. The
expansion order $n=n(S_\Lambda)$ becomes the number of non-null operators in
$S_\Lambda$.

Including the corresponding combinatorial factor, the fixed-length SSE
representation becomes
\begin{equation}\label{eq:sse-fixed-length-propagated}
    Z
    =
    \sum_{S_\Lambda}
    \sum_{\{s_p\}}
    \frac{\beta^n(\Lambda-n)!}{\Lambda!}
    \prod_{p=0}^{\Lambda-1}
    \bra{s_{p+1}}H_{b_p}\ket{s_p},
\end{equation}
with $s_\Lambda=s_0$. 
An SSE configuration can therefore be represented as
\begin{equation}
    \mathcal{C}=(s_0,S_\Lambda),
\end{equation}
where $s_0$ is the initial basis state and $S_\Lambda$ is the fixed-length
operator string. 
 If the local-operator decomposition is chosen so that each $H_{b_p}$ maps a basis state to another basis state up to a scalar, then the propagated states are determined sequentially by
$\ket{s_{p+1}}\propto H_{b_p}\ket{s_p}$. Otherwise, the intermediate states $\{s_p\}$ must be included explicitly in the configuration.
Thus the intermediate
states $\{s_p\}$ are fixed by the configuration rather than sampled as
independent variables.

To compare the two finite-temperature approaches, note that path-integral QMC samples histories of basis states, whereas SSE samples the operator strings that generate such histories. In this limited sense, the distinction is reminiscent of the difference between a state-history viewpoint and an operator-history viewpoint, loosely analogous to the Schrödinger and Heisenberg pictures. Both methods provide complementary configuration-space representations of the Gibbs operator, with the trace imposing periodicity along the transfer direction.

Detailed algorithmic aspects of path-integral QMC and SSE, including local updates, loop updates, and cluster updates, can be found in the references cited at the beginning of this section. Here, we instead focus on their common configuration-space structure, which provides the foundation for the general density-matrix formulation developed in Sec.~\ref{sec:general-rho}.

% :::::::::::::::::::::::::::::::::::::::::::::::::
%   NEW SUBSECTION
% :::::::::::::::::::::::::::::::::::::::::::::::::
\subsection{Ground-state projector representation}\label{sec:projector-qmc}

\subsubsection{Projection to the ground state}

The Gibbs-state formulations above represent equilibrium density matrices
through the trace of $e^{-\beta H}$. Ground-state properties can be reached
within the same framework by taking the low-temperature limit
$\beta\rightarrow\infty$, where excited-state contributions are exponentially
suppressed. Practically, this requires $\beta \gg \Delta^{-1}$, where $\Delta$ is the many-body excitation gap. In gapless or critical systems, where the gap typically scales as $\Delta \sim L^{-z}$ with dynamical critical exponent $z$, achieving ground-state convergence generally requires $\beta \sim L^z$.

Projector QMC takes a different route and targets the ground state directly at
the level of wave functions~\cite{Liang1990vb-projector,Anders2005vb-projector,Sandvik2007-vb-projector,Sandvik2010vb-projector,Hasting2010vb-swap,Alet2007vb-ee,Kallin2009vb-ee,Melko2013sse,Kaul2013lattice-qmc}. Starting from a trial state $\ket{\Psi_T}$ with
nonzero overlap with the ground state, one repeatedly applies a projection
operator constructed from the Hamiltonian,
\begin{equation}
    \ket{\Psi_0}
    \propto
    \lim_{m\to\infty}
    (-H)^m\ket{\Psi_{T}}.
\end{equation}
In practice, $m$ is chosen sufficiently large, and a constant shift of the
Hamiltonian may be introduced so that the ground-state eigenvalue has the
largest magnitude. Repeated application of $(-H)$ then amplifies the
ground-state component of $\ket{\Psi_T}$ while suppressing excited-state
components.

The projection length $m$ plays a role analogous to the inverse temperature
$\beta$ in imaginary-time formulations. Ground-state convergence requires the
excited-state components to be sufficiently suppressed, with the suppression
controlled by the spectral gap $\Delta$. For local many-body Hamiltonians, the
typical operator-string length needed to resolve the extensive energy scale
grows with the system size $N$, leading to the estimate$m \sim {N}/{\Delta}$.
For a $d$-dimensional gapless system with $\Delta\sim L^{-z}$ and
$N\sim L^d$, this gives $m \sim L^{d+z}$,
where $z$ is the dynamical critical exponent.

% ----------------------------------------------------------------
%   NEW SUBSECTION
% ----------------------------------------------------------------
\subsubsection{Configuration-space representation and choice of basis}
We now describe the configuration-space representation of projector QMC. 
Let $\{\ket{\alpha}\}$ be a set of basis states, which may be orthonormal,
nonorthogonal, or overcomplete.
The trial state must have nonzero overlap with the target ground state
$\ket{\Psi_0}$. For simplicity, we take the
trial state to be a single basis state $\ket{\alpha_0}$.

The projected-state normalization factor is
\begin{equation}\label{eq:Zm}
    Z_m[\alpha_0]
    =
    \bra{\alpha_0}(-H)^m(-H)^m\ket{\alpha_0}.
\end{equation}
As in SSE in Sec.~\ref{sec:sse}, we decompose the Hamiltonian into local operators. Expanding one projector gives
\begin{equation}
    (-H)^m
    =
    \sum_{\{b_p\}_{p=0}^{m-1}}
    H_{b_{m-1}}\cdots H_{b_1}H_{b_0}.
\end{equation}
Each choice of $\{b_p\}$ defines an operator string of length $m$.

Practically, the basis and the operator decomposition
are chosen so that the action of a local operator on a basis state is
proportional to another basis state,
\begin{equation}
    H_{b_p}\ket{\alpha_p}
    =
    w_p(\alpha_p,b_p)\ket{\alpha_{p+1}}.
\end{equation}
Therefore, once the initial state $\ket{\alpha_0}$ and the operator string are
specified, the intermediate states are generated deterministically. For the
right projector, this gives
\begin{equation}
    H_{b_{m-1}}\cdots H_{b_1}H_{b_0}\ket{\alpha_0}
    =
    W_R\ket{\alpha_R},
\end{equation}
where
\begin{equation}
    \ket{\alpha_R}\equiv \ket{\alpha_m},
    \qquad
    W_R
    =
    \prod_{p=0}^{m-1} w_p(\alpha_p,b_p).
\end{equation}

Similarly, the left projector acts on the bra state. For another operator
string $\{b'_q\}$, we write
\begin{equation}
    \bra{\alpha_0}
 H_{b'_0}  H_{b'_1}\cdots   H_{b'_{m-1}}
    \equiv 
    W_L\bra{\alpha_L}.
\end{equation}
Here $\bra{\alpha_L}$ is the basis bra obtained after propagating
$\bra{\alpha_0}$ through the left operator string, and $W_L$ is the product of
the corresponding local propagation factors. 
More generally, the bra and ket boundary states need not be the same.

Combining the two projected states, the normalization becomes
\begin{equation}\label{eq:proj-norm}
    Z_m[\alpha_0]
    =
    \sum_{\{b_p\},\{b'_q\}}
    W_L W_R
    \langle \alpha_L|\alpha_R\rangle .
\end{equation}
Thus a projector-QMC configuration may be specified by the initial trial basis
state and the two operator strings,
\begin{equation}
    \mathcal C
    =
    \left(
        \alpha_0,
        \{b_p\}_{p=0}^{m-1},
        \{b'_q\}_{q=0}^{m-1}
    \right),
\end{equation}
where $\alpha_0$ is fixed, and we obtain the weight 
\begin{equation}
    W(\mathcal C)
    =
    W_L W_R
    \langle \alpha_L|\alpha_R\rangle .
\end{equation}
When this weight is nonnegative, it can be used directly as a Monte Carlo
sampling weight.

This representation emphasizes the open-boundary nature of ground-state
projection: the two halves of the projector generate a ket $\ket{\alpha_R}$ and a bra $\bra{\alpha_L}$, 
and the configuration weight is obtained from their overlap.

When the bases are nonorthogonal and overcomplete, one should not insert complete sets of basis states as in an
orthonormal basis. Instead, the overlap
$\langle \alpha_L|\alpha_R\rangle$ is an essential part of the weight. A
prominent example is the valence-bond basis for quantum spin systems. The
propagated states are valence-bond coverings, denoted by $\ket{V_R}$ and
$\bra{V_L}$, and the overlap $\langle V_L|V_R\rangle$ can be evaluated
geometrically from the transition graph formed by superimposing the two
valence-bond configurations, as discussed in Sec.~\ref{sec:vb_basis}.

The orthonormal-basis representation is recovered as a special case. If
$\{\ket{\alpha}\}$ is orthonormal, then the overlap between the two propagated
middle states reduces to
\begin{equation}
    \langle \alpha_L|\alpha_R\rangle
    =
    \delta_{\alpha_L,\alpha_R}.
\end{equation}
Equivalently, one may insert complete sets of basis states between all $2m$ local
operators and write the configuration weight as a product of local matrix
elements along the full operator string.

\subsubsection{Ground-state estimators}
We next discuss how observables are measured in the projector representation.
For a fixed trial basis state $\ket{\alpha_0}$, the projected-state estimator
of an operator $O$ is
\begin{equation}
    \langle O\rangle_m
    =
    \frac{
        \bra{\alpha_0}(-H)^m O (-H)^m\ket{\alpha_0}
    }{
        \bra{\alpha_0}(-H)^m (-H)^m\ket{\alpha_0}
    }.
\end{equation}
If the trial state has nonzero overlap with the ground state and $m$ is large
enough, this converges to the ground-state expectation value,
\begin{equation}
    \langle O\rangle_m
    \longrightarrow
    \frac{\bra{\tilde\Psi_0}O\ket{\tilde\Psi_0}}{\braket{\tilde\Psi_0|\tilde\Psi_0}}.
\end{equation}

Using the same left and right propagated states introduced above, the numerator
can be written as
\begin{align}
     &  \bra{\alpha_0}(-H)^m O (-H)^m\ket{\alpha_0} \label{eq:proj-middle}
   \\  =\;&
    \sum_{\{b_p\},\{b'_q\}}
    W_L W_R
    \bra{\alpha_L}O\ket{\alpha_R}.
\end{align}
Dividing by the normalization factor~\eqref{eq:proj-norm} gives
\begin{equation}
    \langle O\rangle_m
    =
    \frac{
        \sum_{\mathcal C}
        W_L W_R
        \bra{\alpha_L}O\ket{\alpha_R}
    }{
        \sum_{\mathcal C}
        W_L W_R
        \braket{\alpha_L|\alpha_R}
    }.
\end{equation}
Equivalently, define 
the estimator associated with configuration $\mathcal C$ as
\begin{equation}
    O(\mathcal C)
    =
    \frac{
        \bra{\alpha_L}O\ket{\alpha_R}
    }{
        \braket{\alpha_L|\alpha_R}
    }.
\end{equation}
The projected expectation value becomes
\begin{equation}
    \langle O\rangle_m
    =
    \frac{
        \sum_{\mathcal C} W(\mathcal C) O(\mathcal C)
    }{
        \sum_{\mathcal C} W(\mathcal C)
    }.
\end{equation}
Thus the estimator is not simply a diagonal matrix element of $O$. In a
nonorthogonal basis, it is the matrix element of $O$ between the two propagated
basis states, normalized by their overlap. Compared with the finite-temperature
formulation in Sec.~\ref{sec:Gibbs-state}, where the trace gives periodic
boundary conditions in imaginary time and permits averaging over time slices,
projector QMC has open boundaries fixed by the trial state. 
Operator insertions near the boundaries retain trial-state effects, whereas
insertions in the central region of the projection direction probe the
projected ground state.

% :::::::::::::::::::::::::::::::::::::::::::::::::
%   NEW SUBSECTION
% :::::::::::::::::::::::::::::::::::::::::::::::::
\subsection{Reduced density matrices (RDMs)}\label{sec:rdm}
\begin{figure}[ht]
    \centering
    \begin{subfigure}[b]{0.45\linewidth}
        \centering
        \includegraphics[width=\linewidth]{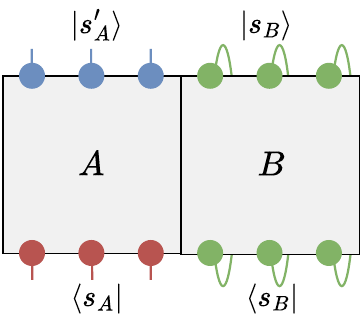}
        \caption{}
    \end{subfigure}
    \hspace{0.01\textwidth}
    \begin{subfigure}[b]{0.45\linewidth}
        \centering
        \includegraphics[width=\linewidth]{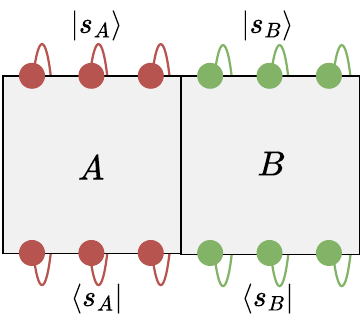}
        \caption{}
    \end{subfigure}
    \caption{
Diagrammatic illustration of a RDM and its trace.
(a) The matrix element $\langle s_A|\tilde\rho_A|s_A'\rangle$ is obtained by
identifying and summing over the $B$ boundary states of the full density
matrix, while keeping the $A$ boundary states open.
(b) Taking $\Tr_A(\tilde\rho_A)$ further identifies and sums over the remaining $A$
boundary states, closing all boundary legs and giving
$\Tr_A(\tilde\rho_A)=\Tr(\tilde\rho)$.
}
    \label{fig:rhoA}
\end{figure}

RDMs provide a simple but deep extension of the open-time-boundary
viewpoint in Sec.~\ref{sec:general-rho}. Suppose the full system is bipartitioned into
subsystems $A$ and $B$, and choose a product basis
\begin{equation}
    \ket{s}
    =
    \ket{s_A}\otimes\ket{s_B}
    \equiv
    \ket{s_A,s_B}.
\end{equation}
The RDM on $A$ is $\tilde\rho_A=\Tr_B(\tilde\rho)$, with matrix elements
\begin{equation}\label{eq:rhoA-matrix-elements}
    \langle s_A|\tilde\rho_A|s_A'\rangle
    =
    \sum_{s_B}
    \langle s_A,s_B|\tilde\rho|s_A',s_B\rangle .
\end{equation}
Thus the partial trace closes only the $B$ boundary indices, while the $A$
indices remain open, as illustrated in Fig.~\ref{fig:rhoA}(a).

Taking the trace over $A$ gives
\begin{align}
    \Tr_A(\tilde\rho_A)
    &=
    \sum_{s_A}
    \langle s_A|\tilde\rho_A|s_A\rangle
    \nonumber \\
    &=
    \sum_{s_A,s_B}
    \langle s_A,s_B|\tilde\rho|s_A,s_B\rangle
     =
    \Tr(\tilde\rho),
\end{align}
as shown in Fig.~\ref{fig:rhoA}(b).
Therefore, $\tilde\rho_A$ carries the same
normalization factor $\Tr(\tilde\rho)$ as the full density matrix, as mentioned in Sec.~\ref{sec:conventions}.

This observation is important for QMC: although $\tilde\rho_A$ acts only on
subsystem $A$, its matrix elements can be obtained from configurations of the
full system. The $B$ degrees of freedom are traced out and sampled
stochastically, while the subsystem-$A$ boundary data are retained as the
external indices of $\tilde\rho_A$~\cite{mao2025samplingrdm,WangTT2025rdm}. Consequently, when measuring observables defined
on $\tilde\rho_A$, such as the two-point correlation function in
Eq.~\eqref{eq:diag-O-example}~\cite{Aditya2026rdm,wangzy2026rdm}, one still simulates
$\Tr_A(\tilde\rho_A)=\Tr(\tilde\rho)$, but the estimator depends only on the subsystem-$A$
boundary data.

% :::::::::::::::::::::::::::::::::::::::::::::::::
%   NEW SUBSECTION
% :::::::::::::::::::::::::::::::::::::::::::::::::
\subsection{Replicated density matrices}\label{sec:rep_states}
\subsubsection{Cyclic gluing and replicated partition functions}
Many quantities in many-body quantum information, such as Rényi entropies, are
nonlinear functions of the density matrix $\tilde\rho$ and therefore cannot be represented
within a single-copy configuration space. This motivates replicated
constructions on enlarged Hilbert spaces.

\begin{figure}[ht!]
    \centering
    \begin{subfigure}[b]{0.72\linewidth}
        \centering
        \includegraphics[width=\linewidth]{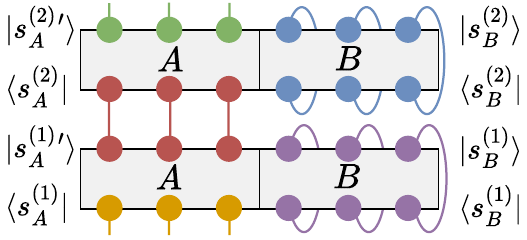}
        \caption{}
    \end{subfigure}
    % \hspace{0.01\textwidth}
    \begin{subfigure}[b]{0.72\linewidth}
        \centering
        \includegraphics[width=\linewidth]{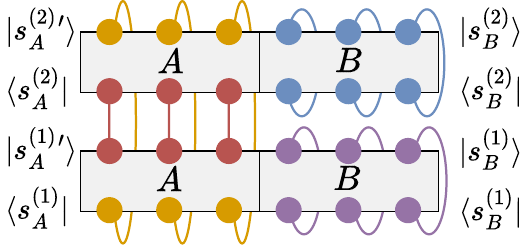}
        \caption{}
    \end{subfigure}
    \caption{
        Diagrammatic illustration of the replicated RDM $\tilde\rho_A^2$
and its trace.
(a) The open-boundary matrix element of $\tilde\rho_A^2$. The two replicas are glued
sequentially in subsystem $A$: the ket boundary state $\ket{s_A^{(1)}{}'}$ of the first replica is
identified with the bra boundary state $\bra{s_A^{(2)}}$ of the second. The $B$ degrees
of freedom are traced within each replica.
(b) The trace $\Tr_A(\tilde\rho_A^2)$, obtained by further identifying and summing
over the remaining open $A$ boundary states $\ket{s_A^{(2)}{}'}$ and $\bra{s_A^{(1)}}$. This closes all $A$ boundary legs.
    }
    \label{fig:rhoA2}
\end{figure}

A central example is the Rényi-$\alpha$ partition function~\cite{Callan1994replica,Christoph1994replica,Calabrese2004-ent-qft}, defined as~\footnote{
    In general, $\alpha$ need not be an integer. Here, however, we restrict our discussion to integer values of $\alpha$, for which the replica construction admits a natural QMC implementation.
}
\begin{equation}\label{eq:rep-pf}
    \Tr(\tilde\rho^\alpha),
    \qquad
    \alpha\in \mathbb{N}^+,
\end{equation}
which couples $\alpha$ replicas of $\tilde\rho$. In the computational basis,
\begin{align}
    \Tr(\tilde\rho^\alpha)
    =
    \sum_{s^{(1)},\dots,s^{(\alpha)}}
    &
    \langle s^{(1)}|\tilde\rho|s^{(2)}\rangle
    \langle s^{(2)}|\tilde\rho|s^{(3)}\rangle
    \cdots
    \nonumber \\
    &\times
    \langle s^{(\alpha)}|\tilde\rho|s^{(1)}\rangle .
\end{align}
Thus, the ket boundary state of each replica is identified with the bra
boundary state of the next replica, while the last replica is connected back
to the first. This produces a cyclic boundary condition across replicas. An
example for the replicated RDM $\tilde\rho_A^2$ and its trace is
shown in Fig.~\ref{fig:rhoA2}.

Provided that $\Tr(\tilde\rho)$, or more generally the open-boundary matrix elements
of $\tilde\rho$, admit a sign-problem-free QMC representation, the replicated
construction requires no new local representation. It enlarges the
configuration space from one copy to $\alpha$ copies and replaces the
single-copy boundary condition by a cyclic gluing condition between replicas.
Each replica carries its own auxiliary variables, while the boundary states are
constrained by the cyclic contraction above. Thus the local weights used to
represent $\tilde\rho$ can be combined replica by replica, with the essential
modification lying in the boundary connectivity. Consequently, QMC
representations of $\Tr(\tilde\rho)$ often extend naturally to Rényi-$\alpha$
partition functions $\Tr(\tilde\rho^\alpha)$.

\subsubsection{SWAP-operator representation}\label{sec:swap}
For $\alpha=2$, the replicated construction admits an equivalent formulation in terms of a SWAP operator. Consider two copies of the original Hilbert space, denoted by $\mathcal H^{[1]}$ and $\mathcal H^{[2]}$, and the doubled Hilbert space $\mathcal H^{[1]}\otimes \mathcal H^{[2]}$. The two-replica density matrix is
\begin{equation}\label{eq:two-copy-notation}
\tilde\rho^{[1]}\otimes \tilde\rho^{[2]}
\equiv
\tilde\rho\otimes\tilde\rho
\in
\mathcal B\left(\mathcal H^{[1]}\otimes\mathcal H^{[2]}\right).
\end{equation}
Here, $\mathcal{B}\!\left(\mathcal{H}^{[1]}\otimes\mathcal{H}^{[2]}\right)$
denotes the corresponding space of bounded operators.
Let $\mathrm{SWAP}\in \mathcal B\left(\mathcal H^{[1]}\otimes\mathcal H^{[2]}\right)$ denote the unitary operator that exchanges the degrees of freedom between the two replicas. 
Then 
\begin{equation}\label{eq:swap-rho2}
\Tr(\tilde\rho^2)
=
\Tr\left[
(\tilde\rho\otimes\tilde\rho)\mathrm{SWAP}
\right].
\end{equation}

In this formulation, the replica gluing is replaced by an operator insertion in the doubled Hilbert space (see Sec.~\ref{sec:rho-insert}). Equivalently, the same quantity can be viewed as the expectation value of the $\mathrm{SWAP}$ operator, rather than as a modification of the boundary connectivity. 

Similarly, restricting the SWAP operator to a subsystem $A$ gives
\begin{equation}\label{eq:swap-rhoA2}
    \Tr(\tilde\rho_A^2)
    =
    \Tr\!\left[
        (\tilde\rho\otimes\tilde\rho)\,\mathrm{SWAP}_A
    \right],
\end{equation}
where $\mathrm{SWAP}_A$ exchanges only the subsystem-$A$ degrees of freedom
between the two replicas and acts trivially on the complement subsystem $B$. 

For a normalized two-copy ensemble, the corresponding expectation value is
   \begin{equation}
       \frac{\Tr[(\tilde\rho\otimes\tilde\rho)\mathrm{SWAP}_A]}
            {[\Tr(\tilde\rho)]^2}
       =
       \Tr(\rho_A^2).
   \end{equation}

This SWAP representation is particularly useful in special bases. In the
valence-bond basis or the Bell basis, discussed in
Secs.~\ref{sec:vb_basis} and~\ref{sec:bell_basis},
Eqs.~\eqref{eq:swap-rho2} and~\eqref{eq:swap-rhoA2} admit direct estimators, and the SWAP operator can be evaluated as an ordinary linear observable in the sense of a two-replica ensemble.

In more general settings, however, the replicated normalization factor must be
evaluated directly. The
Rényi-$\alpha$ entropy of $\tilde\rho_A=\Tr_B(\tilde\rho)$ is
\begin{equation}\label{eq:eg-renyi-ratio}
    S_\alpha(\rho_A)
    =
    \frac{1}{1-\alpha}
    \ln
    \frac{
        \Tr_A(\tilde\rho_A^\alpha)
    }{
        [\Tr(\tilde\rho)]^\alpha
    },
    \qquad
    \alpha\ge 2 \in\mathbb{N}.
\end{equation}
Thus the computation of Rényi entropies reduces to estimating logarithms of
partition-function ratios.
These ratios typically require dedicated estimation methods, which will be introduced in Sec.~\ref{sec:pfr}.

% :::::::::::::::::::::::::::::::::::::::::::::::::
%   NEW SUBSECTION
% :::::::::::::::::::::::::::::::::::::::::::::::::
\subsection{Partially transposed replicated density matrices}
\label{sec:pt_states}

While Rényi-$\alpha$ entanglement entropies characterize bipartite
entanglement in pure states, they are not, in general, faithful measures of
mixed-state entanglement. A central alternative is based on the partial
transpose and the associated positive-partial-transpose (PPT)
criterion~\cite{Peres1996ppt,Horodecki1996separability}. This motivates
replicated constructions of partially transposed density matrices, whose basic
objects are moments of the form
\begin{equation}\label{eq:partial-tranposed-moment}
    \Tr\!\left[
        \left(\tilde\rho_{AB}^{T_B}\right)^\alpha
    \right],\quad \alpha\in\mathbb{N}^+.
\end{equation}
These moments underlie Rényi negativities and related PPT-based diagnostics of
mixed-state entanglement~\cite{Calabrese2012negativity-qft, Calabrese2013negativity-extended}, which will be elaborated in
Sec.~\ref{sec:renyi-negativity}.

For $\alpha=2$, the partially transposed moment
in Eq.~\eqref{eq:partial-tranposed-moment} reduces to the ordinary Rényi-2
partition function,
\begin{equation}
    \Tr\!\left[(\tilde\rho_{AB}^{T_B})^2\right]
    =
    \Tr(\tilde\rho_{AB}^2).
\end{equation}
This follows from the invariance of the Hilbert--Schmidt inner product under
partial transpose. Hence the second moment contains no additional information
beyond the ordinary Rényi-2 partition function~\eqref{eq:rep-pf}.

\begin{figure}[ht!]
    \centering
    \includegraphics[width=0.75\linewidth]{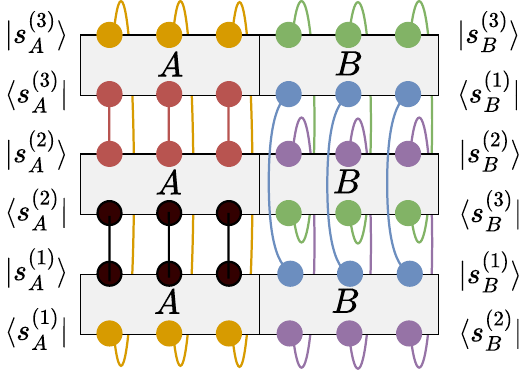}
    \caption{
       Diagrammatic illustration of the partially transposed replicated partition
function $\Tr[(\tilde\rho_{AB}^{T_B})^3]$. The three replicas are glued cyclically in
subsystem $A$, as in the ordinary replicated construction. In subsystem $B$,
however, the partial transpose changes the boundary connectivity, producing a
different gluing pattern between replicas. 
    }
    \label{fig:rhoTB}
\end{figure}

The first nontrivial replicated partially transposed construction appears at
$\alpha=3$. Explicitly,
\begin{widetext}
\begin{align}
    \Tr\!\left[(\tilde\rho_{AB}^{T_B})^3\right]
    &=
    \sum_{\{s_A^{(\gamma)},s_B^{(\gamma)}\}_{\gamma=1}^3}
    \langle s_A^{(1)},s_B^{(1)}|\tilde\rho^{T_B}|s_A^{(2)},s_B^{(2)}\rangle
    \langle s_A^{(2)},s_B^{(2)}|\tilde\rho^{T_B}|s_A^{(3)},s_B^{(3)}\rangle
    \langle s_A^{(3)},s_B^{(3)}|\tilde\rho^{T_B}|s_A^{(1)},s_B^{(1)}\rangle
    \nonumber \\
    &=
    \sum_{\{s_A^{(\gamma)},s_B^{(\gamma)}\}_{\gamma=1}^3}
    \langle s_A^{(1)},s_B^{(2)}|\tilde\rho|s_A^{(2)},s_B^{(1)}\rangle
    \langle s_A^{(2)},s_B^{(3)}|\tilde\rho|s_A^{(3)},s_B^{(2)}\rangle
    \langle s_A^{(3)},s_B^{(1)}|\tilde\rho|s_A^{(1)},s_B^{(3)}\rangle .
\end{align}
\end{widetext}
In this expression, the $A$ indices follow the usual cyclic gluing, whereas the
partial transpose induces a different cyclic connectivity for the $B$ indices,
as illustrated in Fig.~\ref{fig:rhoTB}. This mismatch between the $A$ and $B$
boundary identifications cannot be removed by relabeling for $\alpha=3$, making
it the first genuinely distinct partially transposed replicated geometry.

% :::::::::::::::::::::::::::::::::::::::::::::::::
%   NEW SUBSECTION
% :::::::::::::::::::::::::::::::::::::::::::::::::
\subsection{Operator-inserted density matrices}\label{sec:rho-insert}

The transfer representation introduced in Sec.~\ref{sec:general-rho} also
provides a natural way to incorporate operator insertions. In the SSE
formulation of a Gibbs state, for example, the Taylor expansion of
$e^{-\beta H}$ already represents the transfer process as an operator string
generated by local terms $H_{b_p}$. More generally, additional operators may be
inserted into this transfer process to define new density-matrix-like objects.

More generally, suppose that an unnormalized density matrix $\tilde\rho$ already admits a sign-problem-free QMC representation. Given an operator $O$, such as a local observable, projector, or symmetry operator, one may define the operator-inserted density matrix 
\begin{equation}
    \tilde\rho_O
    :=
    \tilde\rho O .
\end{equation}
Although we use density-matrix notation for convenience, $\tilde\rho_O$ need not be Hermitian or positive, and therefore is not a physical density matrix in general.

Its open-boundary matrix elements are
\begin{equation}
    \langle s|\tilde\rho_O|s'\rangle
    =
    \langle s|\tilde\rho O|s'\rangle .
\end{equation}
If the insertion of $O$ does not introduce sign problems, then $\tilde\rho_O$ can be sampled by QMC within the same framework
as $\tilde\rho$. Operationally, the insertion adds an additional operator slice to
the transfer process.

\begin{figure}[ht!]
    \centering
    \includegraphics[width=0.6\linewidth]{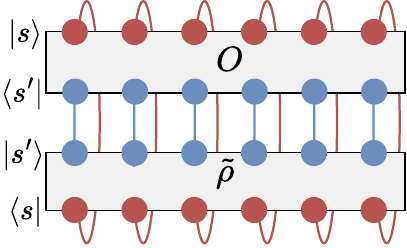}
    \caption{
       Diagrammatic illustration of the operator-inserted partition function
        $Z_O=\Tr(\tilde\rho O)$. The upper replica represents the operator insertion $O$,
        while the lower replica represents the density matrix $\tilde\rho$. Closing the
        boundary indices implements the trace.
    }
    \label{fig:rhoO}
\end{figure}

After
closing the time-boundary indices, the corresponding operator-inserted partition
function is
\begin{equation}\label{eq:Z_O-partition-function}
    Z_O
    =
    \Tr(O\tilde\rho)
    =
    \sum_{s,s'}
    \langle s|\tilde\rho |s'\rangle
    \langle s'| O |s\rangle .
\end{equation}
A diagrammatic representation is shown in Fig.~\ref{fig:rhoO}.

Moreover, operator insertions are not arbitrary: the inserted operator must be compatible
with the sampled configuration space. For example, consider the Gibbs state of
a classical Ising model in the $\sigma^z$ basis. Since $e^{-\beta H}$ is
diagonal in this basis, its configurations have matching bra and ket boundary
states. Inserting $\sigma_i^x$ flips the spin at site $i$ and is therefore
incompatible with the trace boundary condition in $\Tr(\sigma_i^x\tilde\rho)$.
Consequently, the corresponding weight vanishes identically. Thus an
operator-inserted state is meaningful and useful only when the insertion gives a nonzero,
sign-problem-free configuration-space representation.

% :::::::::::::::::::::::::::::::::::::::::::::::::
%   NEW SUBSECTION
% :::::::::::::::::::::::::::::::::::::::::::::::::
\subsection{Density matrices under quantum channels}

Operator-inserted density matrices can be further generalized by allowing a
sum over insertions. This naturally leads to density matrices transformed by
quantum channels, a completely positive trace-preserving
(CPTP) map $\mathcal E$ acting on a density matrix~\cite{NielsenChuang2010}. In Kraus form,
\begin{equation}
    \mathcal E(\tilde\rho)
    =
    \sum_a
    K_a \tilde\rho K_a^\dagger ,
    \qquad
    \sum_a K_a^\dagger K_a = 1 .
\end{equation}
Thus a channel-evolved state is a sum over
operator-inserted contributions, with Kraus operators inserted on both sides of
$\tilde\rho$.

In the open-time-boundary representation, the corresponding matrix elements are
\begin{equation}
    \langle s|\mathcal E(\tilde\rho)|s'\rangle
    =
    \sum_a
    \langle s|K_a\tilde\rho K_a^\dagger|s'\rangle .
\end{equation}
Compared with a single operator-inserted state, the configuration space is
enlarged by the channel label $a$. If $\tilde\rho$ admits a sign-problem-free QMC
representation and the Kraus insertions preserve non-negative real weights, then
$\mathcal E(\tilde\rho)$ can be sampled within the same general framework.

For local channels, the Kraus label is usually local as well. One may then view
the channel as introducing stochastic local insertions, such as dephasing
events, measurement outcomes, decoherence defects, or symmetry-preserving
errors, into the transfer process. The resulting QMC configurations contain
both the auxiliary variables representing $\tilde\rho$ and the additional channel
variables specifying which local Kraus events occur.

This construction is important for mixed-state studies for open quantum systems. It provides a QMC framework for representing decohered density matrices that arise from quantum channels~\cite{Baweja2025post,ding2026mdite,ding2026swssb}. 

In the following, we discuss two
representative examples: a simple dephasing channel in the computational basis
and a strongly $\mathbb Z_2$-symmetric channel.

\subsubsection{Local dephasing channel}\label{sec:dephasing-channel}
A simple example is the local dephasing channel in the computational
($\sigma^z$) basis. For a single qubit $i$, define
\begin{align}
    P_{i,0}
    &=
    \ket{0_i}\bra{0_i}
    =
    \frac{1+\sigma_i^z}{2}
    =
    \begin{bmatrix}
        1 & 0 \\
        0 & 0
    \end{bmatrix},
    \\
    P_{i,1}
    &=
    \ket{1_i}\bra{1_i}
    =
    \frac{1-\sigma_i^z}{2}
    =
    \begin{bmatrix}
        0 & 0 \\
        0 & 1
    \end{bmatrix}.
\end{align}
where $\ket{0_i}\equiv\ket{\uparrow_i}$ and
$\ket{1_i}\equiv\ket{\downarrow_i}$. 
Then the dephasing channel (or projective measurement) is written as 
\begin{align}
    \mathcal E_{p,i}[\tilde\rho_0]
    &=
    \left(1-\frac{p}{2}\right)\tilde\rho_0
    +
    \frac{p}{2}\sigma_i^z\tilde\rho_0\sigma_i^z
    \label{eq:channel:z}
    \\
    &=
    (1-p)\tilde\rho_0
    +
    p\sum_{\sigma=0,1}
    P_{i,\sigma}\tilde\rho_0 P_{i,\sigma},
    \label{eq:channel:z-proj}
\end{align}
where $p\in[0,1]$.
The full channel is obtained by applying this map independently to all qubits,
\begin{equation}\label{eq:full-channel-dephasing}
    \mathcal E_p
    =
    \prod_i
    \mathcal E_{p,i}.
\end{equation}

This channel preserves the diagonal matrix elements of the input density matrix $\tilde\rho_0$ in the
computational basis while suppressing off-diagonal coherences between basis
states that differ on dephased sites. In the fully dephased limit $p=1$, all
local $Z$-basis coherences are removed.

Although Eqs.~\eqref{eq:channel:z} and~\eqref{eq:channel:z-proj} are
mathematically equivalent, they are not equivalent from the perspective of QMC
sampling. Directly inserting $\sigma_i^z$ into the configuration-space
representation can introduce the sign problem, since
$\sigma_i^z\ket{1_i}=-\ket{1_i}$ and local matrix elements such as
$\langle 1_i|\sigma_i^z|1_i\rangle=-1$ are negative. By contrast, the
projector form~\eqref{eq:channel:z-proj} admits a probabilistic interpretation
as stochastic projective measurements and remains sign-problem-free in the
computational basis because the projectors $P_{i,\sigma}$ have non-negative
matrix elements.

\begin{figure}[ht!]
    \centering
    \begin{subfigure}[b]{0.72\linewidth}
        \centering
        \includegraphics[width=\linewidth]{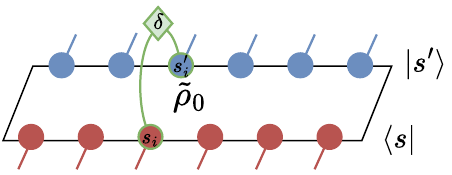}
        \caption{}
    \end{subfigure}
    % \hspace{0.01\textwidth}
    \begin{subfigure}[b]{0.72\linewidth}
        \centering
        \includegraphics[width=\linewidth]{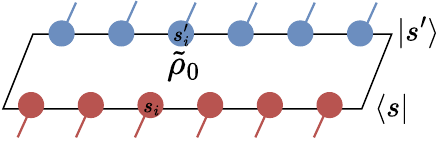}
        \caption{}
    \end{subfigure}
    % \hspace{0.01\textwidth}
    \begin{subfigure}[b]{0.72\linewidth}
        \centering
        \includegraphics[width=\linewidth]{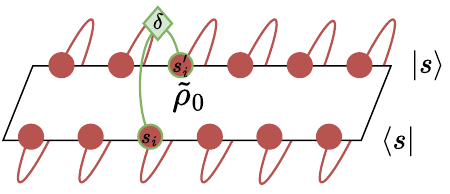}
        \caption{}
    \end{subfigure}
    \caption{
        Diagrammatic illustration of the local dephasing channel applied to the input density matrix $\tilde\rho_0$.
(a) A dephasing event inserts a local projector with probability $p$ to qubit $i$, which enforces the constraint
$s_i=s_i'$ between the bra and ket boundary states at site $i$. Equivalently,
this corresponds to inserting a local Kronecker-delta constraint
$\delta_{s_i,s_i'}^{(2)}$.
(b) With probability $1-p$, no dephasing event occurs and the local boundary
state of qubit $i$ remain unconstrained.
(c) A local dephasing event acting on qubit $i$, followed by tracing over the
boundary indices of the full density matrix.
    }
    \label{fig:dephasing}
\end{figure}

The dephasing channel admits a useful boundary-constraint interpretation. In
the projector representation~\eqref{eq:channel:z-proj}, each qubit carries a
stochastic choice: with probability $p$, a local rank-two Kronecker-delta constraint
$\delta^{(2)}_{s_i,s_i'}$ is inserted between the bra and ket boundary indices;
with probability $1-p$, this constraint is absent, equivalently replaced by
the identity $\mathds{1}$. This construction is illustrated in Figs.~\ref{fig:dephasing}(a) and (b). Additional details can be found in Ref.~\cite{ding2026mdite}.

Thus, for an $N$-qubit system, the channel enlarges the configuration space of
$\tilde\rho_0$ by a binary constraint pattern,
\begin{equation}\label{eq:ext-config-dephasing}
    \mathcal C
    =
    \mathcal C_0 \cup \{c_i\}_{i=1}^N,
    \qquad
    c_i\in\{\mathds{1},\delta^{(2)}\},
\end{equation}
where $\mathcal C_0$ denotes a configuration of the original density matrix
$\tilde\rho_0$, and $c_i$ specifies whether the local boundary indices at qubit $i$
are unconstrained or identified. The $2^N$ choices of $\{c_i\}$ divide the
extended configuration space into $2^N$ constraint sectors.

A QMC simulation of the decohered state under the full channel~\eqref{eq:full-channel-dephasing} can then be organized into two types of
updates. First, for a fixed constraint pattern $\{c_i\}$, one updates
$\mathcal C_0$ using the original QMC updates for $\tilde\rho_0$, subject to the
constraints imposed by the sites with $c_i=\delta^{(2)}$. Second, for fixed
$\mathcal C_0$, one updates the constraint pattern itself. For example, a local
Metropolis update at qubit $i$ may switch between $c_i=\mathds{1}$ and
$c_i=\delta^{(2)}$. If $c_i=\delta^{(2)}$, then compatibility already requires
$s_i=s_i'$, and the update
\begin{equation}
    (\mathcal C_0,\delta^{(2)})
    \to
    (\mathcal C_0,\mathds{1})
\end{equation}
is accepted with probability
\begin{equation}
    P_{\delta^{(2)}\to\mathds{1}}
    =
    \min\left\{
        1,\frac{1-p}{p}
    \right\}.
\end{equation}
Conversely, if $c_i=\mathds{1}$ and the current boundary states satisfy
$s_i=s_i'$, then the update
\begin{equation}
    (\mathcal C_0,\mathds{1})
    \to
    (\mathcal C_0,\delta^{(2)})
\end{equation}
is accepted with probability
\begin{equation}
    P_{\mathds{1}\to\delta^{(2)}}
    =
    \min\left\{
        1,\frac{p}{1-p}
    \right\}.
\end{equation}
If $s_i\ne s_i'$, the latter move is incompatible and is rejected. These
acceptance probabilities satisfy detailed balance between the unconstrained and
constrained sectors when each site is traversed once in every MC step.

We emphasize that, although a local Kronecker-delta constraint identifies the bra and ket states
on a given qubit, it is not a trace over that qubit. The constraint keeps the
diagonal contribution as part of the density matrix, whereas tracing further
sums over the diagonal states and produces either a RDM or,
for a full trace, a scalar normalization factor. 
The
full trace of $\tilde\rho_0$ with a single local dephasing event is illustrated in
Fig.~\ref{fig:dephasing}(c). The same construction naturally extends to the
full channel in Eq.~\eqref{eq:full-channel-dephasing}.

Furthermore, one may use the decohered density matrix itself as the input for
more general state constructions. For example, the dephased state may be
followed by an additional imaginary-time evolution operator $O=e^{-\tau H}$, in a similar spirit as the operator-inserted construction in
Sec.~\ref{sec:rho-insert}, where $\tau$ is the imaginary time and $H$ is a
many-body Hamiltonian. This provides a framework for studying the competition
between decoherence channels and imaginary-time
evolution~\cite{ding2026mdite}.

\subsubsection{$\mathbb Z_2$-symmetric channel}\label{sec:z2-symmetric-chnn}

As a second, more structured example, consider a local channel acting on a pair
of qubits $b=(i,j)$,
\begin{equation}\label{eq:z2-strong-channel}
    \mathcal E_{p,b}[\tilde\rho_0]
    =
    \left(1-\frac{p}{2}\right)\tilde\rho_0
    +
    \frac{p}{2}
    \sigma^z_i\sigma^z_j\tilde\rho_0 \sigma^z_i\sigma^z_j ,
\end{equation}
where $p\in[0,1]$. For lattice models, $b$ denotes a bond, which is
usually taken to connect nearest-neighbor sites. The full channel is then
obtained by applying Eq.~\eqref{eq:z2-strong-channel} over a chosen set of
bonds,
\begin{equation}\label{eq:z2-full-channel}
    \mathcal E_p
    =
    \prod_{b}
    \mathcal E_{p,b} .
\end{equation}

For mixed states, we distinguish two notions of symmetry. Given a
unitary symmetry operator $U$, a density matrix $\rho$ has a \emph{strong
symmetry} if
\begin{equation}
    U\rho = e^{i\theta}\rho,
\end{equation}
and a \emph{weak symmetry} if
\begin{equation}
    U\rho U^\dagger = \rho .
\end{equation}
Strong symmetry is the direct mixed-state analogue of a pure state belonging to
a definite symmetry sector, while weak symmetry is the usual invariance of a
density matrix under conjugation, as in thermal Gibbs states~\cite{Buca2012strongweak,albert2014symmetries,lieu2020strongweak}.

In open quantum systems, this distinction allows a phenomenon with no direct
pure-state counterpart: a strong symmetry may be spontaneously broken down to a weak symmetry. This is referred to as
\emph{strong-to-weak spontaneous symmetry breaking} (SWSSB)~\cite{wang2026swssb-review,leejy2023weakmeasurement,Lessa2025swssb}. 
The channel in Eq.~\eqref{eq:z2-strong-channel} is designed to preserve the strong $\mathbb Z_2$ symmetry generated by the global spin-flip operator $X:=\prod_i \sigma_i^x$, while inducing decoherence through the symmetry-even bond operators $\sigma_i^z\sigma_j^z$. 
We will review the associated nonlinear diagnostics and QMC studies of Ref.~\cite{ding2026swssb} in Sec.~\ref{sec:swssb}.

Again, treating the term
$\sigma_i^z\sigma_j^z\tilde\rho_0\sigma_i^z\sigma_j^z$ in
Eq.~\eqref{eq:z2-strong-channel} as a direct operator insertion may introduce
negative weights. A sign-problem-free representation can instead be obtained by
rewriting the same channel in Kraus form.
Define 
\begin{align}
    M_0
    &=
    \sqrt{1-p}\,\mathds{1}_i\mathds{1}_j,
    \\
    M_1
    &=
    \sqrt{p}\,
    \frac{\mathds{1}_i\mathds{1}_j+\sigma_i^z\sigma_j^z}{2},
    \\
    M_2
    &=
    \sqrt{p}\,
    \frac{\mathds{1}_i\mathds{1}_j-\sigma_i^z\sigma_j^z}{2}.
\end{align}
Then Eq.~\eqref{eq:z2-strong-channel} can be equivalently written as
\begin{equation}
    \mathcal E_{p,b}[\tilde\rho_0]
    =
    \sum_{k=0}^{2}
    M_k\tilde\rho_0 M_k^\dagger .
\end{equation}

\begin{figure*}[ht!]
    \centering
    \begin{subfigure}[b]{0.86\linewidth}
        \centering
        \includegraphics[width=\linewidth]{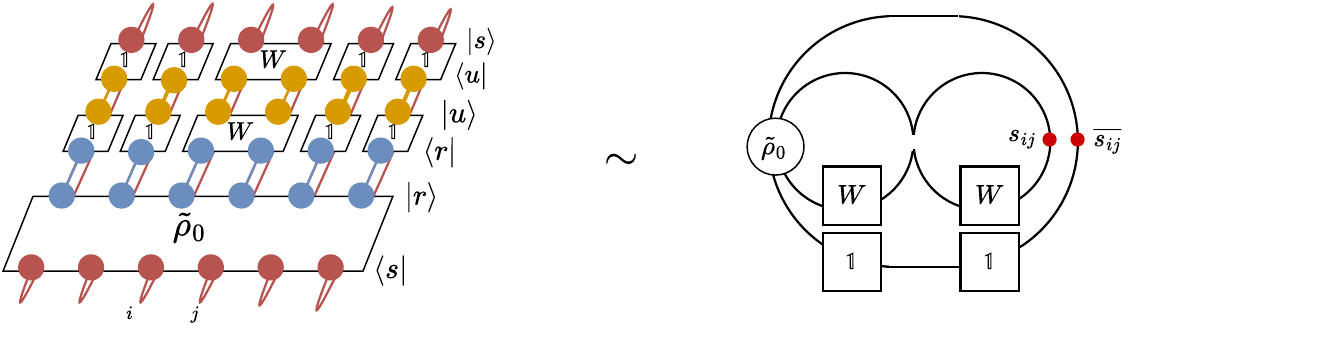}
        \caption{}
    \end{subfigure}
    % \hspace{0.01\textwidth}
    \begin{subfigure}[b]{0.86\linewidth}
        \centering
        \includegraphics[width=\linewidth]{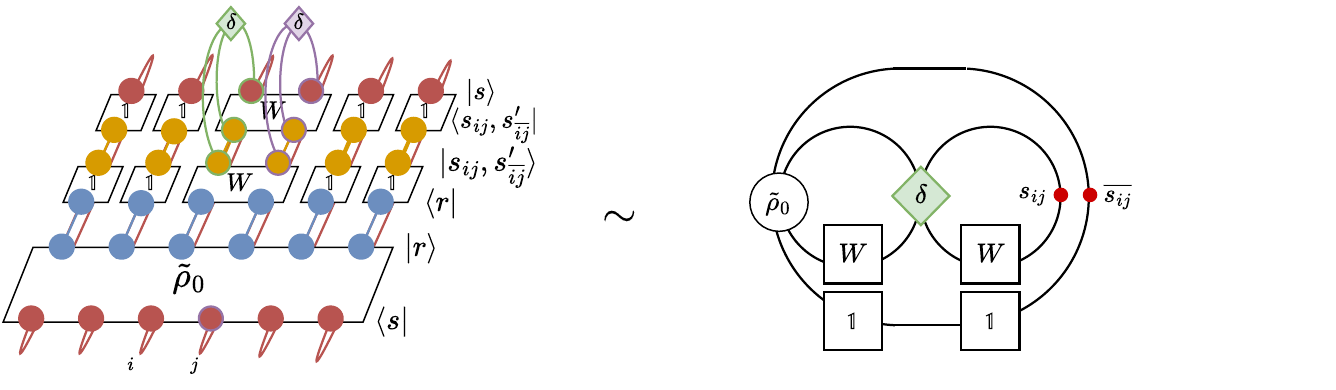}
        \caption{}
    \end{subfigure}
    \caption{
        Diagrammatic representations of the two sign-problem-free sectors generated by the
$\mathbb Z_2$-symmetric channel in
Eq.~\eqref{eq:z2-strong-channel}~\cite{ding2026swssb}.
(a) The first sector $\Tr(\tilde\rho_1)$ in Eq.~\eqref{eq:trace-rho1} corresponds to
a standard operator-inserted construction, with two additional insertions of
$W_b\in\{\mathds{1}_i\mathds{1}_j,\sigma_i^x\sigma_j^x\}$. The right panel shows
an equivalent simplified representation with the same boundary connectivity.
(b) The second sector $\Tr(\tilde\rho_2)$ in Eq.~\eqref{eq:trace-rho2}. In addition
to the operator insertion $W_b$, rank-four Kronecker-delta constraints identify
the boundary spins across four legs at the sites $i$ and $j$. The right panel
shows an equivalent representation in which the constraint
structure is represented more compactly.
    }
    \label{fig:z2-strong}
\end{figure*}

Consider describing the system in the local computational basis
\begin{equation}
    \ket{s}
    =
    \bigotimes_k \ket{s_k}
    =
    \ket{s_{ij}}\otimes\ket{s_{\overline{ij}}},
\end{equation}
where $\ket{s_{ij}}$ denotes the state on the two sites $i,j$, and
$\ket{s_{\overline{ij}}}$ denotes the state on the remaining sites. With this basis, one can prove that the channel can be decomposed into two sign-problem-free sectors~\cite{ding2026swssb},
\begin{equation}
    \mathcal E_{p,b}[\tilde\rho_0]
    =
    \tilde\rho_1+\tilde\rho_2 ,
\end{equation}
where
\begin{align}
    \tilde\rho_1
    &\equiv
    \frac{1-p}{2}
    \sum_{s,s'}
    \sum_{W_b}
    \langle s|\tilde\rho_0 W_b W_b|s'\rangle
    \ket{s}\bra{s'},
    \label{eq:sigma1}
    \\
    \tilde\rho_2
    &\equiv
    p
    \sum_s
    \sum_{s_{\overline{ij}}'}
    \sum_{W_b}
    \langle s|\tilde\rho_0 W_b|s_{ij},s_{\overline{ij}}'\rangle
    \ket{s}\bra{s_{ij},s_{\overline{ij}}'} W_b .
    \label{eq:sigma2}
\end{align}
Here
\begin{equation}
    W_b\in\{\mathds{1}_i\mathds{1}_j,\sigma_i^x\sigma_j^x\},\quad b=(i,j).
\end{equation}

By taking the trace over $\tilde\rho_1$ and $\tilde\rho_2$, one obtains two sector
normalization factors,
\begin{align}
    \Tr(\tilde\rho_1)
    &=
    \frac{1-p}{2}
    \sum_s
    \sum_{W_b}
    \langle s|\tilde\rho_0 W_b W_b|s\rangle
    \nonumber\\
    &=
    \frac{1-p}{2}
    \sum_{s,r,u}
    \sum_{W_b}
    \langle s|\tilde\rho_0|r\rangle
    \langle r|W_b|u\rangle
    \langle u|W_b|s\rangle ,
    \label{eq:trace-rho1}
    \\
    \Tr(\tilde\rho_2)
    &=
    p
    \sum_s
    \sum_{s_{\overline{ij}}'}
    \sum_{W_b}
    \langle s|\tilde\rho_0 W_b|s_{ij},s_{\overline{ij}}'\rangle
    \langle s_{ij},s_{\overline{ij}}'|W_b|s\rangle
    \nonumber\\
    &=
    p
    \sum_{s,r}
    \sum_{s_{\overline{ij}}'}
    \sum_{W_b}
    \langle s|\tilde\rho_0|r\rangle
    \langle r|W_b|s_{ij},s_{\overline{ij}}'\rangle
    \langle s_{ij},s_{\overline{ij}}'|W_b|s\rangle .
    \label{eq:trace-rho2}
\end{align}
Thus the diagonal contributions sampled in $\Tr(\tilde\rho_1)$ and $\Tr(\tilde\rho_2)$ are
decomposed into products of three matrix elements. For $\Tr(\tilde\rho_1)$ in
Eq.~\eqref{eq:trace-rho1}, the construction is a standard operator-inserted
state in the sense of Sec.~\ref{sec:rho-insert}, with two additional insertions
of $W$, as shown in Fig.~\ref{fig:z2-strong}(a).

The second sector, $\Tr(\tilde\rho_2)$ in Eq.~\eqref{eq:trace-rho2}, contains
additional boundary constraints. For the sites $i$ and $j$, the spin
configurations $s_{ij}$ appearing in the bra state of
$\langle s|\tilde\rho_0|r\rangle$, the ket state of
$\langle r|W|s_{ij},s_{\overline{ij}}'\rangle$, and both boundary states of
$\langle s_{ij},s_{\overline{ij}}'|W|s\rangle$ are constrained to be equal.
Equivalently, for each of the two sites $i$ and $j$, one inserts a rank-four
Kronecker-delta tensor $\delta^{(4)}$ that identifies the spins across these four boundary
legs, as illustrated in Fig.~\ref{fig:z2-strong}(b).

The literal diagrams for $\Tr(\tilde\rho_1)$ and especially $\Tr(\tilde\rho_2)$ can become
visually cumbersome, since they contain both operator insertions $W$ and $\delta^{(4)}$. An equivalent but cleaner representation, preserving
the same boundary connectivity, is shown on the right side of
Fig.~\ref{fig:z2-strong}~\cite{ding2026swssb}.

In contrast to the dephasing channel discussed in
Sec.~\ref{sec:dephasing-channel}, the configuration space now contains both
constraint variables and operator-insertion variables. For the full decoherence
channel in Eq.~\eqref{eq:z2-full-channel}, the configuration space is enlarged
as
\begin{equation}
    \mathcal C_0
    \;\to\;
    \mathcal C_0 \cup \{c_{b}\} \cup \{W_{b}\},
\end{equation}
where $\mathcal C_0$ denotes the configuration for simulating the original
density matrix $\tilde\rho_0$. The variable $c_{b}\in\{\mathds{1}_i\mathds{1}_j,\delta_i^{(4)}\delta_j^{(4)}\}$ labels the sector on
bond $b=(i,j)$, i.e., whether the Kronecker-delta constraint associated
with $\tilde\rho_2$ is present, while $W_b\in\{\mathds{1}_i\mathds{1}_j, \sigma_i^x\sigma_j^x\}$
specifies the local operator insertion on that bond. The constraint variables
$\{c_{b}\}$ divide the enlarged configuration space into
$2^{N_b}$ sectors, where $N_b$ is the number of bonds.

The sector updates for $\{c_b\}$ are analogous to those used for
the dephasing channel in Sec.~\ref{sec:dephasing-channel}. Within each sector,
one performs the usual QMC updates of $\mathcal C_0$, subject to the imposed
constraints, together with updates of the bond variables $W_{b}$.
In an SSE representation, the latter can be incorporated into the usual
operator-string updates of $\tilde\rho_0$.

\section{QMC in Special Basis Representations}
\label{sec:special_basis}

In the previous section, we introduced QMC representations of quantum states
beyond conventional Gibbs and ground-state ensembles. These constructions provide the general state-level
framework for evaluating nonlinear information-theoretic diagnostics.
Before turning to the more general approaches in Sec.~\ref{sec:pfr} based on these constructions, we first discuss two
special basis representations in which important nonlinear observables admit
particularly simple estimators. Both examples use bases built from two-spin or
two-copy structures rather than from single-site computational basis states.

The first example is projector QMC in the valence-bond basis~\cite{Liang1990vb-projector,Anders2005vb-projector,Sandvik2007-vb-projector,Sandvik2010vb-projector,Hasting2010vb-swap,Alet2007vb-ee,Kallin2009vb-ee,Melko2013sse,Kaul2013lattice-qmc,Zhou2024swap}. For quantum spin models whose low-energy states are naturally described by singlet structures, such as antiferromagnetic Heisenberg-type systems, valence-bond states provide
an efficient overcomplete basis. Because this basis is adapted to the singlet structure of the ground state, it can improve sampling efficiency and often reduces the projection length $m$ (see Sec.~\ref{sec:projector-qmc}) needed for ground-state convergence. More importantly for many-body quantum information, the valence-bond representation admits a direct estimator for the SWAP operators in Eqs.~\eqref{eq:swap-rho2} and~\eqref{eq:swap-rhoA2}, allowing the Rényi-2 entropy to be estimated without explicitly evaluating a partition-function ratio in Eq.~\eqref{eq:eg-renyi-ratio}, as first proposed in Ref.~\cite{Hasting2010vb-swap}.

The second example is Bell-QMC, proposed in Ref.~\cite{Tarabunga2025bell}, which formulates two-replica QMC directly in
a local Bell basis of the doubled Hilbert space. In this representation, the
SWAP operators in Eqs.~\eqref{eq:swap-rho2} and~\eqref{eq:swap-rhoA2} also
admit simple diagonal estimators. Moreover, all two-copy Pauli operators are diagonal in the Bell basis,
providing direct access to squared Pauli-string expectation values, in contrast to a conventional
single-copy computational basis where only one Pauli component is diagonal.
This makes the Bell-basis formulation particularly useful for diagnostics
involving Pauli-string expectation values, such as quantum magic or
nonstabilizerness, where the distribution of Pauli-string weights encodes magic structures.

% =================================================================================
%   NEW SECTION
% =================================================================================
\subsection{Projector QMC in the valence-bond basis}
\label{sec:vb_basis}

\subsubsection{Valence-bond basis and graph representation}

For an even number $N$ of spin-$1/2$ degrees of freedom, or qubits, a
valence-bond state is a product of $N/2$ singlets,
\begin{equation}
    \ket{V}
    =
    \prod_{\ell=1}^{N/2}
    \ket{i_\ell,j_\ell},
    \qquad
    \ket{(i,j)}
    =
    \frac{
        \ket{0_i 1_j}
        -
        \ket{1_i 0_j}
    }{\sqrt{2}} .
\end{equation}
Here $i$ and $j$ label lattice sites, and each pair $(i_\ell,j_\ell)$ forms one
valence bond. A valence-bond covering pairs every site exactly once.

On a bipartite lattice, we fix the phase convention of each singlet by ordering
it from sublattice I to sublattice II. Since $ \ket{i,j}=-\ket{j,i}$
this convention removes local sign ambiguities in the definition of
valence-bond states. 
With this convention, the bond operators
\begin{equation}
    H_{ij}=\frac{1}{4}-\mathbf S_i\cdot\mathbf S_j
\end{equation}
appearing in the
antiferromagnetic Heisenberg Hamiltonian
$H=-\sum_{\langle ij\rangle}J_{ij}H_{ij}$ 
has non-negative matrix elements in the valence-bond basis. Consequently, the
valence-bond basis provides a sign-problem-free projector representation for
bipartite antiferromagnets, including dimerized Heisenberg models and
$J$-$Q$-class models~\cite{Anders2005vb-projector}.

Though the valence-bond basis is nonorthogonal and massively overcomplete, the operator $H_{ij}$  acting on valence-bond states has simple local
rules~\cite{Sutherland1988vb-overlap,BeachSandvik2006valence}. If $i$ and $j$ already form a singlet,
\begin{equation}
    H_{ij}\ket{\cdots(i,j)\cdots}
    =
    \ket{\cdots(i,j)\cdots}.
\end{equation}
If $i$ and $j$ belong to different singlets, say $(i,k)$ and $(l,j)$ with the
bipartite orientation understood, then
\begin{equation}
    H_{ij}
    \ket{\cdots(i,k)(l,j)\cdots}
    =
    \frac{1}{2}
    \ket{\cdots(i,j)(l,k)\cdots}.
\end{equation}
Thus a bond operator either acts diagonally or reconnects two valence bonds,
with a positive matrix element in the bipartite sign-free case.

According to Sec.~\ref{sec:projector-qmc}, an observable $O$ is evaluated as
\begin{equation}
    \langle O\rangle_m
    =
    \frac{
        \sum_{\mathcal C}
        W_L W_R
        \bra{V_L}O\ket{V_R}
    }{
        \sum_{\mathcal C}
        W_L W_R
        \braket{V_L|V_R}
    }.
\end{equation}
Here $\ket{V_R}$ and $\bra{V_L}$ denote the propagated right and left states in
the valence-bond basis. Therefore, it is essential to know how to evaluate the
overlap between two valence-bond states.

We remark that there is a simple graph representation for evaluating this
overlap. Superimposing two valence-bond coverings gives a transposition graph,
whose connected components are closed loops formed by alternating bonds from
the two coverings. If $N_{\mathrm{loop}}$ is the number of loops and
$N/2$ is the number of valence bonds in one covering, then~\cite{Sutherland1988vb-overlap}
\begin{equation}\label{eq:count-loops}
    \langle V_L|V_R\rangle
    =
    2^{N_{\mathrm{loop}}-N/2}.
\end{equation}

\begin{figure}[ht!]
    \centering
    \begin{subfigure}[b]{0.3\linewidth}
        \centering
        \includegraphics[width=\linewidth]{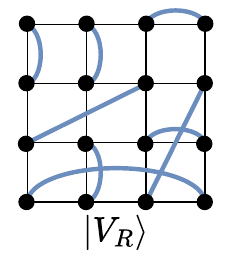}
        \caption{}
    \end{subfigure}
    % \hspace{0.01\textwidth}
    \begin{subfigure}[b]{0.3\linewidth}
        \centering
        \includegraphics[width=\linewidth]{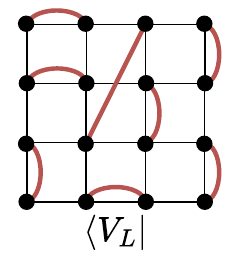}
        \caption{}
    \end{subfigure}
    % \hspace{0.01\textwidth}
    \begin{subfigure}[b]{0.3\linewidth}
        \centering
        \includegraphics[width=\linewidth]{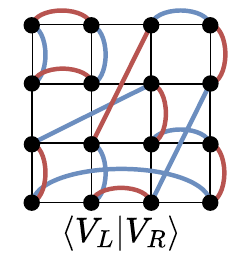}
        \caption{}
    \end{subfigure}
    \caption{
        Graph representation of the overlap between two valence-bond coverings on a $4\times 4$ square lattice. (a) and (b) show two valence-bond states, while (c)
shows the closed loops obtained by superimposing their bond configurations. In
this example, the number of valence bonds is $N/2=8$ and the number of loops is
$N_{\mathrm{loop}}=3$, giving
$\langle V_L|V_R\rangle=2^{N_{\mathrm{loop}}-N/2}=1/32$.
    }
     \label{fig:vb_transition_graph}
\end{figure}

Thus overlaps and many observables can be evaluated geometrically by
counting loops in the graph. 
An example adapted from Ref.~\cite{Anders2005vb-projector} is shown in
Fig.~\ref{fig:vb_transition_graph}.

% --------
% --------

\subsubsection{SWAP estimator for Rényi-2 entropy}

The valence-bond formulation is especially useful for the Rényi-2 entropy,
because the SWAP operator admits a simple geometric estimator~\cite{Sutherland1988vb-overlap,Hasting2010vb-swap}. According to
Sec.~\ref{sec:swap}, for a bipartite system $A\cup B$, the Rényi-2 entropy of a
pure ground state $\ket{\Psi_0}$ can be written as a two-copy expectation value,
\begin{equation}
    \Tr_A(\rho_A^2)
    =
    \frac{
        \bra{\tilde\Psi_0^{[1]}}\otimes\bra{\tilde\Psi_0^{[2]}}
        \mathrm{SWAP}_A
        \ket{\tilde\Psi_0^{[1]}}\otimes\ket{\tilde\Psi_0^{[2]}}
    }{
        \braket{\tilde\Psi_0^{[1]}|\tilde\Psi_0^{[1]}}
        \braket{\tilde\Psi_0^{[2]}|\tilde\Psi_0^{[2]}}
    } .
\end{equation}
Here $\mathrm{SWAP}_A$ exchanges the degrees of freedom in subsystem $A$
between the two copies and acts trivially on $B$. The superscripts $[1]$ and
$[2]$ label the two copies of the ground state.

Each copy of $\ket{\tilde\Psi_0}$ can be represented using valence-bond projector
QMC. Thus, in the two-copy representation, a configuration contains four
propagated valence-bond coverings,
\begin{equation}
    \bra{V_L^{[1]}},\quad \ket{V_R^{[1]}},
    \qquad
    \bra{V_L^{[2]}},\quad \ket{V_R^{[2]}} .
\end{equation}
For compactness, we denote the two-copy propagated ket and bra states by
\begin{equation}
    \ket{\mathcal V_R}
    :=
    \ket{V_R^{[1]}}\otimes \ket{V_R^{[2]}},
    \qquad
    \bra{\mathcal V_L}
    :=
    \bra{V_L^{[1]}}\otimes \bra{V_L^{[2]}} .
\end{equation}
The ordinary two-copy overlap is then
\begin{equation}
    \braket{\mathcal V_L|\mathcal V_R}
    =
    \braket{V_L^{[1]}|V_R^{[1]}}
    \braket{V_L^{[2]}|V_R^{[2]}} .
\end{equation}

\begin{figure}[ht!]
        \includegraphics[height=0.4\linewidth]{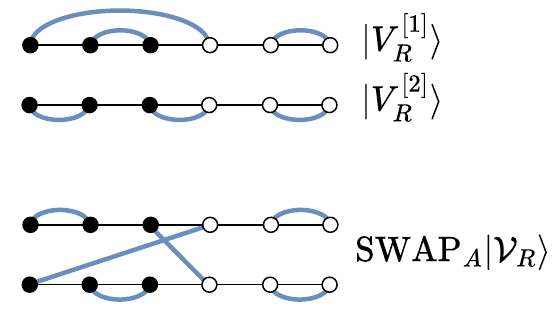}
    \caption{
        Illustration of the SWAP operation for a six-site chain. The left three
        sites form subsystem $A$, while the remaining sites form subsystem $B$.
        A two-copy valence-bond configuration
        $\ket{\mathcal{V}_R}=\ket{V_R^{[1]}}\otimes\ket{V_R^{[2]}}$ and the corresponding configuration $\mathrm{SWAP}_A\ket{\mathcal{V}_R}$  are shown.
    }
    \label{fig:swap-vb}
\end{figure}

Since the SWAP operator exchanges the degrees of freedom in subsystem $A$
between the two copies, its action on $\ket{\mathcal V_R}$ produces another
two-copy valence-bond state,
\begin{equation}
    \mathrm{SWAP}_A\ket{\mathcal V_R}.
\end{equation}
This state may contain valence bonds connecting spins from different copies, as
illustrated in Fig.~\ref{fig:swap-vb}. Therefore, in the valence-bond
representation, the expectation value of the SWAP operator reduces to a
standard overlap problem, but now for valence-bond states defined in the
doubled Hilbert space.

This gives the direct estimator
\begin{equation}
    \widehat{\mathrm{SWAP}}_A(\mathcal C)
    =
    \frac{
        \langle \mathcal V_L|\mathrm{SWAP}_A|\mathcal V_R\rangle
    }{
        \langle \mathcal V_L|\mathcal V_R\rangle
    },
\end{equation}
which can be evaluated using the loop-counting formula in
Eq.~\eqref{eq:count-loops}. The Rényi-2 entropy is then obtained from the
MC average,
\begin{equation}
    S_2(A)
    =
    -\ln
    \left\langle
        \widehat{\mathrm{SWAP}}_A
    \right\rangle .
\end{equation}

\subsubsection{The incremental trick}\label{sec:incremental-trick}

The direct SWAP estimator can become inefficient when
$\langle \mathrm{SWAP}_A\rangle=\Tr_A(\rho_A^2)$ is very small. Since
\begin{equation}
    \langle \mathrm{SWAP}_A\rangle
    =
    e^{-S_2(A)},
\end{equation}
this difficulty becomes severe for large subsystems. In particular, in higher
dimensions $S_2(A)$ typically grows with the boundary size $|\partial A|$, so
the SWAP expectation value can become exponentially small. This leads to a
rare-event problem, even before taking into account the intrinsic variance of
the estimator $\widehat{\mathrm{SWAP}}_A(\mathcal C)$.

A standard improvement is to use a ratio estimator, often called the
\emph{incremental trick}~\cite{Bennett1976interp,Hasting2010vb-swap}. Choose a sequence of nested regions
\begin{equation}
    \varnothing=A_0\subset A_1\subset\cdots\subset A_n=A .
\end{equation}
Then
\begin{equation}
    \langle \mathrm{SWAP}_{A}\rangle
    =
    \prod_{k=0}^{n-1}
    \frac{
        \langle \mathrm{SWAP}_{A_{k+1}}\rangle
    }{
        \langle \mathrm{SWAP}_{A_k}\rangle
    } .
\end{equation}
Equivalently,
\begin{equation}
    S_2(A)
    =
    -\sum_{k=0}^{n-1}
    \ln \mathfrak{R}_k,
    \qquad
    \mathfrak{R}_k
    =
    \frac{
        \langle \mathrm{SWAP}_{A_{k+1}}\rangle
    }{
        \langle \mathrm{SWAP}_{A_k}\rangle
    } .
\end{equation}

A similar idea has also been used in the snake algorithm in lattice gauge
theory, where an exponentially small ratio of partition functions is evaluated
as a product of ratios between neighboring ensembles
\cite{Forcrand2001snake}. In the many-body quantum information context, the same
principle appears naturally beyond this particular setup.

Each ratio $\mathfrak{R}_k$ can be measured in an ensemble reweighted by the denominator
$\mathrm{SWAP}_{A_k}$. In the valence-bond representation, this gives
\begin{equation}
    \mathfrak{R}_k
    =
    \left\langle
    \frac{
        \langle \mathcal V_L|
        \mathrm{SWAP}_{A_{k+1}}
        |\mathcal V_R\rangle
    }{
        \langle \mathcal V_L|
        \mathrm{SWAP}_{A_k}
        |\mathcal V_R\rangle
    }
    \right\rangle_{A_k},
\end{equation}
where $\langle\cdots\rangle_{A_k}$ denotes Monte Carlo averaging in the
ensemble whose weight includes the overlap
$\langle \mathcal V_L|\mathrm{SWAP}_{A_k}|\mathcal V_R\rangle$.

As shown in Ref.~\cite{Hasting2010vb-swap}, this construction reduces the
rare-event problem by comparing nearby subsystem sizes rather than estimating
an exponentially small SWAP expectation value directly. In practice, one adds
only a small number of sites when going from $A_k$ to $A_{k+1}$, so each ratio
$\mathfrak{R}_k$ remains of order unity and can be sampled much more
efficiently.

% =================================================================================
%   NEW SECTION
% =================================================================================
\subsection{Bell sampling in QMC}\label{sec:bell_basis}
\subsubsection{Formulation}

The key idea of Bell sampling is to formulate QMC directly in a two-replica 
Hilbert space. Following the notations in Eq.~\eqref{eq:two-copy-notation}, instead of working with a single density matrix $\tilde\rho$, we consider a two-replica state $\tilde\rho^{[1]}\otimes\tilde\rho^{[2]}$, and the two-replica partition function is therefore 
\begin{equation}\label{eq:two-copy-pt}
    \Tr\!\left(
        \tilde\rho^{[1]}\otimes\tilde\rho^{[2]}
    \right)
    =
    \sum_{\sigma}
    \bra{\sigma}
    \tilde\rho^{[1]}\otimes\tilde\rho^{[2]}
    \ket{\sigma},
\end{equation}
where we have inserted a complete basis $\{\ket{\sigma}\}$ of the doubled Hilbert space $\mathcal H^{[1]} \otimes \mathcal H^{[2]}$.

For each qubit $i$, a straightforward product basis of the doubled
Hilbert space is
\begin{equation}
    \ket{\sigma_i}
    =
    \ket{s_i^{[1]}}\otimes\ket{s_i^{[2]}} .
\end{equation}
Bell sampling instead uses an entangled basis for the pair of qubits formed by site $i$ in the two replicas:
\begin{align}
    \ket{\sigma_i^{00}}
    &=
    \frac{
        \ket{0_i^{[1]}0_i^{[2]}}
        +
        \ket{1_i^{[1]}1_i^{[2]}}
    }{\sqrt{2}},
    \nonumber \\ 
    \ket{\sigma_i^{01}}
    &=
    \frac{
        \ket{0_i^{[1]}1_i^{[2]}}
        +
        \ket{1_i^{[1]}0_i^{[2]}}
    }{\sqrt{2}},
    \nonumber \\
    \ket{\sigma_i^{10}}
    &=
    \frac{
        \ket{0_i^{[1]}0_i^{[2]}}
        -
        \ket{1_i^{[1]}1_i^{[2]}}
    }{\sqrt{2}},
     \nonumber \\
    \ket{\sigma_i^{11}}
    &=
    \frac{
        \ket{0_i^{[1]}1_i^{[2]}}
        -
        \ket{1_i^{[1]}0_i^{[2]}}
    }{\sqrt{2}} \label{eq:bell-basis} .
\end{align}
In the following, we will often omit the replica labels for simplicity when no
confusion can arise.

In short, Bell-QMC is a QMC formulation in which the two-replica
density matrix $\tilde\rho\otimes\tilde\rho$ is represented in the Bell basis. Equivalently,
one samples the two-replica partition function~\eqref{eq:two-copy-pt}, or more generally the
open-boundary matrix elements of $\tilde\rho\otimes\tilde\rho$, using the same
configuration-space logic introduced in Sec.~\ref{sec:general-rho}. Unlike the
valence-bond projector formulation discussed in Sec.~\ref{sec:vb_basis},
Bell-QMC does not assume a particular projector representation; its defining
feature is the choice of the local Bell basis in Eq.~\eqref{eq:bell-basis}. 

\subsubsection{Pauli-string estimators and quantum magic}
An important observation is that the local Bell basis is in one-to-one
correspondence with the Pauli operators through vectorization~\cite{montanaro2017learningstabilizerstatesbell}. For the identity
and Pauli operators acting on qubit $i$, define
\begin{align}
    \sigma^{00}
    &\equiv
    \mathds{1}
    =
    \ket{0}\bra{0}
    +
    \ket{1}\bra{1},
    \nonumber\\
    \sigma^{01}
    &\equiv
    \sigma^x
    =
    \ket{0}\bra{1}
    +
    \ket{1}\bra{0},
    \nonumber\\
    \sigma^{10}
    &\equiv
    \sigma^z
    =
    \ket{0}\bra{0}
    -
    \ket{1}\bra{1},
    \nonumber\\
    \sigma^{11}
    &\equiv
    i\sigma^y
    =
    \ket{0}\bra{1}
    -
    \ket{1}\bra{0}.
\end{align}

Now define the vectorization map
\begin{equation}
    \mathfrak V:
    \mathcal B(\mathcal H_i)
    \to
    \mathcal H_i^{[1]}\otimes\mathcal H_i^{[2]}
\end{equation}
through
\begin{equation}
    \mathfrak V\!\left(
        \ket{s}\bra{s'}
    \right)
    =
    \ket{s^{[1]}}\ket{s'^{[2]}} .
\end{equation} 

Applying the vectorization map to the Pauli operators gives precisely the Bell
basis states,
\begin{equation}
    \sigma^r
    \;\mapsto\;
    \ket{\sigma^r} \equiv \ket{r},
    \qquad
    r\in\{00,01,10,11\},
\end{equation}
up to an overall normalization factor. Thus the Bell basis may equivalently be
viewed as the vectorized Pauli basis in the doubled Hilbert space.

A key consequence is that the two-replica operators
$\sigma_i^r\otimes\sigma_i^r$ are diagonal in the Bell basis. 
Moreover, it is convenient
to separate the two binary components of the label
\begin{equation}
    r=r^a r^b,\quad r^a,r^b\in\{0,1\}.
\end{equation}
For a general Pauli-string operator
\begin{equation}
    \sigma^{\mathbf u}
    =
    \bigotimes_i \sigma_i^{u_i},
    \qquad
    u_i=u_i^a u_i^b,
\end{equation}
the corresponding two-copy observable
$\sigma^{\mathbf u}\otimes\sigma^{\mathbf u}$ therefore admits a diagonal estimator in
the Bell basis.

In a Bell-QMC simulation, each site $i$ carries a Bell-state label
$r_i=r_i^a r_i^b$. Then one obtains~\cite{Tarabunga2025bell}
\begin{align}
    \frac{
        \Tr\!\left[
            (\tilde\rho\otimes\tilde\rho)
            (\sigma^{\mathbf u}\otimes\sigma^{\mathbf u})
        \right]
    }{
        \Tr(\tilde\rho\otimes\tilde\rho)
    }
    &=
    \frac{
        \left[
            \Tr(\tilde\rho\,\sigma^{\mathbf u})
        \right]^2
    }{
        [\Tr(\tilde\rho)]^2
    }
    \nonumber\\
    &=
    \left\langle
        \prod_i
        (-1)^{
            u_i^b r_i^a
            -
            u_i^a r_i^b
        }
    \right\rangle,
\end{align}
where the expectation value is taken over Bell-QMC configurations.
Therefore,
\begin{equation}
    \left|
        \frac{
            \Tr(\tilde\rho\,\sigma^{\mathbf u})
        }{
            \Tr(\tilde\rho)
        }
    \right|
    =
    \sqrt{
        \left\langle
            \prod_i
            (-1)^{
                u_i^b r_i^a
                -
                u_i^a r_i^b
            }
        \right\rangle
    } .
\end{equation}
This determines the magnitude of an arbitrary Pauli-string expectation value through diagonal two-copy measurements. From these expectation values, one may further construct information-theoretic quantities such as the stabilizer entropy~\cite{leone2022sre,HaugLeeKim2024stabilizer}, discussed in Sec.~\ref{sec:stabilizer-entropy}.

Moreover, when $\rho=\ket{\Psi}\bra{\Psi}$ is a normalized pure state with real coefficients, Bell sampling on
$\ket{\Psi}\otimes\ket{\Psi}$ produces an outcome $\mathbf r$ with probability
\begin{equation}\label{eq:characteristic}
    \Xi_{\mathbf r}(\ket{\Psi})
    :=
        \frac{
            \left|
                \bra{ \Psi}
                \sigma^{\mathbf r}
                \ket{ \Psi}
            \right|^2
        }{
            2^N
        } ,
\end{equation}
which is known as the characteristic function of quantum
magic~\cite{montanaro2017learningstabilizerstatesbell,Gross2021charac}. The
distribution $\Xi_{\mathbf r}$ characterizes the Pauli spectrum of the quantum
state and encodes its nonstabilizer structure.

\subsubsection{SWAP estimator and Rényi-2 entropy}

Another useful feature of the Bell basis is that each local SWAP operator is
diagonal. For each qubit $i$, the operator $\mathrm{SWAP}_i$ exchanges the two
replicas at site $i$. Therefore, for a subsystem $A$, we have~\cite{Tarabunga2025bell} 
\begin{equation}\label{eq:swap-bell-estimator}
    \frac{
       \Tr\!\left[
           (\tilde\rho\otimes\tilde\rho)\mathrm{SWAP}_A
       \right]
   }{
       \Tr(\tilde\rho\otimes\tilde\rho)
   }
    =
    \left\langle
        \prod_{i\in A}
        (-1)^{r_i^a r_i^b}
    \right\rangle .
\end{equation}
For a normalized pure state, this gives the Rényi-2 entanglement entropy
\begin{equation}
    S_2(A)
    =
    -\ln
    \left\langle
        \prod_{i\in A}
        (-1)^{r_i^a r_i^b}
    \right\rangle .
\end{equation}

The estimator for $\Tr(\rho_A^2)$ is bounded because it takes values $\pm1$.
Its variance is therefore
\begin{equation}
    \mathrm{Var}\!\left[
        \widehat{\Tr(\rho_A^2)}
    \right]
    =
    1-\left[\Tr(\rho_A^2)\right]^2
    \le 1 ,
\end{equation}
where
$\widehat{\Tr(\rho_A^2)}$
denotes the estimator on the right-hand side of
Eq.~\eqref{eq:swap-bell-estimator}. Thus the purity~\eqref{eq:swap-bell-estimator} itself can be estimated
with bounded variance.

For the entropy $S_2=-\ln\Tr(\rho_A^2)$, error propagation gives
\begin{equation}
    \mathrm{Var}(S_2)
    \sim 
    \frac{
        1-\left[\Tr(\rho_A^2)\right]^2
    }{
        \left[\Tr(\rho_A^2)\right]^2
    }
    =
        e^{2S_2}-1.
\end{equation}
Consequently,
\begin{equation}
    \mathrm{Var}(S_2)
    \lesssim
    \mathcal O(\mathrm{poly}(N)),
\end{equation}
so the estimator remains polynomially efficient. In higher dimensions,
however, the Rényi entropy generically satisfies an area law
$S_2\sim |\partial A|$, leading to an exponential scaling
\begin{equation}
    \mathrm{Var}(S_2)
    \sim
    \mathcal O(e^{c|\partial A|}),
\end{equation}
with the subsystem boundary area $|\partial A|$. 

For $(1+1)$D systems with local Hamiltonians, the Rényi-2 entropy grows
at most logarithmically with subsystem size, while in gapped phases it obeys an
area law and remains constant. 
This feature mitigates the limitations of DMRG when applied to periodic chains. As demonstrated in Ref.~\cite{Tarabunga2025bell}, this approach allows efficient computation of systems with over 1000 qubits.
Furthermore, even such
simple diagonal measurements, without employing advanced
techniques such as the partition-function-ratio methods discussed in
Sec.~\ref{sec:pfr}, already enable Bell-QMC simulations of the $(2+1)$D
$\mathbb Z_2$ gauge-theory ground state on systems as large as
$20\times20=400$ qubits.
\section{Generalized Partition-Function Ratios}\label{sec:pfr}

In this section, we review arguably the most fundamental framework for computing nonlinear observables within QMC, namely the formulation in terms of \emph{partition-function ratios}. 
This framework underlies a wide range of methods and serves as a conceptual foundation for more advanced techniques, as it expresses nonlinear observables as ratios of normalization constants associated with distinct ensembles. 
% Throughout this section, we use the second-order Rényi entanglement entropy as a guiding example to illustrate the general structure. 
The present section develops the formalism, while its broader applications will be discussed in Sec.~\ref{sec:nonlinear}.

% ================================================
%   NEW SUBSECTION
% ================================================
\subsection{The partition-function-ratio problem}

We consider a general setting in which a family of (possibly unnormalized) weights 
$W_{\theta}(\mathcal{C}) \in\mathbb{R}^+$ defines a collection of ensembles labeled by a parameter 
$\theta$, with normalization constant, or (generalized) partition function,
\begin{equation}
    Z(\theta) = \sum_{\mathcal{C}\in\Omega} W_{\theta}(\mathcal{C}),
\end{equation}
where $\theta$ need not be a scalar parameter, but may also encode more general features such as the geometry or boundary conditions of the system like that in Sec.~\ref{sec:qmc_form}.

The \emph{partition-function-ratio problem} is to evaluate
\begin{equation}\label{eq:ratio}
    \mathfrak{R}(\theta_0,\theta_1) = \frac{Z(\theta_0)}{Z(\theta_1)},
\end{equation}
or, equivalently, its logarithmic form,
\begin{equation}
    \log \mathfrak{R}(\theta_0,\theta_1)
    = \ln Z(\theta_0) - \ln Z(\theta_1),
\end{equation}
given the ability to sample configurations from one or more of the corresponding ensembles.

Such ratios arise whenever quantities of interest depend on normalization constants rather than expectation values within a single ensemble. In particular, many nonlinear observables in quantum information can be expressed in this form.

A representative example is again the entanglement Rényi-$\alpha$ entropy
\begin{equation}
   S_{\alpha}(\rho_A)
    = \frac{1}{1-\alpha}\,\ln \frac{\Tr(\tilde\rho_A^{\alpha})}{[\Tr(\tilde\rho_A)]^{\alpha}} ,
\end{equation}
mentioned in Sec.~\ref{sec:swap},
where $\tilde\rho_A = \Tr_B(\tilde\rho)$.

By defining the partition functions 
\begin{equation}
    Z^{(\alpha)}_A := \Tr(\tilde\rho_A^{\alpha}),\qquad 
    Z^{\alpha} := [\Tr(\tilde\rho_A)]^{\alpha} \equiv Z^{(\alpha)}_{\varnothing},
\end{equation}
we can write
\begin{equation}
    S_{\alpha}(\rho_A)
    = \frac{1}{1-\alpha}\,\ln \frac{Z^{(\alpha)}_A}{Z^{(\alpha)}_{\varnothing}},
\end{equation}
thereby providing a concrete realization of the partition-function-ratio problem for entanglement Rényi entropies.

As discussed in Sec.~\ref{sec:rep_states}, $Z^{(\alpha)}_A$ and $Z^{(\alpha)}_{\varnothing}$ correspond to partition functions defined on $\alpha$-replicated systems. They share the same underlying system, but differ in how the trace is implemented: $Z^{(\alpha)}_A$ involves cyclic gluing of replicas across subsystem $A$, whereas $Z^{(\alpha)}_{\varnothing}$ corresponds to independent traces over each replica. 
In this setting, the parameter $\theta$ in Eq.~\eqref{eq:ratio} can be interpreted as specifying the subsystem on which the replica gluing is imposed.

Historically, the evaluation of partition-function ratios has been extensively studied in classical statistical physics, molecular dynamics, and statistics, where it arises in the computation of free-energy differences and normalization constants. Early developments date back to the late 1970s, which establish the foundational toolkit for estimating partition-function ratios in the context of many-body quantum information.
Comprehensive reviews of these methods in classical settings can be found in Refs.~\cite{neal1993probabilistic,XLMeng1998normalizing,FRENKEL2002167freeEnergyBook}. 
We note that biased MC sampling methods, such as Wang--Landau sampling~\cite{WangLandau2001,Troyer2003quantumwang-landau,Mendes-Santos2020wang-landau,Inglis2013wang-landau,ding2024rweight}, are not discussed here.

We will next revisit these foundational ideas through the lens of the entanglement Rényi-2 entropy $S_2(\rho_A)$, which serves as a concrete example throughout this section. 
While the underlying principles remain unchanged, their implementation in quantum settings can require nontrivial adaptations, including the construction of the more intricate states introduced in Sec.~\ref{sec:qmc_form}. 
The following sections therefore introduce both the classical foundations and their modern extensions. 

% ================================================
%   NEW SUBSECTION
% ================================================
\subsection{Extended ensemble and direct estimators of the ratio}
\label{sec:direct-estimator}
One approach is to formulate the ratio $Z_A^{(2)}/Z^{(2)}_{\varnothing}$ within an extended ensemble $Z^{(2)}_A \cup Z^{(2)}_{\varnothing}$ that samples configurations associated with both sectors. The ratio can then be obtained from the relative occupation probabilities of the two sectors~\cite{Stephan2012eeratio}. This construction can be viewed as a natural extension of the classical Bennett acceptance-ratio method~\cite{Bennett1976interp}.

Specifically, given a QMC representation, we write
\begin{equation}
    Z^{(2)}_\theta
    =
    \sum_{\mathcal{C} \in \Omega_\theta}
    W_\theta(\mathcal{C}),
    \qquad
    \theta \in \{A,\varnothing\},
\end{equation}
assuming that the two sectors admit a compatible configuration representation (typically a shared configuration space). 

We then introduce an extended configuration space
\begin{equation}
    \Omega_{\mathrm{ext}}
    =
    \bigl\{
    (\mathcal{C},\theta)
    \;\big|\;
    \mathcal{C} \in \Omega_\theta,\;
    \theta \in \{A,\varnothing\}
    \bigr\},
\end{equation}
with total partition function
\begin{equation}
    Z^{(2)}_{\mathrm{ext}}
    =
    \sum_{\theta}
    \sum_{\mathcal{C} \in \Omega_\theta}
    W_\theta(\mathcal{C})
    =
    Z_A^{(2)} + Z_{\varnothing}^{(2)}.
\end{equation}
In this extended ensemble, a configuration $(\mathcal{C},\theta)$ is assigned weight $W_{\theta}(\mathcal{C})$, with normalization factor $Z^{(2)}_{\mathrm{ext}}$.

Sampling is performed using standard QMC updates within each sector, supplemented by Metropolis--Hasting updates that propose switches between sectors. 
For example, a proposed transition $(\mathcal{C},\varnothing)\to(\mathcal{C},A)$ is accepted with probability
\begin{equation}\label{eq:transition-direct-estimator}
    P\bigl((\mathcal{C},\varnothing)\to(\mathcal{C},A)\bigr)
    =
    \min\!\left\{1,\,
    \frac{W_A(\mathcal{C})}{W_{\varnothing}(\mathcal{C})}
    \right\},
\end{equation}
with the reverse move defined analogously. This expression assumes symmetric forward and reverse proposal probabilities.

\begin{figure}[ht]
    \centering
    \begin{subfigure}[b]{0.35\textwidth}
        \centering
        \includegraphics[width=\linewidth]{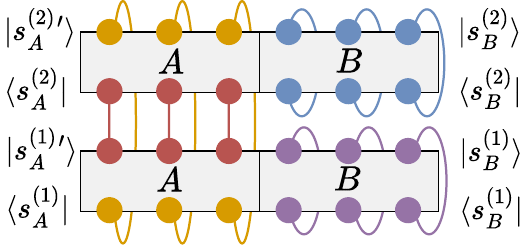}
        \caption{}
    \end{subfigure}
    \hspace{0.05\textwidth}
    \begin{subfigure}[b]{0.35\textwidth}
        \centering
        \includegraphics[width=\linewidth]{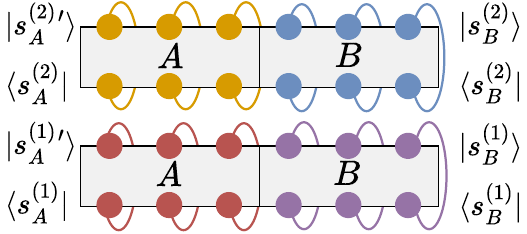}
        \caption{}
    \end{subfigure}
    \caption{
    Diagrammatic representations of (a) $Z^{(2)}_A$ and (b) $Z^{(2)}_{\varnothing}\equiv Z^2$.
    Here $s$ and $r$ denote qubit states in the two replicas, with subscripts indicating the bipartition into regions $A$ and $B$, respectively.
    In (a), the gluing condition enforces $s_A^{(1)}{}' \equiv s_A^{(2)}$.
    }
    \label{fig:S2_GPF}
\end{figure}

A necessary condition for such a transition to be admissible is that the configuration $\mathcal{C}$ belongs simultaneously to both configuration spaces, i.e.,
$\mathcal{C} \in \Omega_A \cap \Omega_{\varnothing}$.
As illustrated in Fig.~\ref{fig:S2_GPF}, this compatibility condition requires that the qubit states on subsystem $A$ match across the replica interface, namely
$s_A^{(1)}{}' = s_A^{(2)}$.

The marginal probability of the sector label $\theta$ in the extended ensemble is given by $P(\theta)
    =
    Z_\theta^{(2)}/{Z^{(2)}_{\mathrm{ext}}}$. 
In practice, during a QMC simulation, one records the number of visits to each sector, denoted by $N_A$ and $N_{\varnothing}$. The ratio is then estimated by~\cite{Stephan2012eeratio}
\begin{equation}\label{eq:direct_estimator}
    \frac{Z_A^{(2)}}{Z_{\varnothing}^{(2)}} = \frac{P(\theta = A)}{P(\theta = \varnothing)}
    \approx
    \frac{N_A}{N_{\varnothing}},
\end{equation}
which becomes exact in the long-sampling limit.

Since ${Z_A^{(2)}}/{Z_{\varnothing}^{(2)}}
    = e^{-S_2(\rho_A)}$,
and the Rényi-2 entropy $S_2(\rho_A)$ typically obeys an area law for ground states of local Hamiltonians and a volume law for generic quantum states, the ratio becomes exponentially small with increasing system size.

From a configuration-space perspective, the direct estimator~\eqref{eq:direct_estimator} becomes increasingly inefficient with system size due to a vanishing overlap between the two sectors. As the subsystem $A$ grows, the subset of configurations that satisfy the gluing constraint constitutes an exponentially small fraction of the total configuration space. Consequently, transitions between the two sectors become exponentially rare, leading to a rapid degradation of sampling efficiency.

To overcome this difficulty, one commonly employs the incremental trick, also introduced in Sec.~\ref{sec:incremental-trick}. We decompose the subsystem $A$ into a sequence of nested regions
\begin{equation}
   A_0 \subset A_1 \subset \cdots \subset A_{|A|} = A,
   \qquad |A_i| = i,
\end{equation}
and express the ratio as a product of incremental contributions,
\begin{equation}\label{eq:discrete-incremental}
    \frac{Z_A^{(2)}}{Z_{\varnothing}^{(2)}}
    =
    \prod_{i=1}^{N}
    \frac{Z_{A_i}^{(2)}}{Z_{A_{i-1}}^{(2)}}.
\end{equation}
Similar extended ensemble can be constructed to evaluate each ratios on the right-hand side. 
Each ratio has a smaller change between the two ensembles and therefore improves the overlap relative to the direct estimation of $Z_A^{(2)}/Z_{\varnothing}^{(2)}$. This can substantially mitigate the severe signal-to-noise problem and make the estimator practical for larger subsystems~\cite{Stephan2012eeratio}.

% ================================================
%   NEW SUBSECTION
% ================================================
\subsection{Reweight-annealing method}\label{sec:reweight-annealing}
Another approach, often referred to as the \emph{reweight-annealing
method}, combines reweighting with simulated annealing
\cite{ding2024rweight,Xuxy2025reweight,Wang2025reweight,ding2025negativity,Wang2026reweight-off,Ma2024sign}. The core 
\emph{reweighting trick} has a long history in MC simulations, dating
back at least to early applications in statistical mechanics
\cite{Zwanzig1954perturbation,Karplus1959reweighting} and later formalized in the widely used
histogram-reweighting framework of Ferrenberg and Swendsen
\cite{Ferrenberg1988reweighting}. It has since become a standard tool in both classical MC and QMC.

In contrast to the preceding subsection, where $\theta$ parametrizes boundary conditions, here we consider a scalar parameter $\beta$ that continuously controls the statistical weights in the partition function.
This parameter may be interpreted as the inverse temperature or, more generally, as a scalar control parameter, such as a coupling constant in the Hamiltonian.

Accordingly, both the numerator and denominator in the ratio $Z_{A}^{(2)}/Z_{\varnothing}^{(2)}$ become smooth functions of $\beta$.
Suppose there exists a \emph{reference point} $\beta_0$ at which the ratio $Z_{A}^{(2)}(\beta_0)/Z_{\varnothing}^{(2)}(\beta_0)$, or equivalently the Rényi-2 entropy $S_2(\beta_0)$, is known \emph{a priori}~\cite{Wang2025reweight}. 
We then write
\begin{align}
    S_2(\beta) - S_2(\beta_0)
    &=
    \ln\frac{Z_A^{(2)}(\beta_0)}{Z_{A}^{(2)}(\beta)}
    -
    \ln\frac{Z_{\varnothing}^{(2)}(\beta_0)}{Z_{\varnothing}^{(2)}(\beta)}.
\end{align}
Therefore, $S_2$ at a target parameter $\beta$ can be obtained by estimating the two ratios
\begin{equation}
    \frac{Z_A^{(2)}(\beta_0)}{Z_A^{(2)}(\beta)},
    \qquad
    \frac{Z_{\varnothing}^{(2)}(\beta_0)}{Z_{\varnothing}^{(2)}(\beta)},
\end{equation}
independently.

This idea is quite general for extracting observables in QMC, since many quantities can be formulated as ratios of two generalized partition functions, including off-diagonal observables~\cite{Wang2026reweight-off}. By applying reweighting--annealing separately to the numerator and the denominator---a procedure we refer to as bipartite reweight-annealing (BRA)---and fixing their relative normalization through the ratio at a reference point that can be evaluated straightforwardly, one can reconstruct the desired ratio along the entire annealing path~\cite{Wang2026reweight-off,Forcrand2001snake}.

The reference point $\beta_0$ is often simple to choose. 
For example, when $\beta$ is interpreted as the inverse temperature, one may take $\beta_0 \to 0$, where the system is maximally mixed. 
For a qubit system with $|A|$ qubits in subsystem $A$, the normalized RDM is then $\rho_A=\mathds{1}_A/2^{|A|}$, giving
\begin{equation}
    S_2(\rho_A;\beta_0=0)
    =
    -\ln \operatorname{Tr}({\rho}_A^2)
    =
    |A|\ln 2 .
\end{equation}

One advantage of this formulation is that, in contrast to directly evaluating $Z_A^{(2)}/Z_{\varnothing}^{(2)}$ at a fixed $\beta$, it does not require changing the geometry (i.e., replica gluing conditions) of the path integral during the simulation. 
Instead, one performs sampling within a fixed configuration space and relates different $\beta$ values via the reweighting trick.

More concretely, for 
\begin{equation}
    Z_A^{(2)}(\beta)
    =
    \sum_{\mathcal{C} \in \Omega_A}
    W_{\beta}(\mathcal{C}),
\end{equation}
where the configuration space $\Omega_A$ is independent of $\beta$, 
the ratio can be written as
\begin{align}
    \frac{Z_A^{(2)}(\beta_0)}{Z_A^{(2)}(\beta)}
    &=
    \frac{\sum_{\mathcal{C} \in \Omega_A} W_{\beta_0}(\mathcal{C})}{
          \sum_{\mathcal{C} \in \Omega_A} W_{\beta}(\mathcal{C})}
    \nonumber \\
    &=
    \frac{\sum_{\mathcal{C} \in \Omega_A}
    \frac{W_{\beta_0}(\mathcal{C})}{W_{\beta}(\mathcal{C})}
    \, W_{\beta}(\mathcal{C})}{
    \sum_{\mathcal{C} \in \Omega_A} W_{\beta}(\mathcal{C})}
    \nonumber \\
    &=
    \left\langle 
    \frac{W_{\beta_0}(\mathcal{C})}{W_{\beta}(\mathcal{C})}
    \right\rangle_{\beta, A},
\end{align}
where $\langle \cdots \rangle_{\beta, A}$ denotes an expectation value evaluated in the ensemble with weight $W_{\beta}$ restricted to $\Omega_A$. 

In practice, direct reweighting between $\beta_0$ and $\beta$ can also suffer from poor overlap between the corresponding ensembles. 
To ensure stable importance sampling, one introduces an annealing path based on the incremental trick, by partitioning the interval $[\beta_0,\beta]$ into $m$ subintervals (assuming $\beta_0<\beta$ without loss of generality),
\begin{equation}
    \beta_0 < \beta_1 < \cdots < \beta_m \equiv \beta.
\end{equation}
The ratio can then be factorized as
\begin{equation}\label{eq:beta-incremental}
    \frac{Z_A^{(2)}(\beta_0)}{Z_A^{(2)}(\beta)}
    =
    \prod_{i=1}^{m}
    \frac{Z_A^{(2)}(\beta_{i-1})}{Z_A^{(2)}(\beta_i)} 
    =
    \prod_{i=1}^m
    \left\langle 
    \frac{W_{\beta_{i-1}}(\mathcal{C})}{W_{\beta_i}(\mathcal{C})}
    \right\rangle_{\beta_i, A},
\end{equation}
with each factor evaluated via reweighting.
An analogous expression holds for $Z_{\varnothing}^{(2)}$. 
For replicated states constructed from Gibbs states, it has been shown that the number of interpolation points $m$ required to achieve a fixed error tolerance scales only polynomially with the system size~\cite{ding2024rweight,ding2025negativity}.

Compared with the discrete incremental construction in Eq.~\eqref{eq:discrete-incremental}, whose resolution is limited by the number of sites in subsystem $A$, Eq.~\eqref{eq:beta-incremental} allows an arbitrary number of intermediate interpolation points. 
This makes it possible to incorporate simulated annealing~\cite{Kirkpatrick1983SimulatedAnnealing} and evaluate all intermediate ratios sequentially within a single simulation.
Starting from the reference point $\beta_0$, the parameter is gradually tuned along the annealing path 
\begin{equation}
    \beta_0 \to \beta_1 \to \cdots \to \beta_m \equiv \beta ,
\end{equation}
and at each step the system is equilibrated and measurements are performed before proceeding to the next value. 
In this way, all ratios in the product decomposition \eqref{eq:beta-incremental} are obtained sequentially within a single run, thus significantly reducing computational overhead and improves sampling efficiency by maintaining continuity between neighboring ensembles. 

Moreover, in both the discrete incremental construction of Eq.~\eqref{eq:discrete-incremental} in Sec.~\ref{sec:direct-estimator} and the continuous interpolation of Eq.~\eqref{eq:beta-incremental}, the intermediate points have direct physical meaning and are not merely auxiliary. 
In the former case, they provide the Rényi-2 entropy as a function of the subsystem size $|A|$. 
In the present case, they provide a smooth curve showing how $S_2$ varies with some parameter, which is particularly useful for studying the dependence of $S_2$ on a tunable system parameter.

% ================================================
%   NEW SUBSECTION
% ================================================
\subsection{Thermodynamic integration}\label{sec:thermo-int}

The reweight--annealing method described above treats the numerator and denominator of
$Z_A^{(2)}/Z_{\varnothing}^{(2)}$ separately, expressing each partition-function ratio
as a product of small, well-overlapped reweighting factors along the $\beta$ path.
Equivalently, the same continuous interpolation can be formulated in differential
form: one computes $\partial_\beta \ln Z_A^{(2)}$ and
$\partial_\beta \ln Z_{\varnothing}^{(2)}$ as expectation values and integrates
them along the path. This gives the thermodynamic-integration method~\cite{Melko2010thmint-mi,Singh2010thmint-mi}.

Let us first recall the general estimator underlying this method. For a standard
canonical ensemble, one may write
\begin{equation}
    Z(\beta)
    =
    \operatorname{Tr}(e^{-\beta H})
    =
    \sum_{\mathcal C\in\Omega} W_\beta(\mathcal C),
\end{equation}
Typically, the configuration space $\Omega$ is
independent of $\beta$, thus we can differentiate it to obtain
\begin{align}
    \frac{\partial \ln Z(\beta)}{\partial \beta}
    &=
    \frac{1}{Z(\beta)}
    \sum_{\mathcal C\in\Omega}
    \partial_\beta W_\beta(\mathcal C)
    \nonumber \\
    &=
    \frac{1}{Z(\beta)}
    \sum_{\mathcal C\in\Omega}
    W_\beta(\mathcal C)\,
    \partial_\beta \ln W_\beta(\mathcal C)
    \nonumber \\
    &=
    \left\langle
    \partial_\beta \ln W_\beta(\mathcal C)
    \right\rangle_{\beta}.
    \label{eq:effective-E}
\end{align}
On the other hand, using $Z(\beta)=\operatorname{Tr}(e^{-\beta H})$, we have 
\begin{equation}\label{eq:partialZ}
    \frac{\partial \ln Z(\beta)}{\partial \beta}
    =
    -\frac{\operatorname{Tr}(H e^{-\beta H})}{Z(\beta)}
    =
    -E(\beta),
\end{equation}
where $E(\beta)=\langle H\rangle_\beta$ is the internal energy. This means, for the
ordinary canonical partition function, $\partial_\beta \ln Z(\beta)$ is the
negative internal energy.

Note that Eq.~\eqref{eq:effective-E} applies equally to generalized
partition functions such as $Z_A^{(2)}$, provided that $\beta$ only changes the
statistical weights within a fixed configuration space. Writing
\begin{equation}
    Z_A^{(2)}(\beta)
    =
    \sum_{\mathcal C\in\Omega_A}
    W_{A,\beta}(\mathcal C),
\end{equation}
we define an effective internal energy by
\begin{equation}\label{eq:lnzA2}
    \frac{\partial \ln Z_A^{(2)}(\beta)}{\partial \beta}
    =
    \left\langle
    \partial_\beta \ln W_{A,\beta}(\mathcal C)
    \right\rangle_{\beta,A}
    \equiv
    -E_A^{(2)}(\beta).
\end{equation}
Similarly,
\begin{equation}\label{eq:lnZphi2}
    \frac{\partial \ln Z_{\varnothing}^{(2)}(\beta)}{\partial \beta}
    =
    \left\langle
    \partial_\beta \ln W_{\varnothing,\beta}(\mathcal C)
    \right\rangle_{\beta,\varnothing}
    \equiv
    -E_{\varnothing}^{(2)}(\beta).
\end{equation}
Since
\begin{equation}
    S_2(\beta)
    =
    -\ln
    \frac{Z_A^{(2)}(\beta)}
         {Z_{\varnothing}^{(2)}(\beta)},
\end{equation}
one obtains
\begin{equation}\label{eq:derivative-estimator}
    \frac{\partial S_2(\beta)}{\partial \beta}
    =
    E_A^{(2)}(\beta)
    -
    E_{\varnothing}^{(2)}(\beta).
\end{equation}

Therefore,
\begin{equation}\label{eq:thermo-int}
    S_2(\beta)
    =
    S_2(\beta_0)
    +
    \int_{\beta_0}^{\beta}
    d\beta'\,
    \left[
    E_A^{(2)}(\beta')
    -
    E_{\varnothing}^{(2)}(\beta')
    \right].
\end{equation}

Eq.~\eqref{eq:thermo-int} is the thermodynamic-integration formula for computing the Rényi-2 entropy~\cite{Melko2010thmint-mi,Singh2010thmint-mi}.
Compared with the reweight--annealing method, thermodynamic integration replaces the overlap problem by the numerical evaluation of a $1$D integral. 
Its efficiency is therefore controlled by two sources of error: the statistical error in the effective-energy estimators and the systematic error from numerical quadrature. 
For replicated partition functions constructed from local Gibbs weights, the effective energies are extensive observables, whose variances scale at most polynomially with the system size. 
Consequently, for a prescribed error tolerance, the required number of interpolation points, together with the MC statistics needed at each point, scales polynomially with the system size, consistent with the scaling expected for the reweight--annealing method. 

% ================================================
%   NEW SUBSECTION
% ================================================
\subsection{Derivatives estimators}\label{sec:pf_derivative}
As shown in Sec.~\ref{sec:thermo-int}, thermodynamic integration expresses the
variation of a partition function in terms of derivatives of its logarithm.
Eqs~\eqref{eq:partialZ}, \eqref{eq:lnzA2},
and~\eqref{eq:lnZphi2} therefore provide direct estimators for the
first derivatives of the corresponding partition functions. For observables expressed as combinations of partition functions, such as the Rényi entropy $S_2$, this immediately yields direct estimators for their first derivatives, as in Eq.~\eqref{eq:derivative-estimator}~\cite{Wang2025reweight,Wu2020negativity,ding2025magic}.

Compared with thermodynamic integration, which reconstructs the original
function by numerically integrating such derivative estimators over a parameter
path, the derivative estimator itself is local in parameter space. It therefore
does not introduce the systematic discretization error associated with the
integration procedure.

One may similarly consider higher-order derivatives by further differentiating
Eqs~\eqref{eq:partialZ}, \eqref{eq:lnzA2},
and~\eqref{eq:lnZphi2}~\cite{ding2025magic}. These derivatives define
effective fluctuation and response observables associated with the chosen
generalized ensemble. When the parameter is temperature, they play a role
analogous to an effective specific heat; when the parameter is a coupling,
field, or measurement/decoherence strength, they define generalized
susceptibilities or response functions.

In many applications, one is interested not only in reconstructing $S_2$
itself, but also in its response to external parameters such as temperature,
coupling constants, or measurement/decoherence strength. These derivatives
often exhibit singular behavior and universal scaling near phase transitions,
making them useful probes of critical phenomena.

% ================================================
%   NEW SUBSECTION
% ================================================
\subsection{Continuous extended ensemble}
The direct estimator in Eq.~\eqref{eq:direct_estimator} samples an extended
ensemble with two discrete replica sectors, $\theta=A$ and
$\theta=\varnothing$, so that
$Z_A^{(2)}/Z_{\varnothing}^{(2)}$ is obtained from their relative probabilities.
In practice, however, this ratio is often exponentially small in the subsystem
size, making direct sampling inefficient. 
As discussed above, one remedy is to introduce a sequence of discrete intermediate subsystems, as in
Eq.~\eqref{eq:discrete-incremental}, thereby decomposing the ratio into a product
of smaller, better-conditioned ratios.

A further refinement is to replace this discrete interpolation by a continuous
one. 
Specifically, one introduces a continuous interpolation parameter $\lambda$ and
defines an extended partition function $Z_{A,\mathrm{ext}}^{(2)}(\lambda)$ that smoothly connects
the $\varnothing$ and $A$ replica geometries.

We choose the convention
\begin{equation}
    Z_{A,\mathrm{ext}}^{(2)}(0)\equiv Z_{\varnothing}^{(2)},\qquad
    Z_{A,\mathrm{ext}}^{(2)}(1)\equiv Z_A^{(2)},\qquad
    \lambda\in[0,1].
\end{equation}
The interval $[0,1]$ and the assignment of endpoints are conventional: any
monotonic reparametrization of $\lambda$, or an exchange of the two endpoints,
leads to a formulation. 

Under the extended ensemble $Z_{A,\mathrm{ext}}^{(2)}(\lambda)$, we can rewrite the Rényi-2 entropy as 
\begin{equation}\label{eq:S2A-ext-cont}
    S_2(\rho_A)=
    -\ln
    \frac{Z_{A,\mathrm{ext}}^{(2)}(\lambda = 1)}
         {Z_{A,\mathrm{ext}}^{(2)}(\lambda = 0)} .
\end{equation}

The functional form of $Z_{A,\mathrm{ext}}^{(2)}(\lambda)$ is not unique. 
A common choice, introduced in Ref.~\cite{DEmidio2020nonequi-work}, is
\begin{equation}\label{eq:extendedZ2}
    Z_{A,\mathrm{ext}}^{(2)}(\lambda)
    :=
    \sum_{D\subseteq A}
    g_{A;D}(\lambda)\,
    Z_D^{(2)},
\end{equation}
with
\begin{equation}\label{eq:extended_weight}
    g_{A;D}(\lambda)
    =
    \lambda^{|D|}(1-\lambda)^{|A|-|D|}.
\end{equation}
Here $D$ denotes a subsystem of $A$, and
the sum runs over all subsets $D\subseteq A$. 
The quantity $Z_D^{(2)}$ is the analogue of $Z_A^{(2)}$ with the Rényi-2 gluing
condition imposed on $D$ instead of $A$. 
Thus the extended ensemble assigns each choice of gluing subsystem $D$ an
additional $\lambda$-dependent weight $g_{A;D}(\lambda)$.

More explicitly, for $Z_D^{(2)} =
    \sum_{\mathcal C\in\Omega_D}
    W_D(\mathcal C)$,
we have 
\begin{equation}
    Z_{A,\mathrm{ext}}^{(2)}(\lambda)
    =
    \sum_{D\subseteq A}
    \sum_{\mathcal C\in\Omega_D}
    g_{A;D}(\lambda)\,
    W_D(\mathcal C).
\end{equation}
Thus a configuration in the extended ensemble is specified by the gluing
subsystem $D$ together with a microscopic QMC configuration
$\mathcal C$. 

Correspondingly, the extended configuration space is
\begin{equation}
    \Omega_{\rm ext}
    =
    \bigl\{
    (\mathcal C,D)
    \;\big|\;
    D\subseteq A,\;
    \mathcal C\in\Omega_D
    \bigr\}.
\end{equation}

We next describe how to update $D$ in the extended configuration
$(\mathcal C,D)$~\cite{DEmidio2020nonequi-work}. 
For each site in $D$, the imaginary-time boundary condition follows the
$A$-glued geometry shown in Fig.~\ref{fig:S2_GPF}(a), whereas sites in
$A\setminus D$ follow the unglued geometry. 
Thus a local update of $D$ can be performed by changing the gluing condition at a
single site:
\begin{itemize}
    \item A \emph{splitting} update removes one site $i$ from $D$,
\begin{equation}
    D\;\to\; D' = D\setminus\{i\},
    \qquad
    |{D'}|=|D|-1 ,
\end{equation}
while keeping the microscopic configuration $\mathcal C$ fixed. This move is admissible only if $\mathcal C\in\Omega_D\cap\Omega_{D'}$; otherwise it is rejected.
The Metropolis--Hasting acceptance probability is then
\begin{align}
        P_{\mathrm{split}}
    & =
    P\bigl((\mathcal C,D)\to(\mathcal C,D')\bigr)
    \nonumber 
    \\ 
    & =
    \min\left\{
    1,\,
    \frac{1-\lambda}{\lambda}
    \right\}.
\end{align} 

\item 
Conversely, a \emph{joining} update adds one site $i\in A\setminus D$ to $D$,
\begin{equation}
    D\;\to\; D' = D\cup\{i\},
    \qquad
    |{D'}|=|D|+1 .
\end{equation}
provided the microscopic configuration $\mathcal C$ is
compatible with both boundary conditions.
Using the notation of Fig.~\ref{fig:S2_GPF}, this compatibility condition is
$s_i^{(1)}{}'=s_i^{(2)}$. 
The corresponding acceptance probability is
\begin{align}
        P_{\mathrm{join}}
    & =
    P\bigl((\mathcal C,D)\to(\mathcal C,D')\bigr)
    \nonumber \\ 
    & =
    \min\left\{
    1,\,
    \frac{\lambda}{1-\lambda}
    \right\}.
\end{align}

\end{itemize}

Although $\lambda$ is not a physical parameter, it continuously connects the two
endpoint ensembles on $[0,1]$.
This makes it possible to evaluate $S_2(\rho_A)$ in
Eq.~\eqref{eq:S2A-ext-cont} using either reweight--annealing or thermodynamic integration~\cite{DEmidio2024So5,Tarabunga2025bell} in Secs.~\ref{sec:reweight-annealing} and \ref{sec:thermo-int}.

For example, thermodynamic integration~\eqref{eq:thermo-int} gives
\begin{equation}\label{eq:thermo-int-ext}
    S_2(\rho_A)
    =
    -\int_{0}^{1}
    d\lambda\,
    \frac{\partial  \ln Z_{A,\mathrm{ext}}^{(2)}(\lambda)}{\partial \lambda}.
\end{equation}

% ================================================
%   NEW SUBSECTION
% ================================================
\subsection{Nonequilibrium work estimator from Jarzynski's equality}

In addition to the equilibrium approaches discussed above,
Ref.~\cite{DEmidio2020nonequi-work} introduced a nonequilibrium estimator based
on \emph{Jarzynski's equality}~\cite{Jarzynski1997equality}. 
Instead of equilibrating the system at a sequence of fixed $\lambda$ values, as
in reweight--annealing or thermodynamic integration, one drives $\lambda$ along a
prescribed protocol and estimates the desired partition-function ratio from an
exponential average of nonequilibrium work. 
We illustrate the construction using the continuous extended ensemble
$Z_{A,\mathrm{ext}}^{(2)}(\lambda)$ in Eq.~\eqref{eq:extendedZ2}.

Analogous to defining the effective internal energies in
Eqs.~\eqref{eq:lnzA2} and \eqref{eq:lnZphi2}, we first define an effective dimensionless free
energy associated with the extended ensemble,
\begin{equation}
    F_{A,\mathrm{ext}}^{(2)}(\lambda)
    :=
    -
    \ln Z_{A,\mathrm{ext}}^{(2)}(\lambda).
\end{equation}

The partition-function ratio in Eq.~\eqref{eq:S2A-ext-cont} can then be
interpreted as the exponential of an effective free-energy difference between
the two endpoint ensembles,
\begin{align}
   \Delta F_{A,\mathrm{ext}}^{(2)}
    &:=
    F_{A,\mathrm{ext}}^{(2)}(\lambda=1)
    -
    F_{A,\mathrm{ext}}^{(2)}(\lambda=0)
    \nonumber \\
    &=
    -\ln
    \frac{
    Z_{A,\mathrm{ext}}^{(2)}(\lambda=1)
    }{
    Z_{A,\mathrm{ext}}^{(2)}(\lambda=0)
    }
    =
    S_2(\rho_A).
\end{align}
Thus computing the Rényi-2 entropy is equivalent to computing the effective
free-energy difference between $\lambda=0$ and $\lambda=1$.

Jarzynski's equality states that this equilibrium free-energy difference can be
obtained from an exponential average over nonequilibrium trajectories~\cite{Jarzynski1997equality}. The trajectories must be initialized from the equilibrium ensemble at $\lambda=0$, and the stochastic updates used during the protocol must preserve
the instantaneous equilibrium distribution when $\lambda$ is held fixed. This results in 
\begin{equation}\label{eq:Jarzynski}
    \left\langle
    e^{-\mathcal W_{A,\mathrm{ext}}^{(2)}[\Gamma]}
    \right\rangle_{\Gamma}
    =
    e^{-\Delta F_{A,\mathrm{ext}}^{(2)}}
    =
    e^{-S_2(\rho_A)} ,
\end{equation}
Here $\Gamma$ denotes a nonequilibrium trajectory,
$\mathcal W_{A,\mathrm{ext}}^{(2)}[\Gamma]$ is the effective nonequilibrium work
accumulated along that trajectory, and $\langle\cdots\rangle_{\Gamma}$ denotes an
average over trajectories generated by the prescribed protocol.

We now define the nonequilibrium work explicitly. 
Using Eq.~\eqref{eq:extendedZ2}, each extended configuration $(\mathcal C,D)$
has statistical weight $g_{A;D}(\lambda)\,W_D(\mathcal C)$.
Then we can define the corresponding dimensionless effective
energy
\begin{equation}
    E_\lambda(\mathcal C,D)
    :=
    -\ln\!\left[g_{A;D}(\lambda)\,W_D(\mathcal C)\right].
\end{equation}
When $\lambda$ is changed by an infinitesimal amount $d\lambda$ while the
extended configuration $(\mathcal C,D)$ is kept fixed, the infinitesimal
dimensionless work $\delta\mathcal{W}$ is defined as the change of this effective energy,
\begin{align}
    \delta \mathcal W
    =
    \frac{\partial E_\lambda(\mathcal C,D)}
         {\partial \lambda}
    d\lambda
    \nonumber 
    =
    -\frac{\partial \ln g_{A;D}(\lambda)}{\partial \lambda}
    d\lambda ,
\end{align}
where we used the fact that $W_D(\mathcal C)$ is independent of $\lambda$.

To generate a nonequilibrium trajectory, one can introduce a monotonic quench
protocol $\lambda=\lambda(t)$ with $t\in[t_i,t_f]$, satisfying $\lambda(t_i)=0$ and $\lambda(t_f)=1$.

During the evolution, the extended configuration changes with time, so the
gluing subsystem becomes trajectory dependent, $D=D(t)$. 
The effective nonequilibrium work accumulated along a single trajectory
$\Gamma$ is therefore~\cite{DEmidio2020nonequi-work}
\begin{equation}\label{eq:noneq-work}
    \mathcal W_{A,\mathrm{ext}}^{(2)}[\Gamma]
    =
    -
    \int_{t_i}^{t_f}
    dt\,
    \frac{d\lambda}{dt}\,
    \frac{\partial \ln g_{A;D(t)}(\lambda(t))}{\partial\lambda}.
\end{equation}

Substituting the trajectory work into Eq.~\eqref{eq:Jarzynski} gives
\begin{equation}
    S_2(\rho_A)
    =
    -\ln
    \left\langle
    e^{-\mathcal W_{A,\mathrm{ext}}^{(2)}[\Gamma]}
    \right\rangle_{\Gamma} .
\end{equation}
Thus the equilibrium Rényi-2 entropy can be obtained from finite-rate
nonequilibrium trajectories. 

A key advantage of this formulation is that Jarzynski's equality does not
require the protocol to be quasistatic~\cite{Jarzynski1997equality}. 
Even when the system is driven out of equilibrium during the $\lambda$ sweep,
the exponential work average still recovers the equilibrium free-energy
difference between the two endpoint ensembles. 
In the quasistatic limit, the work distribution is narrowly peaked around the
reversible work, and the estimator reduces conceptually to thermodynamic
integration, Eq.~\eqref{eq:thermo-int-ext}. 
Finite-rate protocols reduce the cost per trajectory, but they also broaden the
work distribution and can increase the variance because the exponential average
is dominated by rare low-work trajectories. 
The practical efficiency therefore depends on balancing protocol speed against
statistical variance. 
When this balance is favorable, nonequilibrium work estimators can outperform
equilibrium approaches that require equilibration and sampling at many
intermediate $\lambda$ values. 
For example, Ref.~\cite{DEmidio2020nonequi-work} demonstrated that an optimized
nonequilibrium protocol enables efficient simulations of the square-lattice
Heisenberg antiferromagnet on systems as large as $192\times 96=18432$ qubits. 
Combined with the incremental trick, additional refinements are discussed in Ref.~\cite{Zhao2022nonequi-work}.

\section{Application Map: QMC for Many-Body Quantum Information Diagnostics}\label{sec:nonlinear}

The preceding sections introduced several technical ingredients for studying
many-body quantum information with QMC. We now turn from the construction of
these methods to the many-body information-theoretic diagnostics that they make
accessible.

The goal of this section is to provide an application map of representative nonlinear diagnostics that have been investigated within QMC frameworks. These include entanglement entropies for characterizing pure-state bipartite entanglement (Sec.~\ref{sec:ent-entropy}); the entanglement Rényi negativity for characterizing mixed-state bipartite entanglement (Sec.~\ref{sec:renyi-negativity}); the stabilizer entropy for
quantifying many-body quantum magic, or nonstabilizerness
(Sec.~\ref{sec:stabilizer-entropy}); Rényi-$\alpha$ correlators for diagnosing strong-to-weak spontaneous symmetry breaking (SWSSB; Sec.~\ref{sec:swssb}); and composite quantities, such as quantum mutual information (Sec.~\ref{sec:composite-measure}). 

Moreover, in Sec.~\ref{sec:rdm-data-app}, we discuss applications of the reduced-density-matrix sampling method introduced in Sec.~\ref{sec:sampling-rdm-data}. By providing direct access to RDMs, this approach enables the evaluation of a broad class of density-matrix-based quantities, including the entanglement spectrum, few-body and multipartite entanglement measures, measures of mixed-state nonstabilizerness, and other related diagnostics.

Beyond these representative
examples, a broad range of closely related nonlinear quantities can also be accessed using QMC. Collectively, these developments motivate a more general density-matrix-based perspective on QMC simulations. 

Owing to space limitations, the examples used below to illustrate the application map are necessarily selective rather than exhaustive, and the scope of existing QMC studies extends well beyond those discussed here.

% ---------------------------------
%   NEW SECTION
% ---------------------------------
\subsection{Entanglement entropies}\label{sec:ent-entropy}
\subsubsection{Pure-state bipartite entanglement measures}
Rényi-$\alpha$ entropies have been one of the leading examples throughout this
review. Beyond the general partition-function-ratio formulations for
Rényi-$\alpha$ entropies, the special case that $\alpha=2$ can also be accessed by
measuring SWAP operators directly in valence-bond projector QMC and Bell-basis
QMC, as discussed in Secs.~\ref{sec:vb_basis} and~\ref{sec:bell_basis},
respectively.

Although the
definition has appeared in several places above, we recall it here for
convenience:
\begin{equation}
    S_{\alpha}(\rho_A)
    =
    \frac{1}{1-\alpha}
    \ln
    \frac{
        \Tr_A(\tilde\rho_A^{\alpha})
    }{
        [\Tr_A(\tilde\rho_A)]^{\alpha}
    },
    \qquad
    \alpha\ge 2\in \mathbb{N}.
\end{equation}

In the limit $\alpha\to1$, it reduces to the von Neumann entropy
\begin{align}
    S(\rho_A) & = -\Tr_A(\rho_A\ln\rho_A) \\ 
    & = -\frac{\Tr_A(\tilde\rho_A\ln\tilde\rho_A)}{\Tr_A(\tilde\rho_A)}
    +
    \ln\Tr_A(\tilde\rho_A), \label{eq:vne}
\end{align}
where $\rho_A = \tilde\rho_A/\Tr(\tilde\rho)$.
This form makes clear why the von Neumann entropy is substantially more
difficult to access in QMC than integer Rényi-$\alpha$ entropies. The latter involve
moments such as $\Tr_A(\tilde\rho_A^\alpha)$, with integer $\alpha$, which can be
represented by $\alpha$ replicated copies and suitable boundary gluings, as introduced in Sec.~\ref{sec:rep_states}. By
contrast, Eq.~\eqref{eq:vne} contains the operator function $\ln\tilde\rho_A$, which is highly nonlocal in the degrees of freedom of subsystem $A$ and
does not admit a simple representation in terms of a finite number of replicated
density matrices. Consequently, there is currently no general efficient QMC estimator for evaluating the von Neumann entropy~\eqref{eq:vne} in interacting many-body
systems. An exception may arise when the corresponding entanglement Hamiltonian is explicitly accessible and admits an efficient QMC simulation; in that case, the von Neumann entropy can again be formulated as a generalized partition function ratio problem~\cite{Mendes-Santos2020wang-landau}. In practice, however, constructing and simulating such
an entanglement Hamiltonian is usually highly nontrivial.

For pure states, Rényi entropies faithfully diagnose bipartite entanglement.
Indeed, if the total state is pure,
$\rho_{AB}=\ket{\Psi}\bra{\Psi}$, then
\begin{equation}
    S_\alpha(\rho_A)=0
    \quad \Longleftrightarrow \quad
    \ket{\Psi}_{AB}
    =
    \ket{\Psi_A}\otimes \ket{\Psi_B},
\end{equation}
for any $\alpha>0$, since all Rényi entropies are determined by the
Schmidt spectrum of $\ket{\Psi}_{AB}$. 

In many-body applications, Rényi entropies, especially the Rényi-$2$ entropy $S_2$, are powerful
diagnostics of entanglement structure and often reveal universal information of many-body quantum states. 
For ground states of local Hamiltonians, the
leading behavior of the entanglement entropy is governed by locality, which leads to the area law of entanglement~\cite{Srednicki1993area,Calabrese2009ent-cft,Calabrese2004-ent-qft}. 
For a region $A$ of a many-body ground state, one
generically expects
\begin{equation}
    S_\alpha(A)
    =
    a_\alpha |\partial A|
    +
    S_{\alpha,\mathrm{sub}}(A),
\end{equation}
where the coefficient $a_{\alpha}$ is typically nonuniversal, while the subleading contribution
$S_{\alpha,\mathrm{sub}}(A)$ may contain universal information. 

We next review several important applications of QMC to the study of entanglement entropies. More comprehensive introductions to entanglement entropy and its applications in quantum many-body systems can be found, for example, in
Refs.~\cite{Amico2008ent-review,Eisert2010ent-review,Laflorencie2016review,
Calabrese2009ent-cft,Horodecki2009ent-review,Latorre2009ent-spin-review}.

\subsubsection{Subleading corrections}
As mentioned above, entanglement entropies provide access to universal
information beyond the leading area law. A prominent example is provided by
phases with spontaneous symmetry breaking (SSB) of a continuous symmetry.
In such phases, the associated Goldstone modes give rise to a universal
logarithmic correction to the entanglement R\'enyi entropies.

Specifically, for a system with $O(n)\to O(n-1)$ symmetry breaking and a \emph{smooth} subsystem boundary, i.e., when subsystem $A$ has no corners, one expects
\begin{equation}
    S_{\alpha}(A)
    =
    a_{\alpha} L^{d-1}
    +
   b \ln L^{d-1}
    +
    c_{\alpha},\qquad d>1,
    \label{eq:Sn-Goldstonemode}
\end{equation}
where $b=n_G/2$ with $n_G=n-1$ the number of Goldstone modes~\cite{metlitski2015EE-SSB}. The coefficients
$a_{\alpha}$ and $c_{\alpha}$ are generally dependent on the Rényi index
$\alpha$, whereas the universal logarithmic term provides a characteristic
entanglement signature of phases with SSB of a continuous symmetry.

\begin{figure}[ht!]
    \centering
    \includegraphics[width=0.7\linewidth]{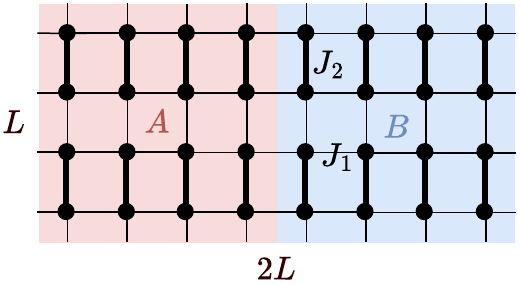}
    \caption{
         $2$D $S=1/2$ dimerized Heisenberg model on a
        $2L\times L$ square lattice with periodic boundary conditions. The
        coupling $J_1$ is assigned to the thin bonds, while $J_2$ is assigned
        to the thick dimer bonds. The system is bipartitioned by a smooth cut
        into two $L\times L$ subsystems, denoted by $A$ and $B$.
    }
    \label{fig:dhm}
\end{figure}

A number of related QMC studies have been performed; see, for example,
Refs.~\cite{Luitz2015Universal,Kulchystkyy2015qmc-goldstone,Wang2025reweight,Zhao2022nonequi-work,DEmidio2020nonequi-work}. A classic example is the $2$D $S=1/2$ dimerized
Heisenberg model, whose Hamiltonian is
\begin{equation}
    H =
    J_1 \sum_{\langle ij\rangle_1} \mathbf{S}_i\cdot\mathbf{S}_j
    +
    J_2 \sum_{\langle ij\rangle_2} \mathbf{S}_i\cdot\mathbf{S}_j ,
    \label{eq:dimerized-heisenberg-model}
\end{equation}
where $\langle ij\rangle_1$ and $\langle ij\rangle_2$ denote two inequivalent
sets of nearest-neighbor bonds, as illustrated in Fig.~\ref{fig:dhm}.
Fixing $J_2=1$, the ground state has a quantum phase transition at
$J_1=J_c=0.52337(3)$, separating a disordered dimer phase for
$J_1<J_c$ from a Néel-ordered phase for $J_1>J_c$. In the Néel phase, the
continuous spin-rotation symmetry is spontaneously broken as
$O(3)\to O(2)$, giving rise to $n_G=2$, or $b=n_G/2=1$. 

Table~\ref{table:wz-nG} summarizes the extracted values of $b(J_1)$ as a
function of the tuning parameter $J_1$. The values are obtained from fits to
the Rényi-$2$ entanglement entropy $S_2(\rho_A)$ in Eq.~\eqref{eq:Sn-Goldstonemode} for the bipartition shown in
Fig.~\ref{fig:dhm}, using the reweight-annealing method introduced in Sec.~\ref{sec:reweight-annealing}. The data are taken
from Ref.~\cite{Wang2025reweight}.
As shown in Table~\ref{table:wz-nG}, for several values of $J_1>J_c$ in the
Néel phase, the fitted coefficient is consistent with the theoretical
prediction $b=1$. At $J_1=J_c$, the fitted logarithmic coefficient vanishes
within error bars, in agreement with the expected scaling at the $O(n)$
Wilson--Fisher quantum critical point, where no logarithmic subleading
correction appears for a smooth entangling boundary~\cite{Metlitski2009EE-ON}. 
Bipartitions with corners can be analyzed in the same way, as demonstrated
in Ref.~\cite{Wang2025reweight}.
\begin{ruledtabular}
    \begin{table}[!h]
        \begin{tabular}{l c c c  c }
                $J_{1}$ 	  & $a_2$ 	  & $b$ 	&$c_2$  \\
                    
            \hline
            $1.0$			& 0.089(2)   	&1.05(4)     &1.61(9)  \\   
            $0.9$			& 0.085(2)  	&1.02(3)     &1.54(7)  \\
            $0.8$			& 0.079(2)   	&1.06(5)     &1.6(1)  \\
            $0.6$			& 0.072(2)   	&1.06(5)     &2.0(2)   \\
            $0.55$			& 0.078(3)   	&0.8(1)      &1.6(2)   \\
            $0.54$			& 0.08(1)   	&0.6(1)      &1.2(2)  \\				
            $J_c=0.52337$	& 0.08(1)   	&0.15(17)    &0.1(5)   \\	
        \end{tabular}
        \caption{
            Fitting results for $S_2(\rho_A)$ of the
            bipartition shown in Fig.~\ref{fig:dhm}. The data are fitted to Eq.~\eqref{eq:Sn-Goldstonemode}, yielding the coefficients
            $a_2$, $b$, and $c_2$. The data are taken from
            Ref.~\cite{Wang2025reweight}. 
        }
    \label{table:wz-nG}
    \end{table}
\end{ruledtabular}

Other universal quantities can also be extracted from entanglement entropies $S_{\alpha}$ using the QMC methods introduced in this review. For
example, in a lattice critical system described by a $(1+1)$D CFT~\cite{Calabrese2009ent-cft}, the scaling of $S_{\alpha}$ provides a direct way to extract
the central charge~\cite{DEmidio2015centralcharge,DEmidio2020nonequi-work,ding2024rweight,Tarabunga2025bell,Kumar2026centralcharge}. In gapped topologically ordered systems, the subleading
constant term gives the topological entanglement entropy~\cite{Isakov2011topo,Selem2013topo-dimer,Pei2014topo,Zhao2022nonequi-work,Tarabunga2025bell}, which is related to the quantum dimension~\cite{Kitaev2006topological,Levin2006topological,Flammia2009renyi-topo}. To extract high-precision logarithmic subleading
correction, a QMC method designed with automatically canceling the leading area-law term has also been developed \cite{liao2024extracting}.

Another recent direction concerns the study of the $J$-$Q$ class of models, which provide promising candidates for realizing deconfined quantum critical points (DQCPs) \cite{sandvik2007evidence,lou2009anti,sen2010example,sandvik2010continuous}. %\ym{references needed}. 
In these models, conventional order parameters do not fully characterize the critical behavior, while entanglement entropies may provide complementary probes of criticality and emergent symmetry-breaking. The earliest QMC work found that the corner correction term of EE scaling violates the unitary CFT  which may point out the DQCP there is not a continuous phase transition \cite{zhao2022scaling}. Furthermore, the anomalous correction has been explained by Goldstone modes due to the emergent SO(5) symmetry-breaking at the phase transition point \cite{deng2024diagnosing}. Thereafter, another work \cite{DEmidio2024So5} thought that the entanglement bipartition used in Refs. \cite{zhao2022scaling,deng2024diagnosing} artificially breaks the emergent SO(5) symmetry at the DQCP, which leads to a biased result. And they found if the entanglement cut guarantees the symmetry of the critical point, the EE scaling is consistent with CFT. Later, the works \cite{zhu2026bipartite,zhu2026criticalityrenyidefects21d} reveal that the entanglement boundary corresponding to different defect CFT needs to be carefully considered at a bulk criticality and only the ordinary entanglement-boundary obeys the previous prediction of field theory.

\subsubsection{Derivatives of entanglement entropy}
In addition to the entanglement entropies themselves, the derivative methods introduced in Sec.~\ref{sec:pf_derivative} enable the calculation of their derivatives with respect to a tuning parameter. The underlying motivation is that, within the replica formulation, entanglement entropies can be interpreted as free-energy-like quantities. Accordingly, their derivatives with respect to Hamiltonian parameters characterize the entanglement response to external perturbations. Analogous to derivatives of the thermodynamic free energy, these quantities may exhibit singularities or universal scaling behavior near quantum critical points, thereby providing useful diagnostics of phase transitions and critical phenomena~\cite{Wang2025DEE}.

As an illustrative example, we consider the entanglement entropy near a $(2+1)$D $O(n)$ quantum critical point. To investigate its critical behavior, we perturb the system slightly away from criticality. 

Let $g$ denote a Hamiltonian parameter and $g_c$ its critical value. For a lattice system with size $L$ under smooth bipartition, $S_{\alpha}$ in the vicinity of the critical point takes the form~\cite{Metlitski2009EE-ON,helmes2014entanglement}
\begin{equation}
    S_{\alpha}(g,L)=a_{\alpha}(g,L)L + \gamma_{\alpha}(g,L),
\end{equation}
where $a_{\alpha}(g,L)$ is an analytic, nonuniversal area-law coefficient. 
Moreover, the singular contribution satisfies
\begin{equation}
    \gamma_{\alpha}(g,L) \sim |g-g_c|^{\nu} L,
\end{equation}
where $\nu$ is the correlation-length critical exponent. 
 At the critical point, $\gamma_{\alpha}(g_c,L\to\infty)$ becomes a universal constant determined by the underlying critical theory and the geometry.

For convenience, we rewrite 
\begin{equation}
    \gamma_{\alpha}(g,L)\equiv \tilde{\gamma}_{\alpha}\left(x\right),\quad 
    x:=(g-g_c)L^{1/\nu}.
\end{equation}
Taking the derivative with respect to $g$ gives
\begin{equation}
    \frac{\partial S_{\alpha}}{\partial g}
    =
    \frac{\partial a_{\alpha}}{\partial g}L
    +
    \widetilde{\gamma}_{\alpha}'(x)L^{1/\nu}.
\end{equation}
At the critical point, the first term scales as $L$, whereas the singular contribution scales as $L^{1/\nu}$. For the $(2+1)$D $O(n)$ universality classes with $n=1,2,3$, one has $\nu<1$ and hence $1/\nu>1$. 

\begin{figure*}[ht!]
    \centering
    \includegraphics[width=0.8\linewidth]{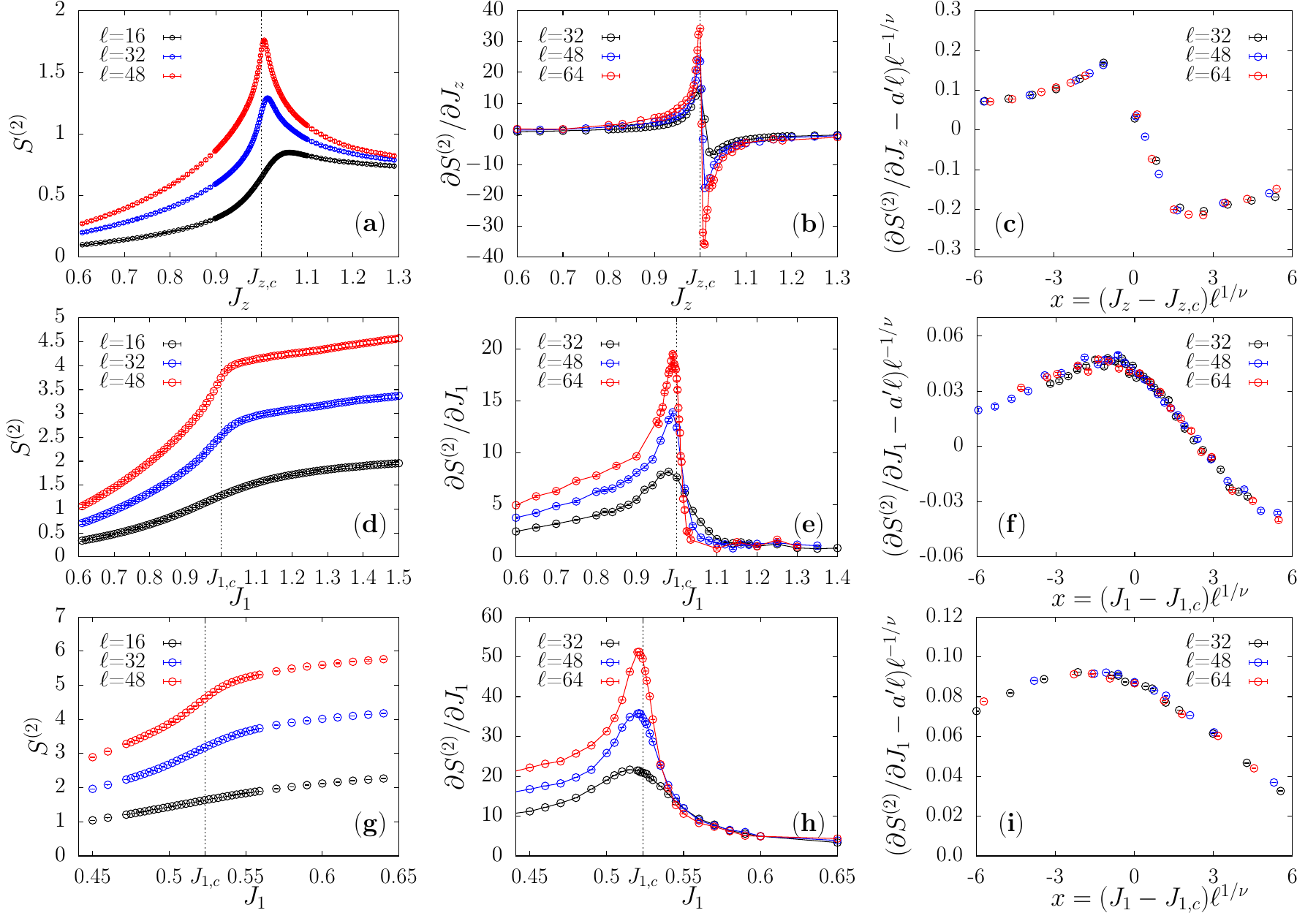}
    \caption{
    Rényi-$2$ entanglement entropy $S_2$, its derivative, and the corresponding scaling function near $(2+1)$D quantum critical points. 
    Here $\ell$ denotes the length of the entanglement boundary, rather than the linear system size $L$. (a--c) Results for an Ising transition, with $J=3.045$ fixed and $J_z$ tuned across the critical point $J_{z,c}=1$.
    (d--f) Results for an $O(2)$ transition with $\Delta=0.9$, where
    $J_2=2.1035$ is fixed and $J_1$ is tuned across the critical point
    $J_{1,c}=1$. (g--i) Results for an $O(3)$ transition, with $J_2=1$ fixed and $J_1$ tuned across the critical point $J_{1,c}=0.52337$.
    Adapted from Ref.~\cite{Wang2025DEE}; the specific lattice models and
Hamiltonian parameters are described therein.
    }
    \label{fig:qcp-DEE}
\end{figure*}

Consequently, for sufficiently large $L$, the singular contribution dominates, yielding
\begin{equation}\label{eq:DEE}
    \left.
    \frac{\partial S_{\alpha}}{\partial g}
    \right|_{g=g_c}
    \sim L^{1/\nu}.
\end{equation}
Ref.~\cite{Wang2025DEE} systematically investigated the derivatives of the
Rényi-$2$ entanglement entropy $S_2$ across several $(2+1)$D $O(n)$ quantum
critical points. Fig.~\ref{fig:qcp-DEE} presents the results for the
$O(1)$, $O(2)$, and $O(3)$ universality classes (from top to bottom),
including $S_2$ and $\partial_g S_2$ as a function of the tuning parameter $g$, and the
corresponding finite-size scaling collapses. The data collapse confirms the
predicted scaling behavior of the first derivative of the universal singular
term $\tilde{\gamma}_2(x)$ in Eq.~\eqref{eq:DEE} and yields accurate estimates of the associated
critical exponents.

% ---------------------------------
%   NEW SECTION
% ---------------------------------
\subsection{Entanglement Rényi negativity}\label{sec:renyi-negativity}

\subsubsection{Mixed-state entanglement and partial transpose}
For mixed states, entanglement entropies are no longer faithful diagnostics of
bipartite quantum entanglement, which generally contains contributions from both quantum
entanglement and classical correlations. A useful way to detect
mixed-state entanglement is instead based on the positive-partial-transpose
(PPT) criterion~\cite{Peres1996ppt,Horodecki1996separability}. It states that
\begin{equation}
    \rho\ \text{separable}
    \quad\Longrightarrow\quad
    \rho^{T_B}\ge 0 ,
\end{equation}
where $\rho\equiv \rho_{AB}$.
In general, this condition is only necessary: there exist entangled states with positive partial transpose, known as bound-entangled states. Nevertheless, the partial transpose provides the fundations for many useful mixed-state entanglement diagnostics that do not require variational minimization procedures, in
contrast to quantities such as the entanglement of formation.

A widely used measure based on the PPT criterion is the logarithmic
negativity~\cite{Vidal2002negativity,Plenio2005log-negativity}. For a normalized
bipartite density matrix $\rho$, it is defined as
\begin{equation}\label{eq:log-negativity}
    \mathcal E(\rho)
    =
    \ln \bigl\|\rho^{T_B}\bigr\|_1,
\end{equation}
where $T_B$ denotes the partial transpose with respect to subsystem $B$, and
$\|X\|_1=\Tr\sqrt{X^\dagger X}$ is the Schatten-$1$ norm, or trace norm.
It is closely related to the negativity
$\mathcal N(\rho)
=[\|\rho^{T_B}\|_1-1]/2$, through
\begin{equation}
    \mathcal E(\rho)
    =
    \ln\!\left[2\mathcal N(\rho)+1\right].
\end{equation}
While the negativity is an entanglement monotone, the logarithmic negativity is
additive under tensor products, making it particularly convenient in many-body
applications. 
This makes $\mathcal E$ an extensive quantity, endowing it with the same scaling behaviour as thermodynamic entropies and enabling its use in finite-size scaling analyses.

However, directly evaluating Eq.~\eqref{eq:log-negativity} is still difficult in QMC because the Schatten-$1$ norm involves the operator square root $\sqrt{(\rho^{T_B})^\dagger{\rho}^{T_B}}$. A more QMC-friendly strategy is to consider integer moments
of the partially transposed density matrix \cite{chung2014entanglement},
\begin{equation}\label{eq:pk-moment}
   p_{\alpha}:= \frac{
        \Tr\!\left[
            \left(\tilde\rho^{T_B}\right)^{\alpha}
        \right]
    }{
        [\Tr(\tilde\rho)]^{\alpha}
    },
    \qquad
    \alpha\in\mathbb N^+ .
\end{equation}
These moments can be represented using replicated partially transposed density
matrices, as introduced in Sec.~\ref{sec:pt_states}.

One useful derived mixed-state entanglement diagnostic is the \emph{Rényi negativity},
\begin{equation}\label{eq:renyi-negativity}
    R_{\alpha}(\rho;B)\equiv R_{\alpha}(\rho)
    :=
    -
    \log
    \frac{
        \Tr\!\left[
            \left(\tilde\rho^{T_B}\right)^{\alpha}
        \right]
    }{
        \Tr(\tilde\rho^{\alpha})
    },
    \qquad
    \alpha\in\mathbb N^+ .
\end{equation}
Equivalently, one can write it as 
\begin{equation}
    R_{\alpha}(\rho)
    =
    -
    \left[
    \log
    \frac{
        \Tr\!\left[
            \left(\tilde\rho^{T_B}\right)^{\alpha}
        \right]
    }{
        [\Tr(\tilde\rho)]^{\alpha}
    }
    -
    \log
    \frac{
        \Tr(\tilde\rho^{\alpha})
    }{
        [\Tr(\tilde\rho)]^{\alpha}
    }
    \right].
\end{equation}
Thus $R_{\alpha}$ compares the replicated partition function of the partially
transposed density matrix with the ordinary replicated partition function. This
makes it naturally suited to the partition-function-ratio methods discussed in Sec.~\ref{sec:pfr}.

More generally, the partial-transpose moments in Eq.~\eqref{eq:pk-moment}, often denoted by $p_k$ in the literature, define a hierarchy of $p_k$-PPT criteria. The simplest nontrivial case, the $p_3$-PPT criterion, detects violations of positivity using only the first few moments of $\tilde\rho^{T_B}$, while higher-order criteria incorporate additional moments and yield progressively stronger constraints~\cite{Neven2021ptmoments,Yu2021ptmoments,Miller2026ptmoments}. Since these moments are partition-function ratios, the corresponding criteria are naturally accessible within QMC formulations.

\subsubsection{Subleading corrections}
For many physically relevant states of local many-body systems, the Rényi negativity is likewise expected to exhibit a leading area-law contribution.
However, compared with entanglement entropies, subleading terms in the Rényi negativity are less systematically understood and remain difficult to characterize in generic many-body systems, although they have been investigated in special settings, including universal scaling functions and unusual finite-size corrections in $(1+1)$D CFTs and critical chains, subleading constants across finite-temperature transitions, and topological constant contributions in exactly solvable topologically ordered phases~\cite{Calabrese2013negativity-extended,Alba2013negativity-cft,Wu2020negativity,Lu2020topo-negativity,Zou2023channel-negativity,ding2025negativity,Liu2026negativity,Wangfh2025negativity,wangfh2025negativity2,Fang2025negativity}.

\begin{figure}[ht!]
    \centering
    \includegraphics[width=\linewidth]{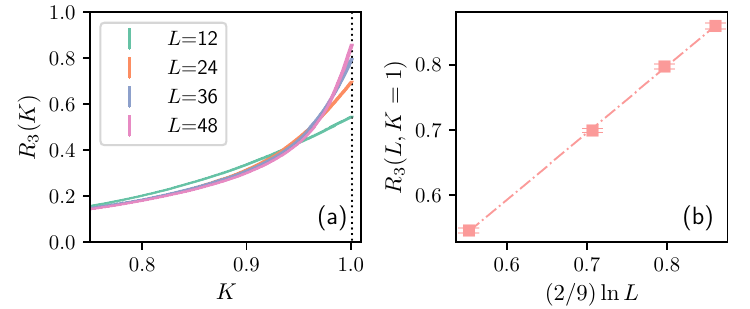}
    \caption{
        (a) Rényi-$3$ negativity $R_3$ as a function of the inter-region coupling $K$.
        (b) QMC finite-size data for $R_3$ at $K=1$, fitted to the $(1+1)$D CFT
        prediction, yielding a central charge $c=1.02(2)\approx 1$.
        Figure reproduced from Ref.~\cite{ding2025negativity}.
    }
    \label{fig:1d-ring-negativity}
\end{figure}

As a QMC illustration, we consider the Rényi negativity of a periodic
$(1+1)$D dimerized dimerized Heisenberg chain $H_{\mathrm{ADHM}}
    =
    \sum_{i\in\mathrm{odd}}\mathbf{S}_i\cdot\mathbf{S}_{i+1}
    +
    K\sum_{i\in\mathrm{even}}\mathbf{S}_i\cdot\mathbf{S}_{i+1}$,
partitioned into three contiguous regions of equal length, $A$, $B$, and
$\overline{A\cup B}$, with the inter-region boundaries chosen to lie on bonds
of strength $K$. 
At $K=0$, the three regions decouple, providing a natural reference point for both the reweight-annealing and thermodynamic-integration methods discussed in Secs.~\ref{sec:reweight-annealing} and~\ref{sec:thermo-int}, respectively, while $K=1$ recovers the uniform Heisenberg chain.
After tracing out $\overline{A\cup B}$, the Rényi-$3$
negativity $R_3$ characterizes the mixed-state entanglement between $A$ and
$B$. As shown in Fig.~\ref{fig:1d-ring-negativity}(a), $R_3$ increases monotonically
with $K$, reflecting the entanglement generated by the boundary interactions.

At $K=1$, the uniform Heisenberg chain is described by a $(1+1)$D CFT. For two adjacent intervals of lengths $\ell_1$ and $\ell_2$ on a ring of circumference $L$, the Rényi negativity~\eqref{eq:renyi-negativity} obeys~\cite{Calabrese2013negativity-extended,ding2025negativity}
\begin{widetext}
\begin{equation}
    R_{\alpha}=
    \begin{cases}
        \displaystyle
        \frac{c}{12}\left(\alpha-\frac{1}{\alpha}\right)
        \ln\!\left[
            \frac{L}{\pi }
            \frac{
                \sin(\pi\ell_1/L)\sin(\pi\ell_2/L)
            }{
                \sin[\pi(\ell_1+\ell_2)/L]
            }
        \right]
        +\mathcal{O}(1),
        & \alpha\ \text{odd}, \\[8pt]
        \displaystyle
        \frac{c}{6}\left(\frac{\alpha}{2}-\frac{2}{\alpha}\right)
        \ln\!\left[
            \frac{L}{\pi }
            \frac{
                \sin(\pi\ell_1/L)\sin(\pi\ell_2/L)
            }{
                \sin[\pi(\ell_1+\ell_2)/L]
            }
        \right]
        +\mathcal{O}(1),
        & \alpha\ \text{even},
    \end{cases}
\end{equation}
\end{widetext}
where $c$ is the central charge and the lattice spacing is taken to be unity. T
For the equal partition $\ell_1=\ell_2=L/3$ and $\alpha=3$, this expression reduces
to
\begin{equation}
    R_3=\frac{2c}{9}\ln L+\mathcal{O}(1).
\end{equation}
The fit in Fig.~\ref{fig:1d-ring-negativity}(b) yields $c=1.02(2)$, in agreement with
the CFT value $c=1$~\cite{Blote1986cft}.

\subsubsection{Derivatives of Rényi negativity}
As for entanglement entropies, derivatives of the Rényi negativity with respect
to a tuning parameter can provide sensitive probes of criticality. At a
finite-temperature transition, the Rényi negativity generally takes the form
\begin{equation}
    R_{\alpha}
    =
    a_{\alpha}(g)L^{d-1}
    +
    \gamma_{\alpha}(g)
    +
    \cdots,
\end{equation}
where $a_{\alpha}(g)$ is the area-law coefficient and $\gamma_{\alpha}(g)$ denotes the
subleading $\mathcal{O}(1)$ contribution. For a smooth entangling boundary in
two spatial dimensions, a nonzero constant term originates from the nonlocal
part of the negativity and therefore signals long-range quantum
entanglement~\cite{Lu2020negativity}. 
By contrast, the area-law coefficient $a_{\alpha}(g)$ predominantly probes
short-range entanglement and may become nonanalytic at a finite-temperature
critical point $T_c$~\cite{Lu2019negativity,Lu2020negativity}. For the Rényi-$3$
negativity defined in Eq.~\eqref{eq:renyi-negativity}, this nonanalyticity is expected to occur at the shifted temperature $3T_c$ rather than at $T_c$. 
Moreover, the temperature derivative of the Rényi-negativity density,
$\partial_T\!\left(R_{\alpha}/|\partial A|\right)$, is expected to exhibit the
same critical scaling as the specific heat~\cite{Wu2020negativity}.

\begin{figure}[ht!]
    \centering
    \includegraphics[width=0.8\linewidth]{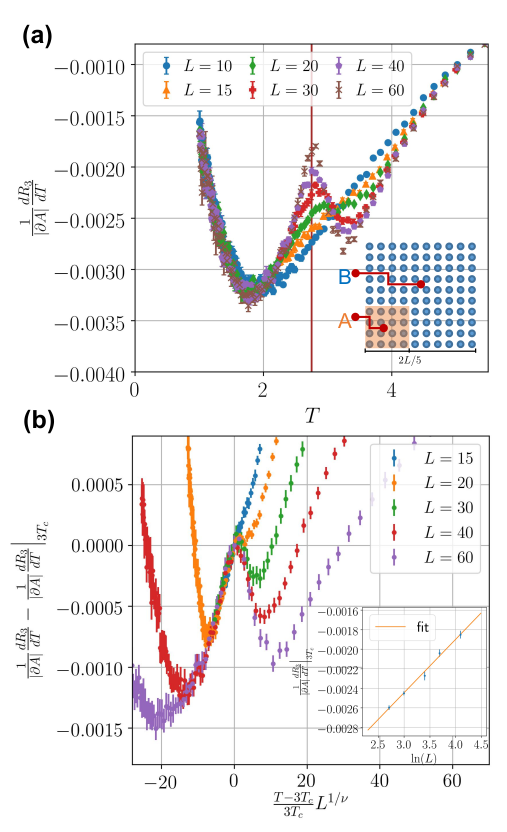}
    \caption{
        (a) Temperature derivative of the Rényi-negativity area-law coefficient
        across the finite-temperature Ising transition. The inset shows the
        bipartition geometry, and the vertical line marks the critical
        temperature. (b) Finite-size scaling collapse, with the inset showing
        the logarithmic growth at criticality. Figure from
        Ref.~\cite{Wu2020negativity}.
    }
    \label{fig:2d-ising-negativity}
\end{figure}

Such derivative singularities and their finite-size scaling have subsequently
been demonstrated in QMC studies of finite-temperature phase
transitions~\cite{Wu2020negativity,Liu2026negativity}. For example,
Ref.~\cite{Wu2020negativity} observed this singularity across a $2$D
finite-temperature Ising transition and confirmed that it exhibits the same
critical scaling as the specific heat, as shown in
Fig.~\ref{fig:2d-ising-negativity}.

% ---------------------------------
%   NEW SECTION
% ---------------------------------
\subsection{Stabilizer entropy}\label{sec:stabilizer-entropy}
\subsubsection{Quantum magic or nonstabilizerness}

Another important information-theoretic diagnostic is quantum magic, or
nonstabilizerness. Its origin lies in the stabilizer formalism and the
Gottesman--Knill theorem~\cite{Gottesman1998heisenberg,Aaronson2004improved},
which states that quantum circuits composed of Clifford gates, stabilizer-state
preparations, and Pauli measurements can be efficiently simulated classically.
Universal quantum computation therefore requires resources outside the
stabilizer sector, such as non-Clifford gates or magic states
\cite{Bravyi2005magic}.

From the perspective of quantum resource theory, magic plays a role analogous
to entanglement, but with respect to a different notion of classical
simulability. Entanglement characterizes the departure from the convex set of
separable states, while magic characterizes the departure from the stabilizer
formalism underlying Clifford quantum computation
\cite{Veitch2014resource,Howard2014magic}. In this sense,
quantum magic captures a form of quantumness that is distinct from
entanglement: a highly entangled stabilizer state may still be efficiently
simulable, whereas a weakly entangled nonstabilizer state may already contain
magic.

In the context of many-body physics, magic has been connected to several
fundamental aspects of quantum many-body states. Recent studies have explored
its relation to quantum criticality
\cite{winter2022mbm,white2021cftmagic,Sarkar_2020_1d_xy_rom,
Haug2023stabilizerentropies,haug2023sremps,lami2023mpssampling_magic,
poetri2023magic,poetri2024criticalbehaviorsof,Tarabunga2024generalizedRK},
quantum chaos
\cite{Leone2021quantumchaosis,leone2022sre,qian2025nonlocal},
CFT
\cite{white2021cftmagic,Hoshino2026magic-cft,Hoshino2026-magic-defect,
Trino2026magic-Shannon},
and holographic duality
\cite{white2021cftmagic,salvatore20221d_tfim_magic,kanato2022magic_rom_chaos}. 

The stabilizer entropy is among the most representative and
computationally accessible measures of magic in many-body
systems~\cite{leone2022sre,leone2024sre}. It is particularly amenable to
many-body and QMC formulations because it can be expressed in terms of moments
of Pauli-string expectation values.

For a $N$-qubit pure-state $\rho$, the Rényi-$\alpha$ stabilizer entropy is defined as
\cite{leone2022sre}
\begin{equation}\label{eq:sre_pure}
    M_{\alpha}(\rho)
    :=
    \frac{1}{1-\alpha}
    \log
    \left[
        \frac{1}{2^N}
        \frac{\sum_{\mathbf r}
        \left|
            \Tr(\tilde\rho \sigma^{\mathbf r})
        \right|^{2\alpha}}{[\Tr(\tilde\rho)]^{2\alpha}}
    \right],
\end{equation}
where the sum runs over all $4^N$ Pauli strings
\begin{equation}
    \sigma^{\mathbf r}
    =
    \bigotimes_{i=1}^N \sigma_i^{r_i},
    \qquad
    r_i\in\{00,01,10,11\},
\end{equation}
following the convention introduced in Sec.~\ref{sec:bell_basis}.

For pure states, the monotonicity under Clifford protocols is ensured for $\alpha\ge 2\in\mathbb{Z}$~\cite{Haug2023stabilizerentropies,leone2024sre}.
This definition can be related to the Rényi-$\alpha$ entropy of the classical characteristic function~\eqref{eq:characteristic}~\cite{zhu2016clifford,leone2022sre}.

Taking $\alpha=2$ as an example, Eq.~\eqref{eq:sre_pure} becomes
\begin{equation}\label{eq:sre2_partition_ratio}
    M_2(\rho)
    =
    -\log
    \left[
        \frac{1}{2^N}
        \frac{Q}{Z^4}
    \right],
\end{equation}
where
\begin{equation}
    Q
    :=
    \sum_{\mathbf r}
    \left[
        \Tr(\tilde\rho \sigma^{\mathbf r})
    \right]^4,
    \qquad
    Z:=\Tr(\tilde\rho).
\end{equation}
Thus, the Rényi-$2$ stabilizer entropy can be viewed as a
generalized partition-function-ratio problem (see Sec.~\ref{sec:pfr}) between $Q$ and $Z^4$.
The central task is then to construct an efficient QMC representation
of the generalized partition function $Q$, assuming that the original ensemble
associated with $Z$ is already accessible.

To see the replicated structure of $Q$, note that
\begin{equation}
    \sum_{\mathbf r}
    \left[
        \Tr(\tilde\rho\sigma^{\mathbf r})
    \right]^4
    =
    \Tr\left[
        \tilde\rho^{\otimes 4}
        \sum_{\mathbf r}
        \left(\sigma^{\mathbf r}\right)^{\otimes 4}
    \right].
\end{equation}
Therefore, $Q$ can be interpreted as a four-copy partition function with an
inserted four-copy Pauli-string operator,
\begin{equation}
     O_{\mathrm{SE}}
    :=
    \sum_{\mathbf r}
    \left(\sigma^{\mathbf r}\right)^{\otimes 4}.
\end{equation}
Equivalently, since the sum is over all Pauli strings, this operator factorizes
site by site as
\begin{equation}
     O_{\mathrm{SE}}
    =
    \bigotimes_{i=1}^N
    \left[
        \sum_{r_i\in\{00,01,10,11\}}
        \left(\sigma_i^{r_i}\right)^{\otimes 4} 
    \right] \equiv \bigotimes_{i=1}^N O_i.
\end{equation}
The operator $O_i$ acts on the local four-copy Hilbert
space and is therefore a $16\times16$ matrix. In the computational basis, one
can verify that its nonzero matrix elements are all equal to $2$. 

More explicitly, for a local four-copy basis state
$\ket{\mathbf s_i}=\ket{s_i^1s_i^2s_i^3s_i^4}$, the single-site operator $O_i$
has nonzero matrix elements only in two cases. The first is the diagonal
contribution, which preserves all four replica indices. The second is the
fully off-diagonal contribution, which flips all four replica indices
simultaneously. Thus one may write
\begin{equation}
    O_i
    =
    O_i^{\mathrm{diag}}
    +
    O_i^{\mathrm{off}},
\end{equation}
with
\begin{equation}
    O_{\mathrm{SE}}
    =
    \bigotimes_{i=1}^N
    \left(
        O_i^{\mathrm{diag}}
        +
        O_i^{\mathrm{off}}
    \right).
\end{equation}
In the computational basis, both $O_i^{\mathrm{diag}}$ and
$O_i^{\mathrm{off}}$ have only nonnegative matrix elements. Therefore, the
operator insertion associated with $O_{\mathrm{SE}}$ does not by itself
introduce a sign problem~\cite{Liu2025magic,ding2025magic}.

This form is useful for QMC because the choice between
$O_i^{\mathrm{diag}}$ and $O_i^{\mathrm{off}}$ can be treated as an additional
local variable in the QMC configuration. Consequently, the generalized
partition function $Q$ can be sampled as an enlarged four-replica ensemble,
where each spacetime configuration of the original model is supplemented by a
site-resolved pattern of diagonal and fully off-diagonal operator insertions (see Sec.~\ref{sec:rho-insert}).

\subsubsection{Subleading corrections}

Quantum-magic diagnostics characterize nonstabilizerness beyond conventional
entanglement measures. Unlike the ground-state entanglement of local 
Hamiltonians, many-body quantum magic typically exhibits volume-law scaling. Its
field-theoretical understanding is also considerably less developed, with the
first systematic treatments appearing only recently
~\cite{Hoshino2026magic-cft}. 

For critical $(1+1)$D systems, the stabilizer
entropy can be related to boundary conformal field theory (BCFT)
~\cite{Hoshino2026magic-cft,Hoshino2026-magic-defect,Trino2026magic-Shannon,benedetti2026universalitymagiclocalquantum}.
In the doubled Hilbert space, the characteristic distribution~\eqref{eq:characteristic} can be interpreted
as the Born-probability distribution obtained by Bell-basis measurements. 
The stabilizer entropy~\eqref{eq:sre_pure} is therefore equivalent to the corresponding participation entropy in the doubled space. 
Applying the replica trick yields a $2\alpha$-component field theory in which the Bell projectors generate an interlayer line defect along an imaginary-time slice.
After folding, this defect becomes a conformal boundary condition, allowing
the universal contributions to be determined from BCFT data.

In particular, for a periodic critical chain of length $L$, the full-state stabilizer entropy scales as
\begin{equation}
    M_\alpha
    =
    m_\alpha L-c_\alpha,
    \qquad
    c_\alpha
    =
    \frac{\ln g_1}{\alpha-1},
\end{equation}
where $m_\alpha$ is nonuniversal, while $g_1$ is the boundary $g$-factor
associated with the Bell-measurement-induced conformal boundary condition.

\begin{figure}[ht!]
    \centering
    \includegraphics[width=0.7\linewidth]{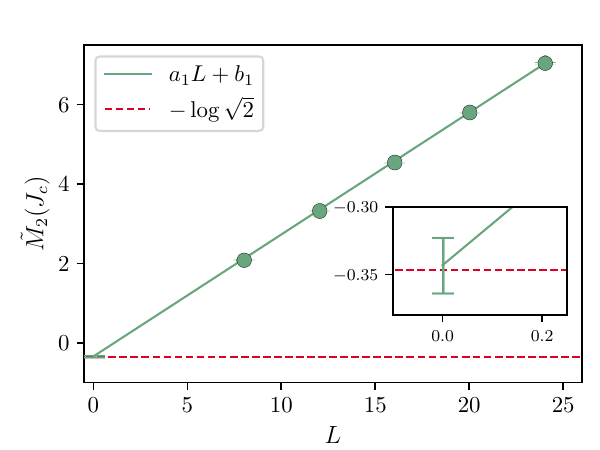}
    \caption{
        Finite-size scaling of the stabilizer entropy for the
        $(1+1)$D Ising CFT at $\alpha=2$. The QMC fit yields the universal
        boundary contribution $c_2 = 0.34(2)\approx \ln\sqrt{2}$. Figure from
        Ref.~\cite{ding2025magic}.
    }
    \label{fig:magic-bcft}
\end{figure}

Figure~\ref{fig:magic-bcft} shows the QMC extraction of the boundary contribution reported
in Ref.~\cite{ding2025magic}. For the $(1+1)$D Ising CFT with $\alpha=2$, the
boundary factor is $g_1=\sqrt{2}$, corresponding to
$c_2=\ln g_1=\ln\sqrt{2}\approx 0.3466$. The QMC estimate,
$c_2=0.34(2) $, is in excellent agreement with this BCFT prediction.

\subsubsection{Derivatives of stabilizer entropies}
\begin{figure}[ht!]
    \centering
    \includegraphics[width=\linewidth]{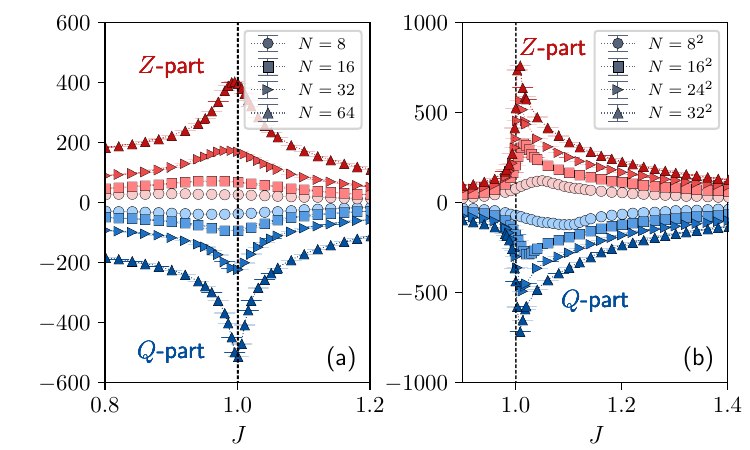}
    \caption{
        Second-order derivatives of the $\ln Q$ contribution to the
        stabilizer entropy $M_2(J)$. According to
        Eq.~\eqref{eq:sre2_partition_ratio}, $\ln Z$ may itself become
        singular at the quantum critical point, so the contributions from
        $\ln Q$ and $\ln Z$ must be separated to isolate the nontrivial
        singularity associated with $\ln Q$. Results are shown for the
        (a) $(1+1)$D and (b) $(2+1)$D transverse-field Ising models, with the
        critical coupling rescaled to $J_c=1$. Figure from
        Ref.~\cite{ding2025magic}.
    }
    \label{fig:magic-derivatives}
\end{figure}

Derivatives of stabilizer entropies remain largely unexplored. Although
universal scaling behavior may be expected near criticality, a corresponding
field-theoretical description is still lacking. Ref.~\cite{ding2025magic}
identified nontrivial singularities in higher-order derivatives of the
stabilizer entropy across the quantum critical points of the $(1+1)$D and
$(2+1)$D transverse-field Ising models, as shown in Fig.~\ref{fig:magic-derivatives}. These
results provide strong numerical evidence that stabilizer entropies exhibit
nonanalytic behavior at quantum critical points.

% ---------------------------------
%   NEW SECTION
% ---------------------------------
\subsection{Composite measures}\label{sec:composite-measure}

So far, we have mainly discussed information-theoretic quantities that can be
expressed in terms of a single entropy, nonlinear observable, or replicated
partition-function ratio. In many applications, however, the relevant
diagnostics are \emph{composite measures}, constructed from combinations of
several entropies, partition-function ratios, or nonlinear observables.

Representative examples include mutual
information~\cite{Melko2010qmi,Singh2011finite-mi,Kallin2011anomaly-mi,
Inglis2013wang-landau-mi}, conditional-mutual-information
combinations~\cite{Tarabunga2025bell,Wang2025analog-tee}, tripartite
information, topological entanglement
entropy~\cite{Isakov2011topo,Wildeboer2017topo,Zhao2022nonequi-work,
Tarabunga2025bell,Matthew2020topo-spin-liquid,Wang2025analog-tee},
mutual stabilizer entropy~\cite{Liu2025magic,ding2025magic,fang2026twopoint-sre}, and other diagnostics. Because their QMC evaluation typically follows by
combining the entropy estimators or generalized partition-function ratios
introduced above, we do not discuss each construction separately and instead
refer the reader to the cited works for technical details and representative
applications. 

% ---------------------------------
%   NEW SECTION
% ---------------------------------
\subsection{Strong-to-weak spontaneous symmetry breaking (SWSSB)}
\label{sec:swssb}

As discussed in Sec.~\ref{sec:z2-symmetric-chnn}, strongly symmetric
decohered states provide a natural setting for investigating mixed-state phases
and phase transitions in open quantum systems. A representative phenomenon is
strong-to-weak spontaneous symmetry breaking
(SWSSB)~\cite{wang2026swssb-review,leejy2023weakmeasurement,Lessa2025swssb}.

A useful diagnostic of SWSSB is the Rényi-$2$ correlator, which was recently evaluated using QMC in Ref.~\cite{ding2026swssb}. Consider a transverse-field Ising model $H=-J\sum_{\langle ij\rangle}\sigma_i^z\sigma_j^z - \sum_i \sigma_i^x$,
with a global $\mathbb{Z}_2$ symmetry generated by $X
    :=
    \prod_{i=1}^{N}\sigma_i^x$.
    
For a strongly $\mathbb{Z}_2$-symmetric mixed state $\tilde\rho$, the Rényi-$2$
order correlator is defined as
\begin{equation}
    C^{(2)}
    =
    \lim_{|i-j|\to\infty}
    \frac{
        \Tr\!\left(
            \tilde\rho\,\sigma_i^z\sigma_j^z\,
            \tilde\rho\,\sigma_i^z\sigma_j^z
        \right)
    }{
        \Tr(\tilde\rho^2)
    } .
    \label{eq:renyi2_order_correlator}
\end{equation}
The state exhibits Rényi-$2$ SWSSB when the Rényi-$2$ correlator remains
finite at long distances, $C^{(2)} \neq 0$,
while the ordinary linear correlator vanishes,
\begin{equation}
    C^{(0)}
    =
    \lim_{|i-j|\to\infty}
    \frac{\Tr\!\left(
        \tilde\rho\,\sigma_i^z\sigma_j^z
    \right)}{\Tr(\tilde\rho)}
    =
    0 .
\end{equation}
Thus, conventional long-range order is absent at the level of linear
expectation values, whereas it survives in the nonlinear Rényi-$2$ diagnostic.

When $\tilde\rho$ admits a sign-problem-free QMC representation in the local
$\sigma^z$ basis, as occurs for the strongly symmetric states generated by the
decoherence channel in Eq.~\eqref{eq:z2-full-channel}, the operator insertions
$\sigma_i^z\sigma_j^z$ are diagonal in the sampling basis. Consequently, the
Rényi-$2$ correlator in Eq.~\eqref{eq:renyi2_order_correlator} can be evaluated
as a diagonal observable in the replicated ensemble, enabling an efficient QMC
investigation of SWSSB. 

\begin{figure}[ht!]
    \centering
    \includegraphics[width=0.8\linewidth]{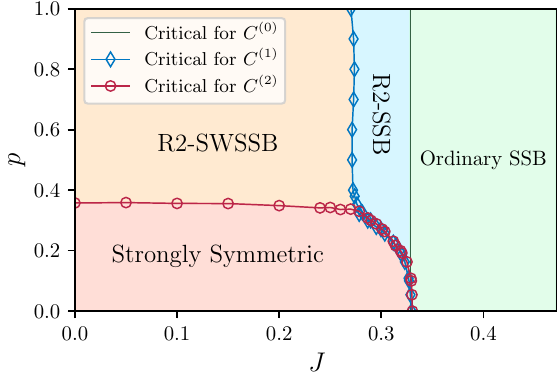}
    \caption{
        Phase diagram of the decohered ground state of the $(2+1)$D
        transverse-field Ising model under the quantum channel $\mathcal{E}$.
        The markers denote critical points extracted from QMC simulations using
        the corresponding order-parameter diagnostics. Further details,
        including the definition of the correlator $C^{(1)}$, can be found in
        Ref.~\cite{ding2026swssb}.
    }
    \label{fig:swssb-phases}
\end{figure}

Fig.~\ref{fig:swssb-phases} shows the phase diagram of the decohered ground
state of the $(2+1)$D transverse-field Ising model under the decoherence
channel in Eq.~\eqref{eq:z2-full-channel}, as obtained from QMC simulations.

% ---------------------------------
%   NEW SECTION
% ---------------------------------
\subsection{Reduced-density-matrix (RDM) data}\label{sec:rdm-data-app}
As introduced in Sec.~\ref{sec:sampling-rdm-data}, one can directly sample and reconstruct the RDM by simulating the open-ensemble representation in Eq.~\eqref{eq:open-ensemble} and tracing out only subsystem $B$, as described in Sec.~\ref{sec:rdm}. This closes the bra and ket indices of subsystem $B$ while leaving those of subsystem $A$ open. Direct access to the resulting RDM enables the evaluation of a broad range of information-theoretic diagnostics. Although the size of subsystem $A$ is restricted by the exponentially growing memory required to store its RDM, the full system remains amenable to polynomial-cost QMC simulation. Consequently, subsystem $B$, and hence the total system size, is not subject to the same exponential memory limitation.

% ---------------------------------
%   NEW SECTION
% ---------------------------------
\subsubsection{Entanglement spectrum}
Denote the normalized RDM of subsystem $A$ reconstructed from the QMC samples by $\rho_{A,\mathrm{MC}}$. In the limit of a sufficiently large number of samples, $\rho_{A,\mathrm{MC}}$ converges to
the exact normalized RDM $\rho_A$ within statistical and numerical precision.
Once the RDM is available, both the Rényi entropies and the von Neumann entanglement entropy can be computed directly from its eigenvalues. Moreover, direct access to the RDM allows one to extract the full entanglement spectrum.

Entanglement entropies provide compact scalar measures of quantum entanglement, but necessarily discard much of the information contained in the RDM by compressing its full eigenvalue distribution into a
single number. Consequently, distinct many-body states, including states belonging to different quantum phases, can in principle exhibit identical or very similar entanglement entropies. Entanglement entropy alone may therefore
be insufficient to distinguish, for example, a $(1+1)$D trivial insulator from a symmetry-protected topological (SPT) phase~\cite{Pollmann2010ee,Pollmann2012ee}.

The entanglement spectrum, by contrast, retains substantially more information.
Diagonalizing the RDM $\rho_A$ yields its eigenvalues $\{\lambda_\alpha\}$. The corresponding entanglement energies are defined as
$\xi_\alpha = -\ln \lambda_\alpha$,
and the collection $\{\xi_\alpha\}$ constitutes the entanglement spectrum.
Equivalently, one may define the \emph{entanglement Hamiltonian} of subsystem $A$ as
\begin{equation}
    H_A
    = -\ln \rho_{A} ,
\end{equation}
whose eigenvalues are precisely the entanglement energies $\{\xi_\alpha\}$.

Compared with entanglement entropies, the full entanglement spectrum additionally resolves the level structure, degeneracies, symmetry quantum numbers, and the transformation properties of $\{\lambda_{\alpha}\}$~\cite{Calabrese2008ee,Li2008li-Haldane}.

% \cite{Yoshino2021ee,Jafari2021ee}

A particularly influential connection between the entanglement spectrum and boundary physics is the Li--Haldane-Poilblanc conjecture~\cite{Li2008li-Haldane,Poilblanc2010ee}. 
In their original study of fractional quantum Hall states, Li and Haldane observed that the low-lying levels of the
orbital entanglement spectrum exhibit the characteristic level counting predicted by the CFT describing the physical edge excitations \cite{Li2008li-Haldane}. Poilblanc subsequently demonstrated an
analogous correspondence in gapped quantum Heisenberg ladders, where the low-lying entanglement spectrum closely reflects the gapless spectrum associated with the effective, or ``virtual,'' edge created by the entanglement cut \cite{Poilblanc2010ee}.

Subsequent studies further developed this bulk--edge correspondence and showed that a universal low-lying sector can, under appropriate conditions, be separated from higher, nonuniversal entanglement levels by an entanglement gap
\cite{Lauchli2010ee-fqhe,Thomale2010ee,Sterdyniak2011ee,Chandran2011bulk-edge,Dubail2012ee-fqhe,Qi2012edge-ee,Swingle2012proof-ee}.
Closely related correspondences have also been established for free-fermion topological insulators and superconductors, disordered Chern insulators, quantum spin ladders, tensor-network representations of $(2+1)$D
quantum states, and topological critical free-fermion systems
\cite{Fidkowski2020ee-topo-ins,
Prodan2010chern-Insulator-ee,
Alexandradinata-ee-topo-insu,Pollmann2010ee,Cirac2011ee-boundary,Guo2026lie-Haldane}.

\begin{figure}[ht!]
    \centering
    \includegraphics[width=0.85\linewidth]{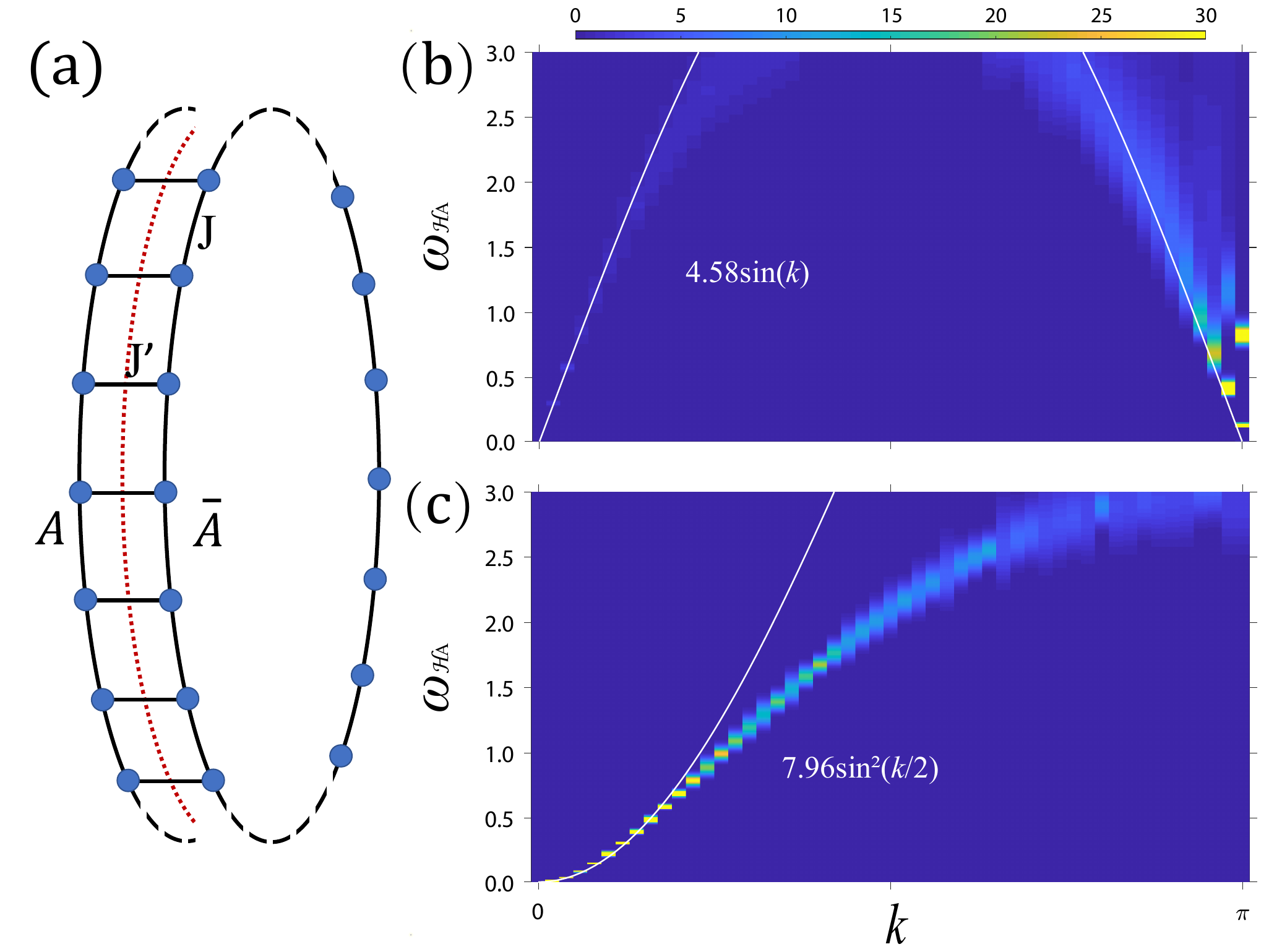}
    \caption{
       {ES of Heisenberg ladder.} (a) Heisenberg spin ladder. The red dashed line cut it into two entangled constituents, the system $A$ and the environment $\overline{A}$. (b) The low-lying ES with $L=100$, $J'=1.732$, $J=1$ and $\beta=100$, $\beta_A=200$. The white line is fitting to the data with the dispersion $4.58 \sin(k)$. (c) The low-lying ES with $L=100$, $J'=1.732$, $J=-1$ and $\beta=100$, $\beta_A=800$. The white line is fitting to the data with the dispersion $7.96\sin^2(k/2)$.
        Figure from Ref.~\cite{Yan2023ee-wormhole}.
    }
    \label{fig:esf}
\end{figure}

An earlier replica-QMC approach reconstructed the leading,
particle-number-resolved levels of the entanglement spectrum from measured moments of the RDM, with the eigenvalues obtained by solving a polynomial root-finding problem~\cite{Chung2014entanglement-spectrum}.

Going beyond the reconstruction of a few leading eigenvalues, Ref.~\cite{Yan2023ee-wormhole} developed a multi-replica QMC approach for bosonic and spin systems that extracts the low-lying entanglement spectral function through stochastic
analytic continuation of imaginary-time correlation functions~\cite{Shao2023sac}. This approach provides spectral information indirectly, through the reconstruction of a dynamical response associated with the entanglement Hamiltonian. The key idea is to simulate $\propto\Tr(\rho_A^n)=\Tr(e^{-n H_A})$, where the power $n$ can be treated as an effective inverse temperature $\beta_A$ for the entanglement Hamiltonian $H_A$, so that the dynamics of $H_A$ is contained in the effective imaginary-time correlators.
As Fig. \ref{fig:esf} shows, a super long entanglement cut can be calculated but the result is limited to the spectral function of the entanglement Hamiltonian instead of a full spectrum. The advantage of this method is that large systems can be simulated, meanwhile, the disadvantage is the rough precision due to the analytical continuation and integer imaginary-time. 
A method based on a similar idea has also been proposed for fermionic systems~\cite{assaad2014entanglement}.

By contrast, Ref.~\cite{mao2025samplingrdm} introduced the
RDM sampling approach discussed in this section. The method
directly reconstructs $\rho_{A,\mathrm{MC}}$, whose diagonalization
resolves the fine, individual levels of the entanglement spectrum. 

A representative application is the N\'eel-ordered phase of the $(2+1)$D square-lattice antiferromagnetic Heisenberg model. More generally, for a finite $d$D quantum $O(N)$ system whose thermodynamic limit spontaneously breaks the continuous $O(N)$ symmetry, the low-lying physical energy spectrum contains an Anderson tower of states (TOS)~\cite{Anderson1952tos,Bernu1992tos-neel},
with
\begin{equation}
    E_S(L)-E_0(L)
    =
    \frac{S(S+N-2)}{2\chi_{\perp}L^d},
    \label{eq:energy-tos}
\end{equation}
where $S$ labels the total $O(N)$ angular-momentum quantum number and $\chi_{\perp}$ is the transverse susceptibility in the thermodynamic limit. For the $(2+1)$D Heisenberg model, for which $N=3$ and $d=2$, Eq.~\eqref{eq:energy-tos} reduces to
\begin{equation}
    E_S(L)-E_0(L)
    =
    \frac{S(S+1)}{2\chi_{\perp}L^2}.
\end{equation}

An analogous TOS structure is expected to appear in the low-lying
entanglement spectrum of ground states exhibiting SSB of a
continuous symmetry
\cite{Alba2013ee,Kolley2013ee-tos}. 

\begin{figure}[ht!]
    \centering
    \includegraphics[width=\linewidth]{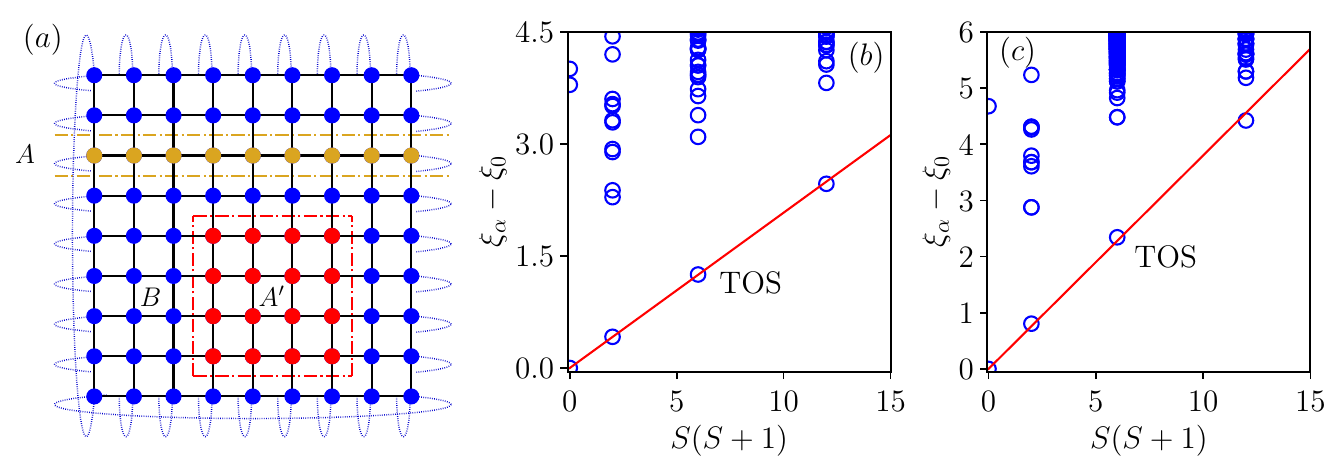}
    \caption{
        Tower-of-states structure in the entanglement spectrum of the
        square-lattice antiferromagnetic Heisenberg model.
        (a) A $20\times20$ lattice with a ring-shaped subsystem $A$
        (yellow) and a $4\times4$ block $A'$ (red).
        (b, c) Entanglement spectra for $A$ and $A'$, respectively, plotted
        against $S(S+1)$. The red lines connect the lowest TOS levels.
        All data were obtained in the total-$S^z=0$ sector.
        Figure from Ref.~\cite{mao2025samplingrdm}.
    }
    \label{fig:tos}
\end{figure}

As shown in Fig.~\ref{fig:tos}, the directly reconstructed entanglement spectrum exhibits this TOS organization for both a ring-shaped subsystem and a square-block subsystem. This example illustrates how direct sampling of the RDM resolves the individual low-lying levels and their multiplet structure, which would not be accessible from a scalar entanglement entropy alone.

A further application of this approach was demonstrated in
Ref.~\cite{Mao2026rdm-tos}, where the tower-of-states structure in the entanglement spectrum of a small subsystem was used to identify the underlying continuous symmetry of a quantum critical point. In particular, this provides
a direct probe of emergent continuous symmetries without requiring prior knowledge of the corresponding order parameters or low-energy effective field theory.

% ---------------------------------
%   NEW SECTION
% ---------------------------------
\subsubsection{Few-site correlation functions, multipartite entanglement, and long-range quantum magic}

Once the RDM of a subsystem has been reconstructed, it can be used to evaluate any equal-time observable supported within that subsystem. In particular, if $A=A_1\cup A_2$ consists of two possibly separated groups of sites, then for arbitrary operators
$O_{A_1}$ and $O_{A_2}$ supported on the respective regions,
\begin{equation}
    \langle O_{A_1}O_{A_2}\rangle
    =
    \Tr\left[
        \rho_{A,\mathrm{MC}}
        \left(O_{A_1}\otimes O_{A_2}\right)
    \right].
\end{equation}
Thus, a single reconstructed RDM contains all equal-time
correlation functions between operators supported on $A_1$ and $A_2$, rather
than only a predetermined set of conventional observables, as exploited in Ref.~\cite{Aditya2026rdm}.
More generally, this approach provides a systematic way to measure off-diagonal operators in QMC, rather than relying on the extraction of Green's functions within a worm-update scheme~\cite{Dorneich2001accessing}.

Direct access to few-site RDMs also enables the study of multipartite entanglement. 
In Ref.~\cite{WangTT2025rdm}, they used reconstructed two- and three-site density matrices to investigate bipartite negativity and genuine tripartite entanglement. This program was subsequently extended to RDMs containing up to four sites, allowing three- and four-party genuine multipartite entanglement to be examined across
the quantum Ising models in one, two, and three spatial dimensions~\cite{Lyu2025rdm}. 

Beyond entanglement, the same framework provides access to nonlinear mixed-state measures of quantum magic or nonstabilizerness. In Ref.~\cite{Hari2025rdm-rom}, RDMs of spatially separated partitions were combined with numerical computations of the robustness of magic~\cite{Howard2017rom} to quantify the long-range quantum magic. At the quantum critical point of the 1D quantum Ising model, this quantity exhibits an algebraic decay with the separation between the partitions, revealing long-range magic correlations.

\subsubsection{Nonequal-Time Observables and the Rényi-1 Correlator for SWSSB in the RDM-QMC}
\label{sec:unequal-time-renyi1}

As discussed in Sec.~\ref{sec:sampling-rdm-data}, the reduced-density-matrix sampling framework can be generalized to operator-inserted RDMs, as introduced in Sec.~\ref{sec:rho-insert}. 
The key idea is to insert the measured operator at a specific imaginary-time slice to construct a generalized RDMs, from which imaginary-time observables can be extracted.
This generalized sampling
scheme was implemented in Ref.~\cite{wangzy2026rdm}, providing QMC access to both equal- and nonequal-imaginary-time observables, including off-diagonal
correlation functions and, through analytic continuation~\cite{Shao2023sac}, dynamical spectral functions.

An important application in many-body quantum information is the
R\'enyi-1, or Wightman, correlator
\cite{Weinstein2025Renyi1,Liu2025renyi1} to Gibbs states,
\begin{equation}
    R_1(i,j)
    =
    \frac{\Tr\!\left(
        \sqrt{\tilde\rho}\,
        O_{ij}\,
        \sqrt{\tilde\rho}\,
        O_{ij}^{\dagger}
    \right)}{\Tr(\tilde\rho)},
    \label{eq:renyi1}
\end{equation}
where $O_{ij}$ is a symmetry-neutral bilocal operator constructed from local
operators carrying nontrivial symmetry charge, for example
$O_{ij}=O_i O_j^{\dagger}$. 

\onecolumngrid

\refstepcounter{figure}
\begin{center}
    \includegraphics[width=0.9\textwidth]{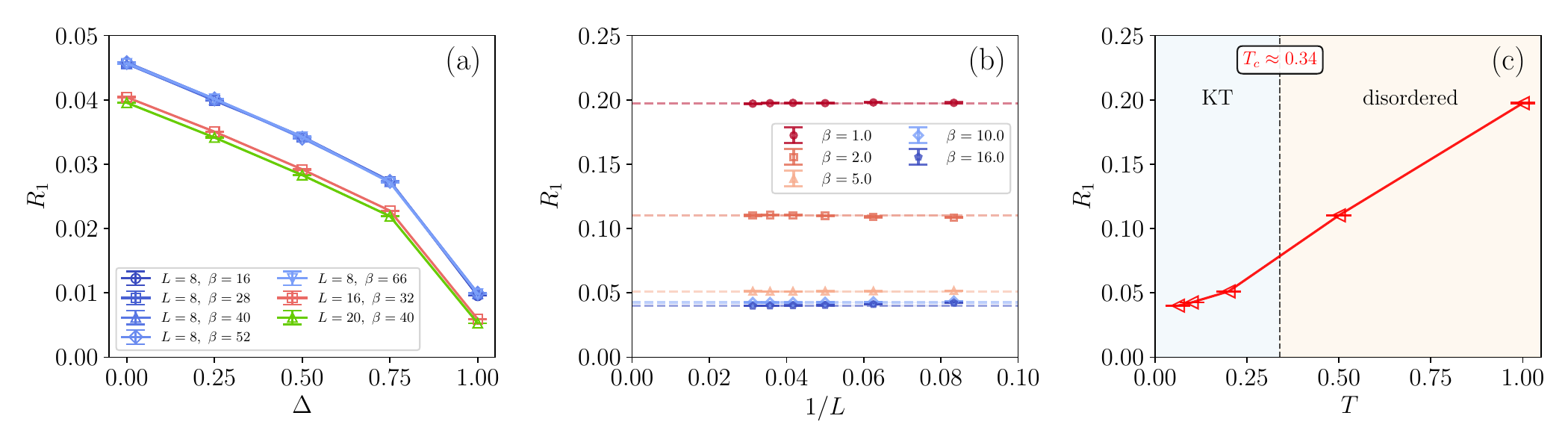}
\end{center}
\noindent {\small{FIG.~\thefigure.} R\'enyi-1 correlator $R_1(r=L/2)$ in the $2$D XXZ model.
(a) $R_1$ as a function of the anisotropy $\Delta$ for square lattices with $L=8, 16, 20$. In the ground-state limit ($\beta \gtrsim 2L$), $R_1$ maintains a finite value throughout the parameter space.
(b) Finite-size scaling of $R_1$ at the pure XY point ($\Delta=0$) across various temperatures. The correlator exhibits a distinct plateau in the large-$L$ regime.
(c) The dashed vertical line marks the Berezinskii-Kosterlitz-Thouless transition temperature $T_c \approx 0.34$. The signal evolves smoothly across $T_c$, and no behavioral changes are observed on either side of the phase transition. All data were computed using the generalized RDM framework. Figure from Ref.~\cite{wangzy2026rdm}.}
\label{fig:r1tot}
\vspace{3.0em}
\twocolumngrid

For a Gibbs state, $\tilde\rho_{\beta}\propto e^{-\beta H}$,
Eq.~\eqref{eq:renyi1} can be written as an nonequal-imaginary-time correlation
function,
\begin{equation}
    R_1(i,j)
    =
    \left\langle
        O_{ij}(\beta/2)
        O_{ij}^{\dagger}(0)
    \right\rangle_\beta .
\end{equation}
It can therefore be evaluated through an operator insertion at
$\tau=\beta/2$, without explicitly constructing $\sqrt{\tilde\rho_\beta}$. In an
appropriate fixed symmetry-charge sector, long-range order in $R_1(i,j)$ can
diagnose strong-to-weak spontaneous symmetry breaking even when the
corresponding conventional two-point correlation function remains
short-ranged. The authors took the 2D quantum XXZ model as an example to demonstrate the SWSSB, as shown in Fig. \ref{fig:r1tot}.
\section{Summary}\label{sec:summary}

In this review, we have presented a unified perspective on how, over nearly two decades, QMC methods have evolved into a general framework for studying many-body quantum information beyond conventional observables.
The organizing principle, developed in Sec.~\ref{sec:qmc_form}, is that the scope of QMC is determined less by the Hamiltonian itself than by the class of density matrices---and generalized density-matrix-like objects---that admit sign-problem-free configuration-space representations. From this perspective, conventional Gibbs-state QMC and ground-state projector QMC emerge as two standard representatives of a much broader family of constructions, including reduced density matrices, replicated states, partially transposed replicated states, operator-inserted density matrices, and states evolving under quantum channels. These seemingly different constructions can be described within a common language of boundary connectivity: traces, partial traces, cyclic replica gluings, and channel-induced constraints correspond to different ways of identifying, closing, or constraining the open transfer-direction boundary indices of a density matrix.

This framework further clarifies how an appropriate choice of basis can transform an otherwise difficult nonlinear quantity into an ordinary QMC observable. In the valence-bond basis and the Bell basis discussed in Sec.~\ref{sec:special_basis}, SWAP operators admit direct estimators, allowing the R\'enyi-2 entropy to be measured without evaluating partition-function ratios. In the Bell basis, moreover, the full Pauli spectrum can be sampled directly, providing access to the characteristic distribution used in
stabilizer-entropy diagnostics of quantum magic.

The second pillar of the modern formulation, reviewed in Sec.~\ref{sec:pfr}, is the toolkit for evaluating generalized partition-function ratios, to which Rényi-type quantities can ultimately be reduced. This toolkit includes extended ensembles with sector-switch updates, reweighting--annealing schemes, thermodynamic integration, derivative estimators, continuous extended ensembles, and nonequilibrium work estimators based on Jarzynski's equality. Together with the incremental trick, which avoids the direct estimation of exponentially small ratios, these techniques can recast certain nonlinear observables into more stable sampling problems. In favorable cases, this can substantially reduce the computational cost and make QMC calculations feasible for system sizes beyond those accessible by direct estimators.

The applications reviewed in Sec.~\ref{sec:nonlinear} demonstrate the broad reach of these developments. QMC now provides high-precision access to entanglement entropies and their universal subleading contributions, including Goldstone-mode logarithmic corrections, central charges, topological entanglement entropy, and entanglement derivatives with characteristic critical scaling. It also enables the study of mixed-state entanglement through Rényi negativities, with quantitative tests of CFT predictions and finite-temperature critical behavior; many-body magic through stabilizer entropies and their connections to BCFT; composite information-theoretic quantities such as mutual information; and SWSSB in decohered quantum states. More recently, direct sampling of reduced density matrices has opened access to the entanglement spectrum and its tower-of-states structure and emergent symmetries, as well as multipartite entanglement, long-range mixed-state magic, and the Rényi-1 correlator. Taken together, these developments substantiate a density-matrix-centered formulation of QMC in which statistical sampling and quantum-information structure are treated on equal footing.

Finally, the developments reviewed here point to a deeper convergence among
classical simulation, quantum information, and many-body physics. Concepts
imported from quantum information---Bell sampling and the stabilizer formalism in particular---have become native tools in QMC, while QMC in turn provides a classical laboratory for information-theoretic diagnostics at system sizes far beyond current quantum devices. We hope that the unified viewpoint presented in this review will serve as a useful guide and entry point for readers from the QMC, quantum-information, many-body, and condensed-matter communities, and will stimulate further developments at their interfaces.

\section*{Acknowledgements}
We are grateful to Zhe Wang, Yanzhang Zhu, and Yan Liu for carefully reading the manuscript and providing valuable comments and suggestions.
We also thank our collaborators Zhiyan Wang, Zenan Liu, Xu Tian, Poetri Sonya Tarabunga, Jiarui Zhao, Gaopei Pan, and Weilun Jiang for their contributions to the related algorithmic works reviewed here and for many insightful discussions throughout these projects.
We further thank Zi Yang Meng, Yan-Cheng Wang, Sylvain Capponi, Fabien Alet, Wenan Guo and Youjin Deng for helpful discussions on related algorithms.

\bibliography{
    bibs/reviews,
    bibs/qmc,
    bibs/sign,
    bibs/entropy,
    bibs/negativity,
    bibs/magic,
    bibs/qinfo,
    bibs/freeenergy,
    bibs/decoherence,
    bibs/rdm,
    bibs/tn,
    bibs/correlations,
    bibs/ReAn, 
    bibs/others,
    bibs/swssb,
    bibs/ee,
    bibs/composite
}

\end{document}